\pdfoutput=1
\documentclass[a4paper,11pt]{article}

\usepackage{jcappub} 

\usepackage[T1]{fontenc} 
\usepackage{natbib}
\usepackage{amsmath,amssymb,bm}
\usepackage{enumitem,units,multirow}
\usepackage{hyperref}
\usepackage{color}

\usepackage{graphicx}
\usepackage{rotating}
\usepackage{epsfig}
\usepackage{bm}
\usepackage{latexsym,amssymb,amsmath,amsfonts,amssymb,txfonts,pxfonts,wasysym,float}
\usepackage{multirow}
\usepackage{enumitem}
\usepackage{color}
\usepackage{multirow}
\usepackage{array}
\newcolumntype{C}[1]{>{\centering\let\newline\\\arraybackslash\hspace{0pt}}m{#1}}
\newcommand{\beq}[1]{\begin{equation}\label{#1}}
\newcommand{\eeq}{\end{equation}}
\newcommand{\bea}[1]{\begin{eqnarray} \label{#1}}
\newcommand{\eea}{\end{eqnarray}}
\newcommand{\ba}{\begin{array}}
\newcommand{\ea}{\end{array}}

\def\be{\begin{equation}}
\def\ee{\end{equation}}
\def\gs{\mathrel{
   \rlap{\raise 0.511ex \hbox{$>$}}{\lower 0.511ex \hbox{$\sim$}}}}
\def\ls{\mathrel{
   \rlap{\raise 0.511ex \hbox{$<$}}{\lower 0.511ex \hbox{$\sim$}}}}

\newcommand{\taon}{$\tau$-lepton }

\newcommand{\nutau}{$\nu_{\tau}$}

\newcommand{\postscript}[2]{\setlength{\epsfxsize}{#2\hsize}
\centerline{\epsfbox{#1}}}

\newcommand{\comment}[1]{}

\newcommand{\tauon}{$\tau$-lepton }
\newcommand{\tauons}{$\tau$-leptons }
\newcommand{\xmax}{\ensuremath{X_{\rm max}}}

\newcommand{\lsim}{\mathrel{\hbox{\rlap{\lower.75ex \hbox{$\sim$}} \kern-.3em \raise.4ex \hbox{$<$}}}}
\newcommand{\gsim}{\mathrel{\hbox{\rlap{\lower.75ex \hbox{$\sim$}} \kern-.3em \raise.4ex \hbox{$>$}}}}

\usepackage[usenames,dvipsnames]{xcolor}
\definecolor{orange}{cmyk}{0,0.5,1,0}
\definecolor{rossoCP3}{cmyk}{0,.88,.77,.40}
\definecolor{graa}{rgb}{0.8,0.8,0.8}
\definecolor{blaa}{rgb}{0.2,0.2,0.6}
\definecolor{bluexx}{rgb}{0.23,0.77,0.82}

\definecolor{pink1}{RGB}{226, 24, 166}

\begin{document}
\title{The POEMMA  (Probe of Extreme Multi-Messenger Astrophysics) Observatory}

\author[1]{A.~V.~Olinto,\note{Corresponding author.}}
\author[2,3]{J.~Krizmanic,}
\author[4]{J.~H.~Adams,}
\author[5]{R.~Aloisio,}
\author[6]{L.~A.~Anchordoqui,}
\author[7,8]{A.~Anzalone,}
\author[9]{M.~Bagheri,}
\author[10]{D.~Barghini,}
\author[10]{M.~Battisti,}
\author[11]{D.~R.~Bergman,}
\author[10]{M.~E.~Bertaina,}
\author[12]{P.~F.~Bertone,}
\author[13]{F.~Bisconti,}
\author[14]{M.~Bustamante,}
\author[15]{F.~Cafagna,}
\author[16,8]{R.~Caruso,}
\author[17,18]{M.~Casolino,}
\author[19]{K.~\v{C}ern\'{y},}
\author[12]{M.~J.~Christl,}
\author[5]{A.~L.~Cummings,}
\author[5]{I.~De~Mitri,}
\author[1]{R.~Diesing,}
\author[20]{R.~Engel,}
\author[1]{J.~Eser,}
\author[21]{K.~Fang,}
\author[10]{F.~Fenu,}
\author[22]{G.~Filippatos,}
\author[9]{E.~Gazda,}
\author[23]{C.~Guepin,}
\author[20]{A.~Haungs,}
\author[2]{E.~A.~Hays,}
\author[24]{E.~G.~Judd,}
\author[25]{P.~Klimov,}
\author[22]{V.~Kungel,}
\author[4]{E.~Kuznetsov,}
\author[26]{\v{S}.~Mackovjak,}
\author[27]{D.~Mand\'{a}t,}
\author[18]{L.~Marcelli,}
\author[2]{J.~McEnery,}
\author[28]{G.~Medina-Tanco,}
\author[22]{K.-D.~Merenda,}
\author[1]{S.~S.~Meyer,}
\author[2]{J.~W.~Mitchell,}
\author[10]{H.~Miyamoto,}
\author[29]{J.~M.~Nachtman,}
\author[30]{A.~Neronov,}
\author[31]{F.~Oikonomou,}
\author[29]{Y.~Onel,}
\author[32]{G.~Osteria,}
\author[9]{A.~N.~Otte,}
\author[33]{E.~Parizot,}
\author[6]{T.~Paul,}
\author[27]{M.~Pech,}
\author[2]{J.~S.~Perkins,}
\author[18,34]{P.~Picozza,}
\author[35]{L.W.~Piotrowski,}
\author[10]{Z.~Plebaniak,}
\author[33]{G.~Pr\'ev\^ot,}
\author[4]{P.~Reardon,}
\author[29]{M.~H.~Reno,}
\author[36]{M.~Ricci,}
\author[9]{O.~Romero Matamala,}
\author[22]{F.~Sarazin,}
\author[27]{P.~Schov\'{a}nek,}
\author[32,37]{V.~Scotti,}
\author[38]{K.~Shinozaki,}
\author[6]{J.~F.~Soriano,}
\author[2]{F.~Stecker,}
\author[17]{Y.~Takizawa,}
\author[20]{R.~Ulrich,}
\author[20]{M.~Unger,}
\author[2]{T.~M.~Venters,}
\author[22]{L.~Wiencke,}
\author[29]{D.~Winn,}
\author[12]{R.~M.~Young,}
\author[25]{M.~Zotov}

\affiliation[1]{The University of Chicago, Chicago, IL, USA}
\affiliation[2]{NASA Goddard Space Flight Center, Greenbelt, MD, USA}
\affiliation[3]{Center for Space Science \& Technology, University of Maryland, Baltimore County, Baltimore, MD, USA}
\affiliation[4]{University of Alabama in Huntsville, Huntsville, AL, USA}
\affiliation[5]{Gran Sasso Science Institute, L'Aquila, Italy}
\affiliation[6]{City University of New York, Lehman College, NY, USA}
\affiliation[7]{Istituto Nazionale di Astrofisica INAF-IASF, Palermo, Italy}
\affiliation[8]{Istituto Nazionale di Fisica Nucleare, Catania, Italy}
\affiliation[9]{Georgia Institute of Technology, Atlanta, GA, USA}
\affiliation[10]{Universita' di Torino, Torino, Italy}
\affiliation[11]{University of Utah, Salt Lake City, Utah, USA}
\affiliation[12]{NASA Marshall Space Flight Center, Huntsville, AL, USA}
\affiliation[13]{Istituto Nazionale di Fisica Nucleare, Turin, Italy}
\affiliation[14]{Niels Bohr Institute, University of Copenhagen, DK-2100 Copenhagen, Denmark}
\affiliation[15]{Istituto Nazionale di Fisica Nucleare, Bari, Italy}
\affiliation[16]{Universita' di Catania, Catania Italy}
\affiliation[17]{RIKEN, Wako, Japan}
\affiliation[18]{Istituto Nazionale di Fisica Nucleare, Section of Roma Tor Vergata, Italy}
\affiliation[19]{Joint Laboratory of Optics, Faculty of Science, Palack\'{y} University, Olomouc, Czech Republic}
\affiliation[20]{Karlsruhe Institute of Technology, Karlsruhe, Germany}
\affiliation[21]{Kavli Institute for Particle Astrophysics and Cosmology, Stanford University, Stanford, CA 94305, USA}
\affiliation[22]{Colorado School of Mines, Golden, CO, USA}
\affiliation[23]{Department of Astronomy, University of Maryland, College Park, MD, USA}
\affiliation[24]{Space Sciences Laboratory, University of California, Berkeley, CA, USA}
\affiliation[25]{Skobeltsyn Institute of Nuclear Physics, Lomonosov Moscow State University, Moscow, Russia}
\affiliation[26]{Institute of Experimental Physics, Slovak Academy of Sciences, Kosice, Slovakia}
\affiliation[27]{Institute of Physics of the Czech Academy of Sciences, Prague, Czech Republic}
\affiliation[28]{Instituto de Ciencias Nucleares, UNAM, CDMX, Mexico}
\affiliation[29]{University of Iowa, Iowa City, IA, USA}
\affiliation[30]{University of Geneva, Geneva, Switzerland}
\affiliation[31]{Institutt for fysikk, NTNU, Trondheim, Norway}
\affiliation[32]{Istituto Nazionale di Fisica Nucleare, Napoli, Italy}
\affiliation[33]{Universit\'e de Paris, CNRS, Astroparticule et Cosmologie, F-75013 Paris, France}
\affiliation[34]{Universita di Roma Tor Vergata, Italy}
\affiliation[35]{Faculty of Physics, University of Warsaw, Warsaw, Poland}
\affiliation[36]{Istituto Nazionale di Fisica Nucleare - Laboratori Nazionali di Frascati, Frascati, Italy}
\affiliation[37]{Universita' di Napoli Federico II, Napoli, Italy}
\affiliation[38]{National Centre for Nuclear Research, Lodz, Poland}

\abstract{The Probe Of Extreme Multi-Messenger Astrophysics (POEMMA) is designed to accurately observe ultra-high-energy cosmic rays (UHECRs) and cosmic neutrinos from space with sensitivity over the full celestial sky. POEMMA will observe the air fluorescence produced by extensive air showers (EASs) from UHECRs and potentially UHE neutrinos above 20 EeV. Additionally,  POEMMA has the ability to observe the Cherenkov signal from upward-moving EASs induced by Earth-interacting tau neutrinos above 20 PeV. The POEMMA spacecraft are designed to quickly re-orientate to follow up transient neutrino sources and obtain currently  unparalleled neutrino flux sensitivity. Developed as a NASA Astrophysics Probe-class mission,  POEMMA consists of two identical satellites flying in loose formation in 525 km altitude orbits. Each POEMMA instrument incorporates a wide field-of-view (45$^\circ$) Schmidt telescope with an optical collecting area of over 6 m$^2$. The hybrid focal surface of each telescope includes a fast (1~$\mu$s) near-ultraviolet camera for EAS fluorescence observations and an ultrafast (10~ns) optical camera for Cherenkov EAS observations. In a 5-year mission, POEMMA will provide measurements that open new multi-messenger windows onto the most energetic events in the universe, enabling the study of new astrophysics and particle physics at these extreme energies.}
\arxivnumber{} 

\maketitle
\flushbottom


\section{POEMMA Overview}

The Probe Of Extreme Multi-Messenger Astrophysics (POEMMA) is designed to observe  ultra-high-energy cosmic rays (UHECRs) and cosmic neutrinos using space-based measurements of extensive air showers (EASs) \cite{Olinto:2019mjh,Olinto:2019euf,Krizmanic:2019hiq,Olinto:2020vid}.  POEMMA will monitor colossal volumes of the Earth's atmosphere to observe the development of EASs produced by UHECRs and UHE neutrinos above 20 EeV and observe the Cherenkov signal from cosmic neutrinos above 20 PeV, in particular from astrophysical transient events. In neutrino target-of-opportunity (ToO) operational mode, the POEMMA telescopes quickly slew (90$^\circ$ in 500~s) to observe and then follow transient sources. In ToO mode, POEMMA will utilize an external alert of transient events from gravitational waves, electromagnetic transients, or signals detected by other neutrino observatories. POEMMA will have significant sensitivity over the entire celestial sphere for UHECRs above 20 EeV and cosmic neutrinos above 20 PeV. 

\begin{figure}[tbp]
    \postscript{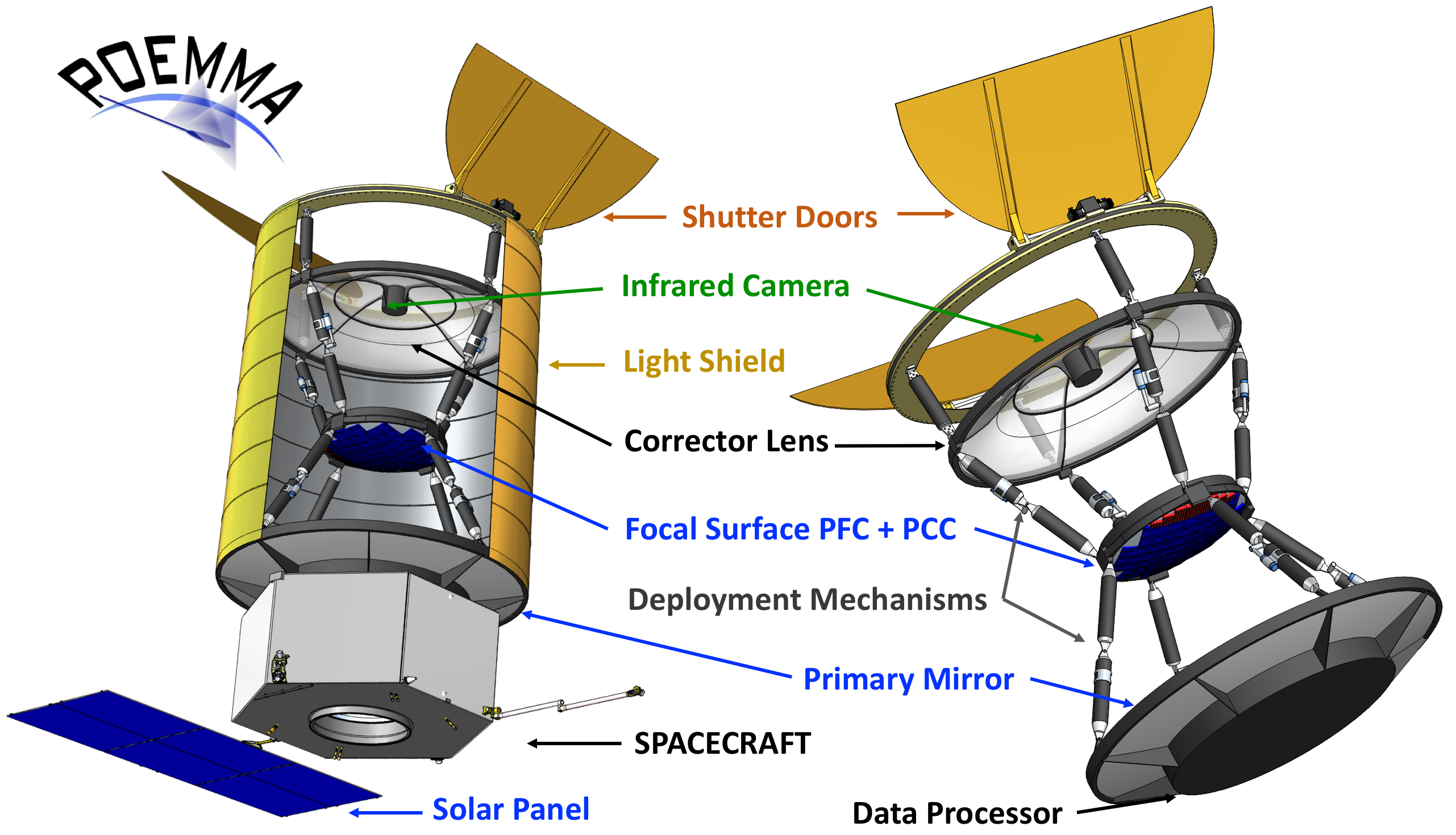}{1}
    \caption{Concept of POEMMA telescope (left) and instrument (right) with major components identified. Adapted from Ref. \cite{Olinto:2019mjh}.}
    \label{fig1}
\end{figure}

The POEMMA conceptual design was developed under the  NASA 2020 Astrophysics Decadal Probe-class mission Studies \cite{POEMMAnasaReport}.
The observatory is comprised of two identical telescopes flying in a loose formation on a  low Earth orbit (LEO) at an altitude of 525~km with a 28.5$^\circ$-inclination. The two spacecraft fly in tandem separated by less than  $\sim$300~km. Each POEMMA telescope is composed of a wide (45$^{\circ}$) field of view (FoV) Schmidt optical system with an optical collecting area of over 6~m$^2$  (see Figure \ref{fig1}) and a spacecraft bus. The focal surface of each telescope has an innovative hybrid design enabling two types of camera capabilities, each optimized for either fluorescence or Cherenkov EAS measurements. The POEMMA Fluorescence Camera (PFC) covers $\sim$80\% of the focal surface and records the near-UV ($300\text{--}500$~nm) signals to measure the waxing and waning of a UHECR EAS development on a 1 $\mu$s time scale. The POEMMA Cherenkov Camera (PCC) is located near the edge of the focal surface and is optimized to observe the ultrafast (10 ns) EAS optical ($300$--$1000$ nm) Cherenkov  signals from  EASs generated by  $\nu_\tau$ interactions in the Earth from below the limb and by cosmic rays from above the limb of the Earth.

Two science observation modes are envisioned for POEMMA: one optimized for stereo UHECR fluorescence observations, and another optimized for Cherenkov emissions from EASs induced by \tauons produced by cosmic $\nu_{\tau}$ interactions in the Earth. The latter observation mode requires a $\sim$45$^\circ$ tilt from the nadir to monitor below the Earth's limb (located at 67.5$^\circ$). In the first phase of the mission, the observatory will be in the quasi-nadir configuration, denoted as {\bf POEMMA-Stereo}, optimized for the fluorescence stereo observation of EASs. The POEMMA instruments are slightly tilted toward one another to monitor a common atmospheric volume between the instruments, which lowers the UHECR measurement energy threshold to $E_{\rm CR} \sim 20$ EeV \cite{Anchordoqui:2019omw}.  In this POEMMA-Stereo science mode, each pixel in the PFC will view $\sim$0.6 km$^2$ on the ground, and the spacecraft altitudes do not change.  Since the spacecraft travel at $\sim$7.6 km/s in their orbits, the atmospheric volume viewed by a PFC pixel will be fully traversed in $\sim$100 ms. 

The second science observation mode is optimized for the detection of transient neutrino events for $E_\nu \gsim 20$ PeV using the EAS Cherenkov signal. This configuration, denoted as {\bf POEMMA-Limb},  will occur once a ToO alert is received and the POEMMA instruments are tilted to follow the transient event near the Earth's limb~\cite{Venters:2019xwi}.  Cherenkov signals from UHECRs ($\lsim 2^\circ$ above the limb) will also be measured for calibration and background measurements.  Concurrently, UHECR fluorescence observations will  continue with an enhanced acceptance above about 100 EeV  and less precise reconstruction capabilities \cite{Anchordoqui:2019omw}. Thus, POEMMA will provide a significant increase in the statistics of observed UHECRs at the highest energies with exposure over the entire sky and will have a target of opportunity (ToO) follow-up program for cosmic neutrinos from extremely energetic transient astrophysical events. 

\medskip
\noindent {\bf POEMMA} is designed to:

$\bullet$ {\bf Discover the nature and origin of the highest-energy particles in the universe.} Where do UHECRs come from?  What are these extreme cosmic accelerators and how do they accelerate to such high energies? What is the UHECR composition at the highest energies? What are the magnetic fields in the galactic and extragalactic media? How do UHECRs interact in the source, in galactic and extragalactic space, and in the atmosphere of the Earth?

$\bullet$ {\bf Discover neutrino emission above 20 PeV associated with extreme astrophysical transients.} What is the high-energy neutrino emission of gravitational wave events? Do binaries with black holes and neutron stars produce high-energy neutrinos when they coalesce? Neutrino observations will elucidate the underlying dynamics of gravitational wave events, gamma-ray bursts, newborn pulsars, tidal disruption events, and other transient events as seen by neutrinos;

$\bullet$ {\bf Probe particle interactions at extreme energies.} POEMMA can test models with physics beyond the Standard Model (BSM) through cosmic neutrino observations from hundreds of PeV to tens of ZeV;

$\bullet$ {\bf Observe Transient Luminous Events} and study the dynamics of the Earth's atmosphere, including extreme storms;

$\bullet$ {\bf Observe meteors}, thereby contributing to the understanding of the dynamics of meteors in the Solar System;

$\bullet$ {\bf Search for exotic particles} such as nuclearites.

\medskip

Here we describe the science impact, the detailed instrument, and the mission design of POEMMA. In \S 2, we discuss the extreme multi-messenger science goals and capabilities of POEMMA. The design of the POEMMA Observatory is described in \S 3 and the mission in \S 4. We end with a technology roadmap in \S5 and summarize in \S 6.

POEMMA will have access to {\bf multi-messenger signals} from the most energetic environments and events in the universe, enabling the study of astrophysics and particle physics at extreme energy scales over the full sky.

\bigskip

\begin{table}
\centering
\caption{POEMMA Observatory Specifications}
\label{tab-1}  \
\begin{tabular}{lllllllll}
\hline
\hline
Telescope: & Instrument &  &$\ \ $& Spacecraft  & \\ \hline
Optics &  Schmidt & 45$^\circ$ full FoV && Slew rate & 90$^\circ$  in 8 min \\
 & Primary Mirror & 4 m diam. && Pointing Res. & 0.1$^\circ$ \\
 & Corrector Lens & 3.3 m diam. && Pointing Know. & 0.01$^\circ$ \\
 & Focal Surface & 1.6 m diam. && Clock synch. & 10 ns \\  
 & Pixel Size & $3 \times 3$ mm$^2$  && Data Storage & 7 days \\ 
& Pixel FoV & 0.084$^\circ$ && Communication & S-band \\
PFC & MAPMT (1$\mu$s)& 126,720 pixels  && Wet Mass & 3,450 kg \\
PCC & SiPM (20 ns)& 15,360 pixels  && Power (w/cont)& 550 W \\ \hline
Observatory & \multicolumn{3}{l}{Each Telescope}  &Mission  & (2 Telescopes)\\ \hline
 & Mass & 1,550 kg  && Lifetime & 3 year  (5 year goal)\\
 & Power (w/cont) & 700 W   && Orbit & 525 km, 28.5$^\circ$ Inc \\
 & Data & $<$ 1 GB/day && Orbit Period & 95 min \\
 & & && Telescope Sep. & $\sim$25 - 1000 km
\\\hline
\hline
\end{tabular}
 \
\center{POEMMA Observatory = Two Telescopes; Each Telescope = Instrument + Spacecraft}

\end{table}

\section{POEMMA Extreme Multi-Messenger Science}

\subsection{Introduction}

The main scientific goals of POEMMA  are to discover the elusive sources of UHECRs
and to observe cosmic neutrinos from multi-messenger transients. POEMMA exploits the tremendous gains in both UHECR and cosmic neutrino exposures offered by space-based measurements, {\it including full-sky coverage of the celestial sphere}.  For cosmic rays with energies $E \gtrsim 20~{\rm EeV}$, POEMMA offers the potential to enable charged-particle astronomy by obtaining definitive measurements of the UHECR spectrum,  composition,
and likely source locations. For multi-messenger transients, POEMMA will follow ToOs to potentially detect the first cosmic neutrino emission with energies $E_{\nu} \gtrsim$ 20 PeV from astrophysical transients. POEMMA also has sensitivity to neutrinos with energies above 20~EeV through fluorescence observations of neutrino-induced EASs. Supplementary science capabilities of POEMMA include probes of physics beyond the Standard Model of particle physics, the study of atmospheric transient luminous events (TLEs), and the search for meteors and nuclearites.

\begin{figure}[ht]
    \postscript{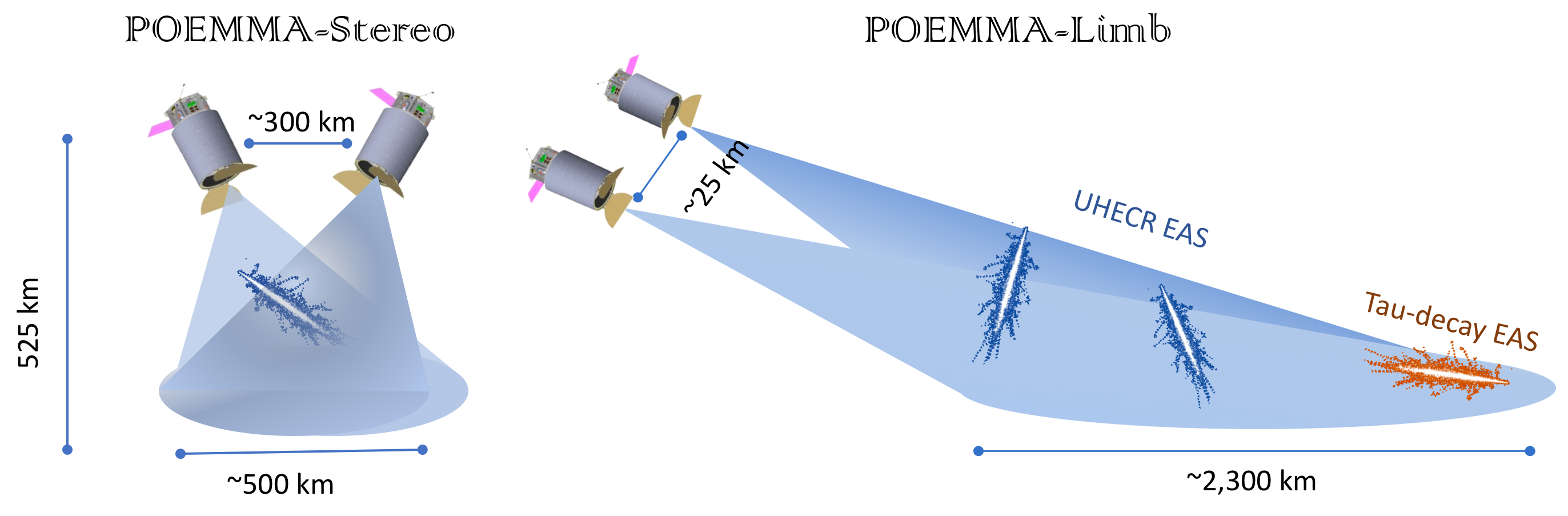}{0.99}
\caption{POEMMA observing modes (see also \S4). {\it Left:} POEMMA-Stereo mode to observe fluorescence from UHE cosmic rays and neutrinos in stereo. (Telescope separation $\sim$300 km and pointing close to nadir for the most precise measurements at 10s of EeV.) {\it Right:} POEMMA-Limb mode to observe Cherenkov from cosmic neutrinos just below the limb of the Earth and fluorescence from UHECRs throughout the volume. (Telescope separation $\sim$25 km and pointing towards rising or setting source for ToO-stereo mode.)
\label{fig2}}
\end{figure}

These groundbreaking measurements are obtained by operating POEMMA's two telescopes (described in Figure~\ref{fig1} and Table~1) in different orientation modes. The first is POEMMA-Stereo, a quasi-nadir  configuration, optimized for stereo fluorescence observations of UHECR  and UHE neutrino (shown in the left panel of Figure~\ref{fig2}). The second is POEMMA-Limb, a tilted  configuration pointed towards the Earth's limb as shown in the right panel of Figure~\ref{fig2}. POEMMA-Limb is designed to simultaneously search for cosmic neutrinos from just below the limb via the EAS Cherenkov signals and for UHECRs via fluorescence in the angular range from below the limb to $\sim$45$^\circ$ from nadir. In addition, observations of cosmic ray Cherenkov signals just above the limb can also be made for calibration and background estimation.

To follow up ToO transient alerts, the observatory is swiftly positioned in POEMMA-Limb mode pointing towards the rising or setting source position  to search for neutrino emission associated with the astrophysical event. If transient neutrino events lasting longer than a day are uncovered, the spacecraft propulsion systems will bring the POEMMA telescopes closer together to observe the ToO source with overlapping instrument light pools, lowering the energy threshold for neutrino detection via the use of time coincidence (denoted {\bf ToO-stereo} configuration). For shorter-duration transients, the two POEMMA telescopes will conduct independent observations of the source in separate light pools. This {\bf ToO-dual} configuration doubles the effective area for observations while increasing the neutrino energy threshold to reduce the night sky air glow background effects. It should be noted that the space-based POEMMA observatory has sensitivity for neutrinos above 20 PeV over the full-sky in a 180-day time span, without the blind spots inherent to ground-based experiments, see Figs. 7 and 8 in Ref. \cite{Venters:2019xwi}.

In the POEMMA-Stereo configuration, the two wide-angle (45$^{\circ}$) Schmidt telescopes with several square meters of effective photon collecting area view a common, immense atmospheric volume corresponding to approximately $10^{4}$ gigatons of atmosphere. This stereo mode yields a factor of $4$--$18$ increase in yearly UHECR exposure compared to that obtainable by current ground observatory arrays and a factor of $40$--$180$ compared to current ground fluorescence observations. In all of the limb-viewing configurations, POEMMA searches for optical Cherenkov signals of upward-moving EASs generated by \tauon decays produced by $\nu_\tau$ interactions in the Earth. The terrestrial neutrino target monitored by POEMMA  reaches nearly $10^{10}$ gigatons. 
In the POEMMA-Limb configuration, an even more extensive volume of the atmosphere is monitored for UHECR fluorescence observations. Thus, POEMMA uses the Earth and its atmosphere as a giant high-energy physics detector and astrophysics observatory, monitoring the atmospheric volume over nearly $2 \times 10^5$ (POEMMA-Stereo) to $1.2 \times 10^6$ km$^2$ (POEMMA-Limb) area which is nearly 70 $--$ 400 larger than that for Auger.

\subsection{UHECR Science}

\begin{figure}[ht]
    \postscript{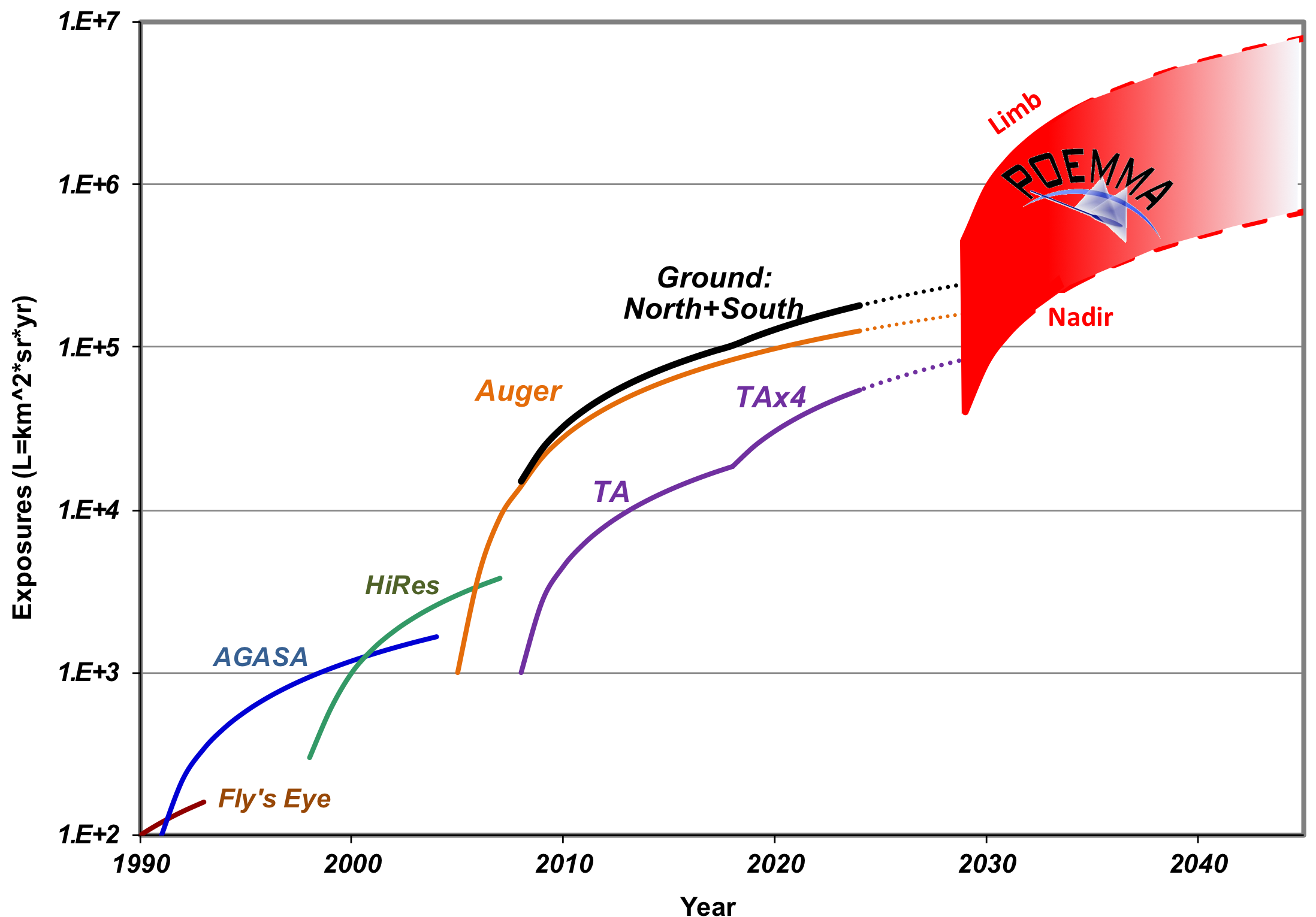}{0.8}
\caption{The range of POEMMA's exposure growth in time as compared
to current ground-based UHECR experiments depending on observation mode: from nadir to limb observations. Dotted lines extrapolate Auger and TAx4 observations to 2030.
\label{fig3}}
\end{figure}

Over a half-century since John Linsley reported the observation of a 10$^{20}$ eV (= 100 EeV) EAS \cite{Linsley1963}, the astrophysical sources of these extremely energetic cosmic rays remain unknown. UHECRs with energies $\ge 100 \ {\rm EeV}$ have energies over 7 orders of magnitude higher than terrestrial accelerators can currently achieve.  
A succession of increasingly large ground-based experiments (Fly's Eye \cite{Bird:1994wp}, AGASA \cite{Takeda:2002at}, and HiRes \cite{AbuZayyad:2002sf}) paved the way for the two leading ground-based observatories currently in operation: the Pierre Auger
Observatory~\cite{Abraham:2010zz,Abraham:2009pm} in the Southern
Hemisphere, with $\sim$80,000~${\rm km^2 \ sr \ yr}$ exposure in 14
years of operation~\cite{Aab:2017njo,Verzi2019}, and the Telescope Array (TA)~\cite{AbuZayyad:2012kk,Tokuno:2012mi} in the Northern Hemisphere, with $\sim$14,000$~{\rm km}^2 \, {\rm sr} \, {\rm yr}$ exposure \cite{Ivanov2019,IvanovPriv}
in 11 years (see Figure \ref{fig3}). Much has been learned with these ground observatories, but the nature of the astrophysical sources of UHECRs remains a mystery~\cite{Kotera:2011cp,Anchordoqui:2018qom,AlvesBatista:2019tlv}.  
Proposed sources include extremely fast-spinning
young pulsars~\cite{Blasi:2000xm,Fang:2012rx,Fang:2013cba},
active galactic nuclei (AGN)~\cite{Biermann:1987ep,Rachen:1992pg,Eichmann:2017iyr},
starburst galaxies (SBGs)~\cite{Anchordoqui:1999cu,Anchordoqui:2018vji,Anchordoqui:2020otc}, and
gamma-ray bursts
(GRBs)~\cite{Waxman:1995vg,Vietri:1995hs}, among others. Some of these models can partially accommodate current Auger and TA observations, but the scarcity of observed events above tens of EeV has hindered a clear identification of the sources. 

The POEMMA observatory will perform precision UHECR measurements in POEMMA-Stereo mode that provide a significant increase in UHECR exposure with all-sky sensitivity starting at 20 EeV. A comprehensive study of the  performance of POEMMA for measurements of UHECR showers is detailed in Ref. \cite{Anchordoqui:2019omw}. POEMMA-Stereo performance can be summarized as follows: energy resolution:  $\le 18\%$ above 50 EeV; angular resolution:  $< 1.5^\circ$ above 40 EeV; and $X_\text{max}$  resolution of $< 40$ g/cm$^2$ above 40 EeV. All of these improve as UHECR energy increases.  At 40 EeV, the yearly exposure of POEMMA-Stereo is more than 4 times higher than that of the Auger ground array, and 18 times higher compared to the TA ground array. Above 100 EeV, the POEMMA gain in exposure increases nearly twofold to a factor of 7 (30) compared to the Auger (TA) ground array and 70 (300) compared to Auger (TA) fluorescence observations. In POEMMA-Limb mode, the yearly exposure can be further increased by  another factor of 6, albeit with less precise UHECR measurements. In Ref. \cite{Anchordoqui:2019omw} we conservatively quantified the POEMMA-limb performance assuming two monocular EAS observations. This yields an energy resolution of $\le 30\%$ above 50 EeV; an angular resolution of $< 10^\circ$ above 40 EeV; and a $ X_\text{max}$ resolution of  $\le 100$ g/cm$^2$, above 40 EeV.
In a 5-year long mission, POEMMA will achieve a UHECR fluorescence exposure that is $4$--$18$ times larger than that for current surface array observations\footnote{as reported at ICRC2019} and over $40$--$300$ times larger than current fluorescence exposure for UHECRs above 100 EeV in the more precise POEMMA-Stereo configuration and about 6 times more exposure in the POEMMA-Limb configuration (see Figure \ref{fig3}). The POEMMA UHECR performance is discussed in \S2.3.

Both the Auger and TA observatories are giant ground arrays of particle detectors (3,000 km$^2$ for Auger and 700 km$^2$ for TA) overseen by fluorescence telescopes (4 telescope sites for Auger and 3 for TA). These detectors observe EASs via the shower particles that hit the ground array detectors at any time of the day and, $\sim$10\% of the time (on dark moonless nights), they observe the faint ultraviolet (UV) fluorescence emission induced by excitation of atmospheric nitrogen by EAS particles. The Auger Observatory is expected to continue observations over the next decade with enhanced detector units for the ground array, an upgrade named AugerPrime \cite{Aab:2016vlz}. TA is now being upgraded to TAx4 which entails an expansion of its ground array four-fold to reach the Auger ground-array scale \cite{Kido:2017nhz}. 

Further in the future, the Giant Radio Array for Neutrino Detection (GRAND) collaboration is developing a ground-based radio array that  will have sensitivity to both UHE cosmic neutrinos and UHECRs \cite{Alvarez-Muniz:2018bhp}. In its largest configuration, GRAND200k, the modeled response to UHECRs above 1 EeV predicts an aperture of 107,000 km$^2$ sr and a five-year exposure of 535,000 km$^2$ sr, which is within the exposure band predicted for POEMMA (see Figure~\ref{fig3}). GRAND200k is anticipated to have sub-degree angular resolution and an $X_{\rm max}$ resolution goal of $\sim$20~g/cm$^2$, leading to similar UHECR composition resolution to that of POEMMA. GRAND200k will be sensitive to UHECRs in the zenith angular range $65^\circ$--$85^\circ$ and will be sensitive to cosmic UHECR sources in the celestial declination band in the approximate range from $-40^\circ$ to $70^\circ$, in contrast to the full-sky coverage for POEMMA. There will be a subset of co-measured UHECRs from the space-based POEMMA EAS fluorescence measurements and ground-based GRAND radio UHECR measurements, which can be used to understand the systematic errors of the two complimentary UHECR measurement techniques.

\begin{figure}[ht]
    \postscript{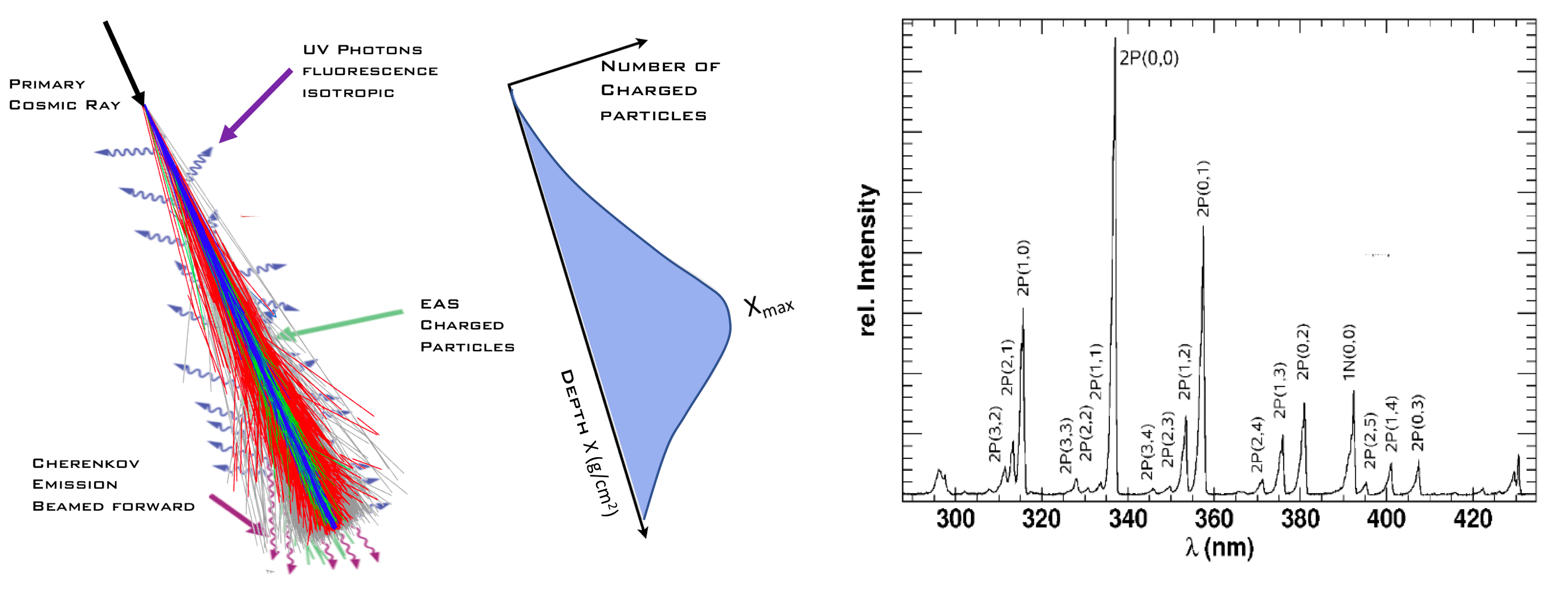}{0.99}
\caption{{\it Left:} Simulation of the development of an extensive air shower, with a schematic of the number of charged particles as a function of depth and shower maximum indicated by $X_{\rm max}$~\cite{Engel,Bietenholz}. {\it Right:} Measurement of the nitrogen fluorescence spectrum of dry air showing the relative intensity of lines in the UV range \cite{Ave:2007xh}.
\label{fig4}}
\end{figure}

Figure \ref{fig4} left shows an example of an EAS  with a schematic development of the number of charged particles with atmospheric depth ($X$ in g/cm$^2$). The EAS particles excite atmospheric nitrogen that fluoresces in the UV isotropically. The fluorescence signal of extremely energetic EASs can be observed hundreds of km  away from the shower axis for space-based instruments. Figure \ref{fig4} right shows the relative intensity of the UV lines in the air nitrogen fluorescence spectrum. In the forward direction of the EAS development, beamed Cherenkov photons are also emitted. POEMMA is designed to observe both the fluorescence UV emission and the Cherenkov emission of EASs.

Auger and TA have measured key features of UHECRs: the energy
spectrum (up to $\sim$ 100~EeV shown in Figures \ref{figSpec} right and \ref{figNumbCover}), the composition (up to $\sim$ 50 EeV shown in Figure \ref{figXmax}), and the sky distribution of their arrival directions. The UHECR spectrum  exhibits an {\it ankle} feature at $\sim$ 5~EeV and a suppression of the UHECR flux above $\sim$
40~EeV~\cite{Abbasi:2007sv,Abraham:2008ru,Abraham:2010mj,AbuZayyad:2012ru}. The suppression is consistent with the predicted  Greisen-Zatsepin-Kuzmin (GZK) effect~\cite{Greisen:1966jv,Zatsepin:1966jv} caused by interactions of UHECRs with the cosmic microwave background as they travel astronomical distances from extragalactic sources to Earth. The same spectral feature may  be produced by the maximum energy of the sources in models that also predict  a change to heavier composition at the highest energies~\cite{Allard:2008gj}.

In addition to spectral and composition measurements, a crucial step in unveiling the origin of UHECRs is the localization of sources in the sky distribution of their arrival directions. Since UHECRs are charged and magnetic fields fill the galactic and extragalactic media, pointing to sources is best achieved at the highest energies (or rigidities). The typical deflection of a UHECR of energy $E$ and charge $Z$ (in units of proton charge) in an extragalactic magnetic field  $B \sim 1~{\rm nG}$~\cite{Pshirkov:2015tua} is  $\delta \theta \approx 1.5^\circ \, Z  \,(10 ~{\rm EeV}/E)$, for a source at 4 Mpc and a magnetic field coherence length of about 100 kpc~\cite{Waxman:1996zn,Farrar:2012gm}. The deflections when crossing the Galaxy can be somewhat larger, $\delta \theta \sim 3^\circ \, Z \, (100 ~{\rm EeV}/E)$, depending on the UHECR arrival trajectory through the Galaxy~\cite{Aartsen:2015dml}.\footnote{It is interesting to note that experiments like POEMMA which experience large exposures beyond $10^{20.3}~{\rm eV}$ could see a directional neutron signal from nearby sources like Cen A. This is because the neutron decay length is $\lambda (E) = 0.9 (E/10^{20}~{\rm eV})~{\rm Mpc}$.  Because of the exponential depletion, about 2\% of the neutrons would survive the trip at $10^{20}~{\rm eV}$, and about 15\% at $10^{20.3}~{\rm eV}$~\cite{Anchordoqui:2001nt}.} These deflections suggest that large statistics of events above tens of EeV are necessary to observe small-scale anisotropies around source positions in the sky, although this depends on the nuclear composition of the UHECRs. 

To date the only high-significance departure from an isotropic sky
distribution of UHECRs is a dipole anisotropy, reported by the Auger
collaboration, above $8~{\rm EeV}$ with an amplitude $A =
(6.5^{+1.3}_{-0.9})\%$ pointing in the direction $(l, b) = (233^\circ,
-13^\circ) \pm 10^\circ$ in galactic coordinates
\cite{Aab:2017tyv}. This important milestone confirms the expectation
that the sources of UHECRs are extragalactic, as the dipole shows no
correlation with the galactic plane. Hints of clustering in the sky
distribution have been reported for energies above $\sim 40$ to 60
EeV. TA reports a hot spot in its sky distribution
\cite{Abbasi:2014lda,Kawata:2015whq} with an origin point that is consistent with the starburst galaxy M82 being a source of these events~\cite{Abbasi:2020fxl}. Auger finds a significant
correlation ($4.5\sigma$) with SBGs and a weaker association ($3.1\sigma$) with gamma ray-emitting active galactic nuclei
($\gamma$AGN)~\cite{Aab:2018chp,Aab:2019ogu}. These hints can reach 5$\sigma$ significance with a dramatic increase in exposure above 60 EeV~\cite{Biteau:2019aaf,Sarazin:2019fjz}. 
  
\begin{figure}[ht]
\begin{center}
\includegraphics[trim = 25mm 40mm 15mm 35mm, clip, width=0.5\textwidth]{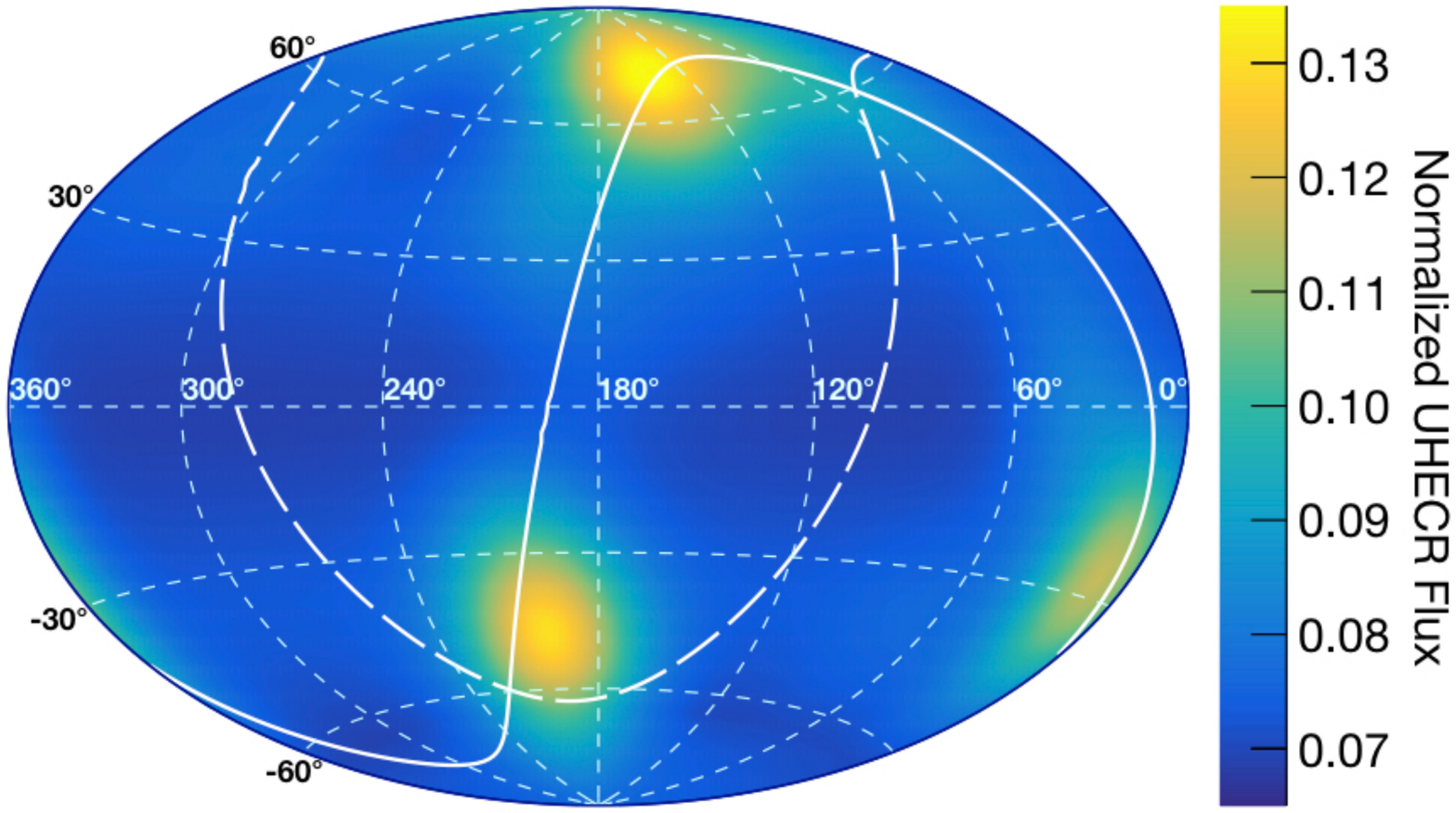}\includegraphics[trim = 25mm 40mm 15mm 35mm, clip, width=0.5\textwidth]{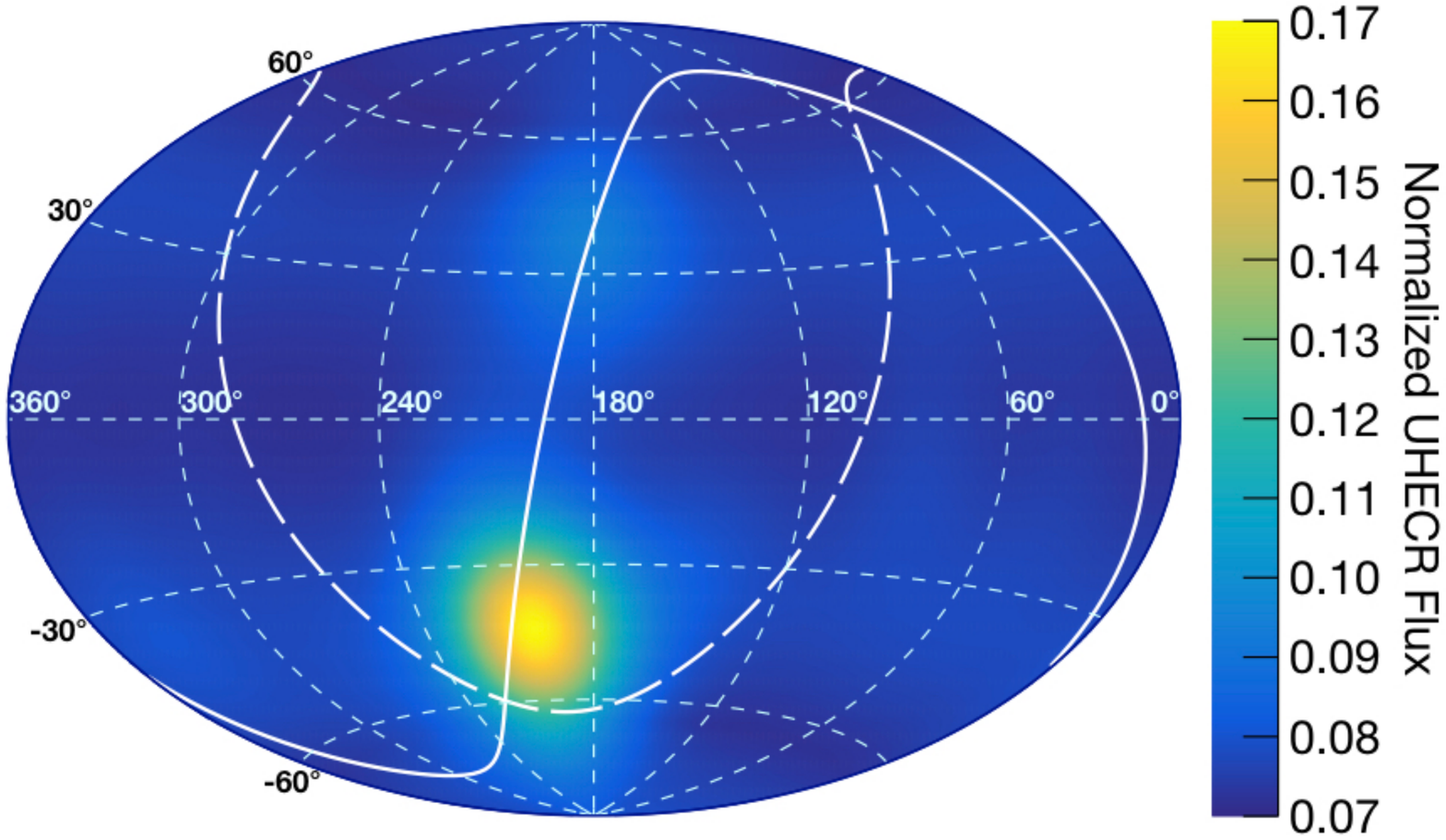}
\includegraphics[trim = 25mm 40mm 15mm 35mm, clip, width=0.5\textwidth]{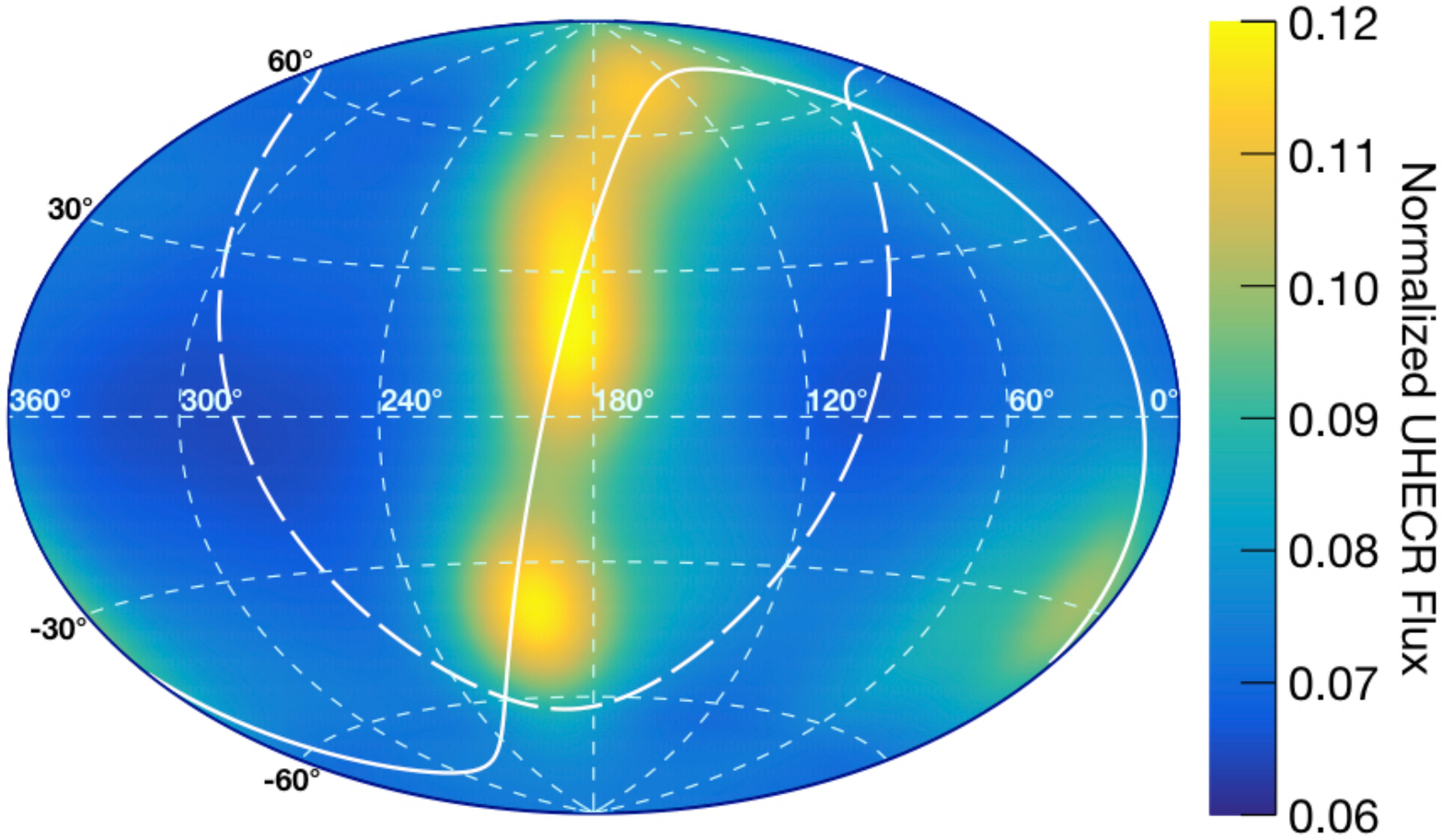}
\caption{Sky maps of the normalized UHECR flux in equatorial coordinates for several astrophysical catalogs with parameters selected to coincide with the best-fit parameters from Ref.~\cite{Aab:2019ogu}. \textit{Top left:} Starburst galaxies with $11$\% anisotropy fraction and angular spread of $15^{\circ}$. \textit{Top right:} \textit{Swift}-BAT AGNs with $8$\% anisotropy fraction and angular spread of $15^{\circ}$. \textit{Bottom:} 2MRS galaxies with $19$\% anisotropy fraction and angular spread of $15^{\circ}$. The dashed white line indicates the Galactic Plane. The solid white line indicates the Supergalactic Plane.}
\label{figAniso}
\end{center}
\end{figure}

  POEMMA will collect a dataset larger than the current statistics of Auger and TA combined (see Figure~\ref{fig3})~\cite{Anchordoqui:2019omw}. With full-sky coverage, POEMMA will observe the UHECR source distribution over the full celestial sphere, eliminating the need for cross calibration between two different experiments with only partial-sky coverage. Figure~\ref{figAniso} shows the relative UHECR flux as a function of position on the sky for three different astrophysical scenarios, revealing hot spots in the Northern and/or Southern Celestial Hemispheres that would be observable by either Auger or TA, but not both simultaneously. In contrast, POEMMA will be able to observe all of the hot spots predicted in these scenarios. Simulations of UHECR arrival directions for these and similar astrophysical scenarios demonstrate that within its nominal mission lifetime, POEMMA will be capable of detecting anisotropy at the level of $5\sigma$ for cross-correlation search parameters within the vicinity of the signal regions for the anisotropy hints reported by TA and Auger~\cite{Anchordoqui:2019omw}. Thus, POEMMA will turn the TA and Auger anisotropy hints (and/or other anisotropy signals yet to be discovered) into significant detections to finally discover the locations of the UHECRs sources.

\subsection{POEMMA UHECR Performance}

POEMMA addresses the challenges of discovering the sources of UHECRs by the dramatic increase in exposure enabled by space-based platforms. Figure~\ref{fig3} shows the POEMMA mission exposure range for observations where the instruments point close to the nadir out to the Earth's limb and compared to extrapolations of the current ground-based experiments. 
POEMMA-Stereo quasi-nadir mode (Figure \ref{fig2} left) enables excellent angular, energy, and composition resolution, while the POEMMA-Limb observing mode (Figure \ref{fig2} right) gives unparalleled volumes of monitored atmosphere for orders of magnitude increase in the statistics of observed UHECRs. 

\begin{figure}[ht]
\begin{minipage}[t]{0.47\textwidth}
  \postscript{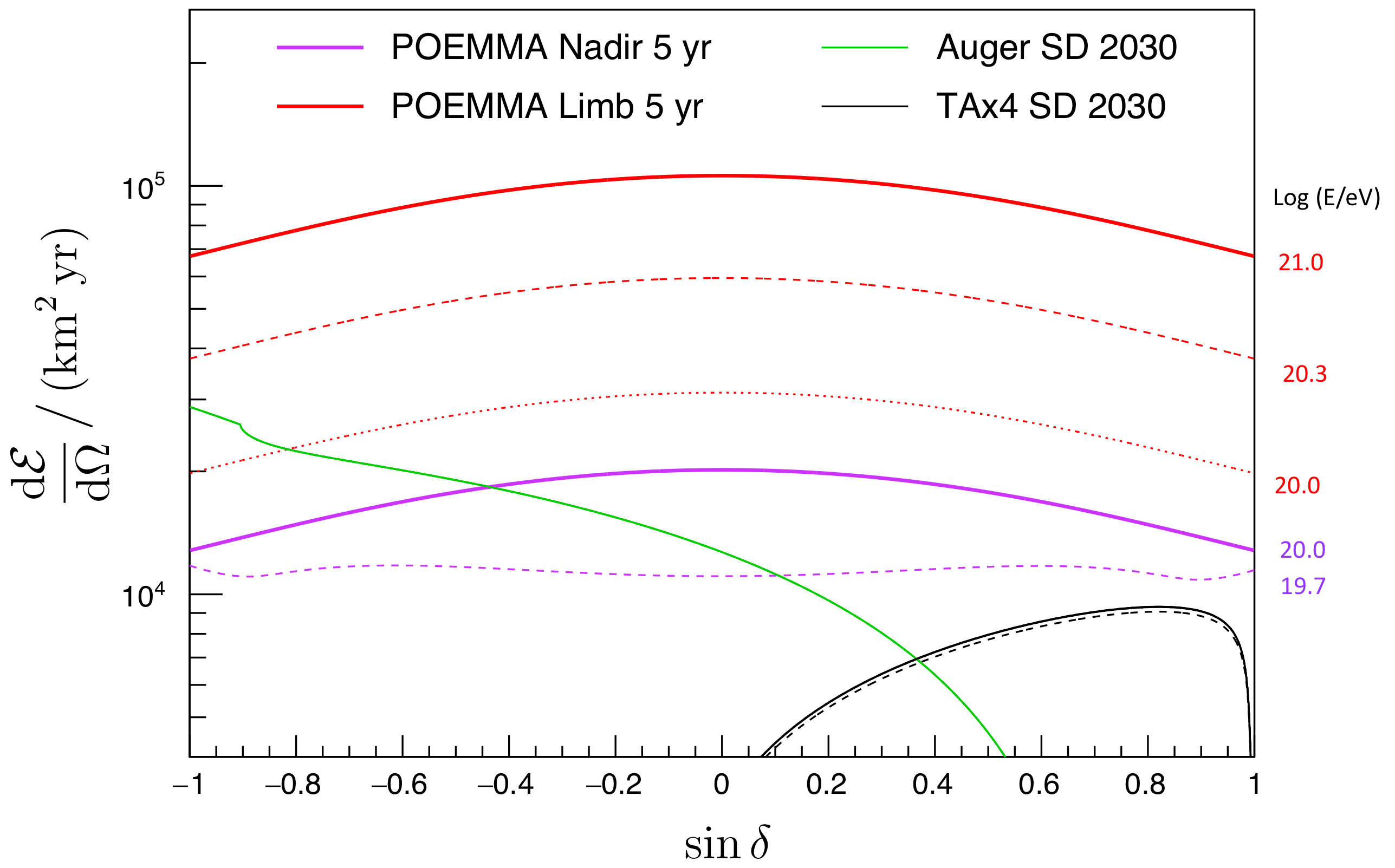}{1.0}
\end{minipage}
\hfill 
\begin{minipage}[t]{0.48\textwidth}
   \postscript{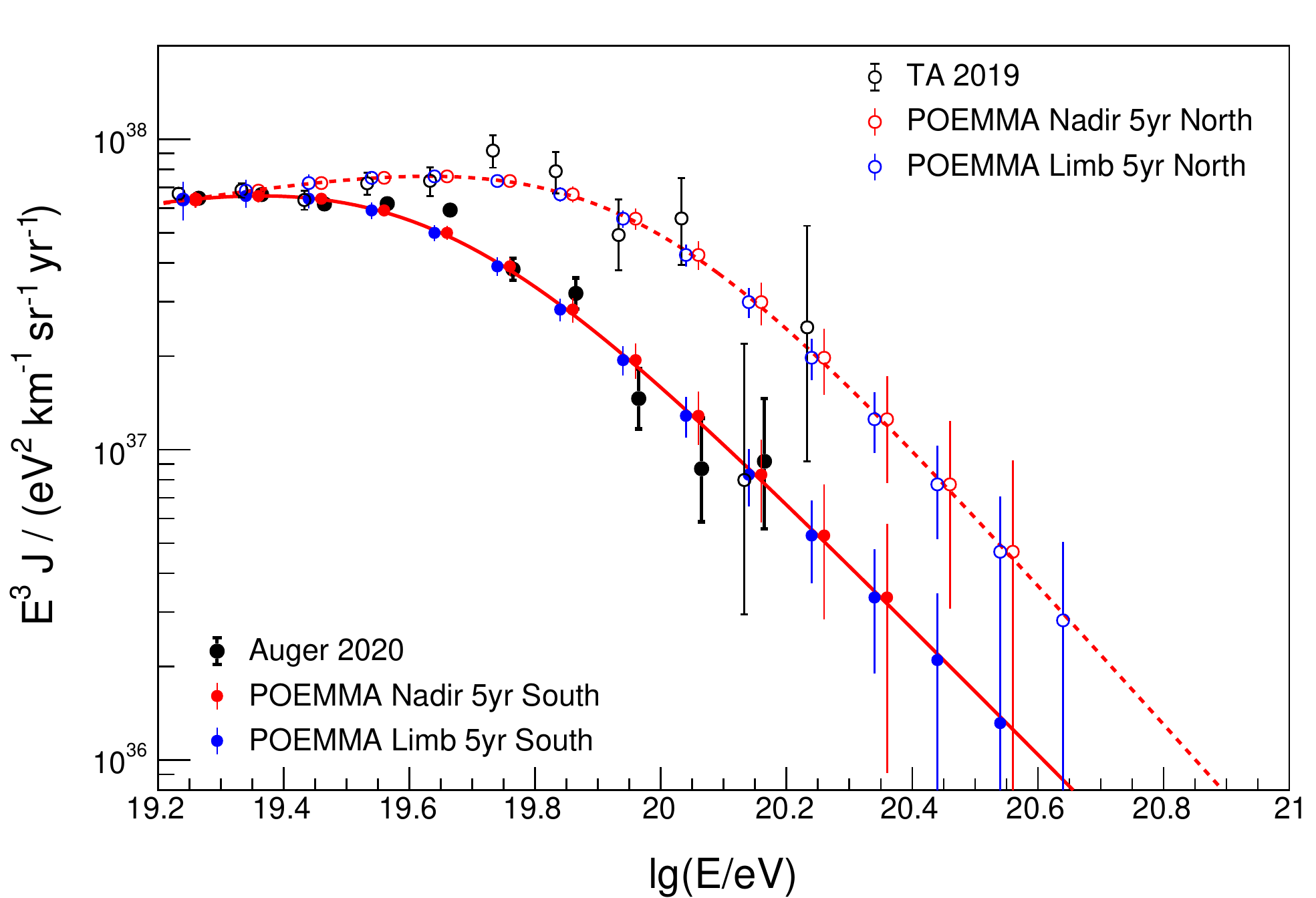}{1.0}
\end{minipage}
    \caption{{\it Left:} 
    Differential exposure as a function of declination for five years of POEMMA operations in POEMMA-Stereo mode (purple lines) and two energies for EASs: 10$^{19.7}$ eV (dotted) and 10$^{20}$ eV (solid); and for POEMMA-Limb mode (red lines) and three energies: 10$^{20}$ eV (dotted), 10$^{20.3}$ eV (dashed), and 10$^{21}$ eV (solid). 
  The exposures of Auger and TA (including the TAx4 upgrade) are shown as black (TA) and green (Auger) lines respectively assuming operations until 2030.
 {\it Right:} Simulated POEMMA spectra extrapolation compared with Auger 2020 spectrum (black dots and solid line) from Ref.~\cite{Aab:2020rhr} and the TA 2019 spectrum (black open circles and dotted line) from Ref. \cite{Ivanov2019} for both POEMMA-Stereo (red) and POEMMA-Limb (blue) observations, shown for energies above 10$^{19.2}$ eV. Adapted from Ref. \cite{Anchordoqui:2019omw} with more recent results.
 \label{figSpec}}
\end{figure}

POEMMA is designed to obtain definitive measurements of the UHECR spectrum, composition, and source locations for energies $E \gtrsim 20~{\rm  EeV}$. EAS fluorescence signals are observed as video recordings with 1$\mu$s frame rate.  Each telescope records an EAS trace in its focal plane (see Figure \ref{fig9}, right), which defines an observer-EAS plane. In UHECR stereo mode, the intersection of the two observer-EAS planes defines the geometry of the EAS trajectory. 
Precise reconstruction of the EAS is achieved for opening angles between these two planes larger than $\sim$5$^\circ$.
In mono observations, the distance to the EAS in the observer-EAS plane is determined by the evolution in time of the EAS and a model of the atmosphere.

Detailed simulations of POEMMA's UHECR exposure, angular resolution,
and composition ($X_{\rm max}$) resolution were performed using POEMMA's instrument design~\cite{Anchordoqui:2019omw}. 
Figure~\ref{fig14} shows POEMMA's stereo reconstructed angular resolution, which is $\lesssim 1.5^\circ$ above 30~EeV.  The stereo trigger condition in each satellite leads to a highly efficient reconstruction fraction of $\sim$80\%, with losses due mainly to the requirement that the opening angle between each EAS geometrical plane be $\ge 5^\circ$.
The fine angular resolution from stereo reconstruction leads to accurate 3-dimensional EAS reconstruction with energy resolution of $< 20\%$ and $X_{\rm max}$ resolution of $\sim$30~${\rm g/cm^2}$ above $\sim$50~${\rm EeV}$ improving to $\sim$20~${\rm g/cm^2}$ above $\sim$100~${\rm EeV}$. 

The UHECR energy threshold is set by the brightness of the EASs with respect
to the dark-sky airglow background in the EAS fluorescence band of $300 <
\lambda/{\rm nm} < 500$ at the POEMMA focal surface. The near UV filter over
the MAPMTs in the POEMMA Fluorescence Camera combined with the $\sim$6~${\rm m^2}$ optical collecting area of each POEMMA instrument yields an UHECR energy threshold of $\sim$20~${\rm EeV}$.

POEMMA's UHECR exposure enables precise measurements of the spectrum at energies higher than those reached by current observatories. Figure \ref{figSpec} right shows simulated POEMMA spectra based on extrapolation of the Auger spectrum to higher energies (filled circles following the solid line). If the  extrapolation is based on the TA spectrum (black open circles and dotted line), the POEMMA measurement will reach higher energies for both POEMMA-Stereo (red) and POEMMA-Limb (blue) observations. The impact of the POEMMA exposure in the spectrum is clear for energies above $\sim$100 EeV, where new spectral features can signal source signatures such as the effect of the closest sources in a given hemisphere \cite{Kotera:2011cp,Anchordoqui:2018qom,AlvesBatista:2019tlv}.

\begin{figure}[ht]
    \postscript{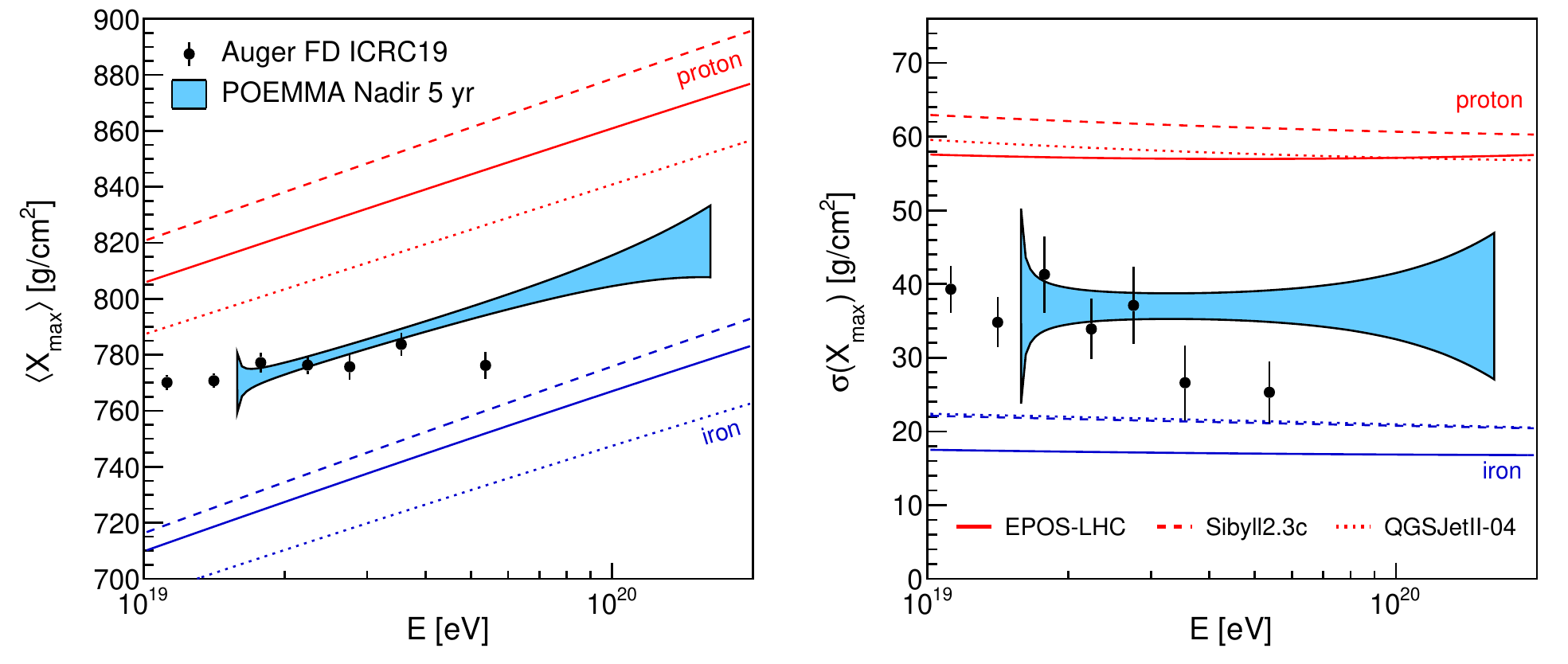}{0.99}
\caption{Capability of POEMMA to measure $\langle X_\text{max}\rangle$
  and $\sigma(X_\text{max})$ for composition studies at UHEs. The width
  of the blue band illustrates the expected statistical uncertainties in five
  years of POEMMA-Stereo (nadir) operations given the number of events per 0.1 in the logarithm of energy, the $X_\text{max}$ resolution and efficiency for $\theta < 70^\circ$, and the intrinsic  shower-to-shower fluctuations of 40~g/cm$^2$. The band spans the energy range for which more than 10 events are within a 0.1 decade bin (assuming the Auger spectrum). The black dots are fluorescence data from Auger ICRC 2019~\cite{Yushkov:2020nhr} and the blue bands are from Ref.~\cite{Anchordoqui:2019omw}.
\label{figXmax}}
\end{figure}

\begin{figure*}[ht]
\begin{minipage}[t]{0.49\textwidth}
  \postscript{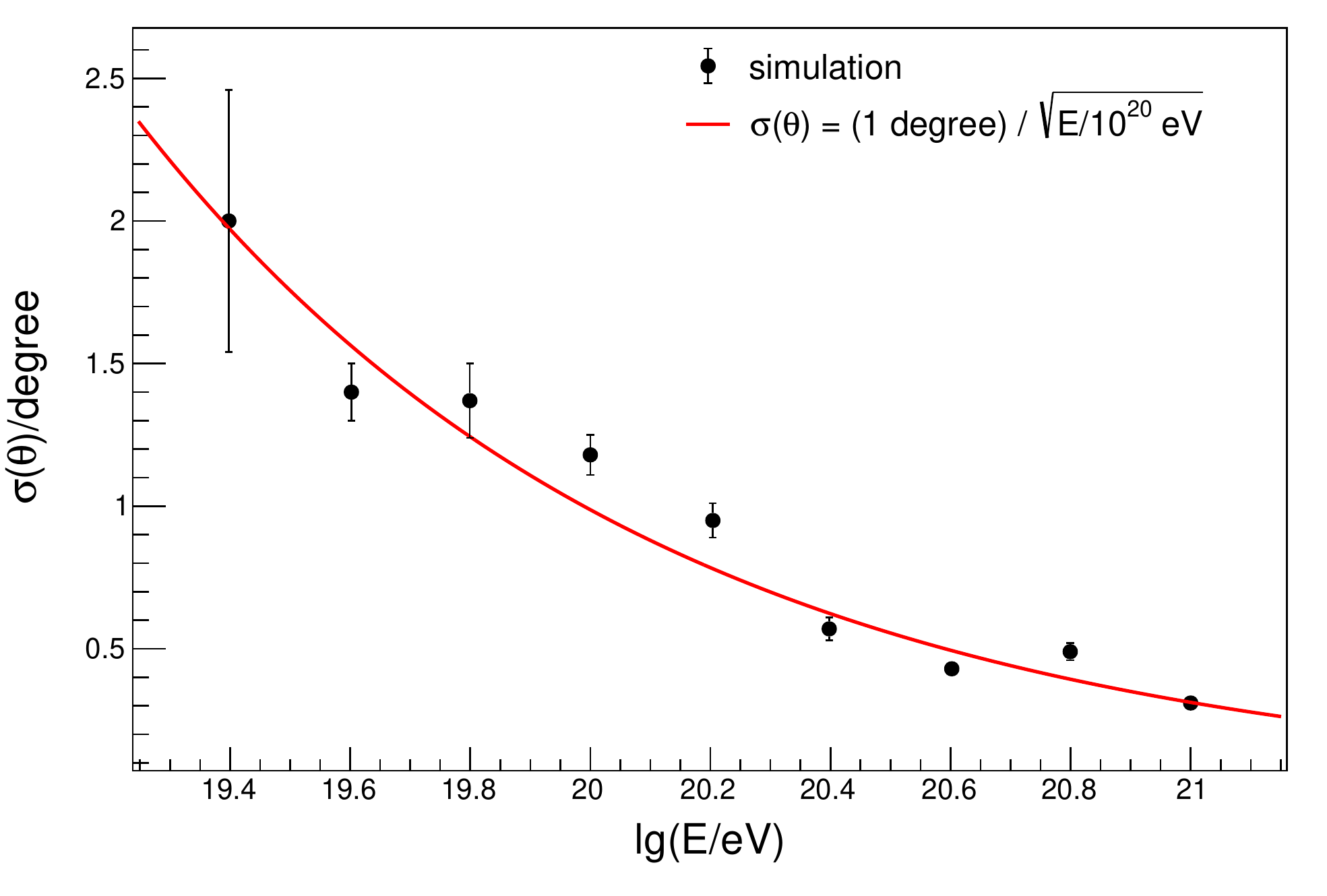}{0.99}
\end{minipage}
\hfill \begin{minipage}[t]{0.49\textwidth}
   \postscript{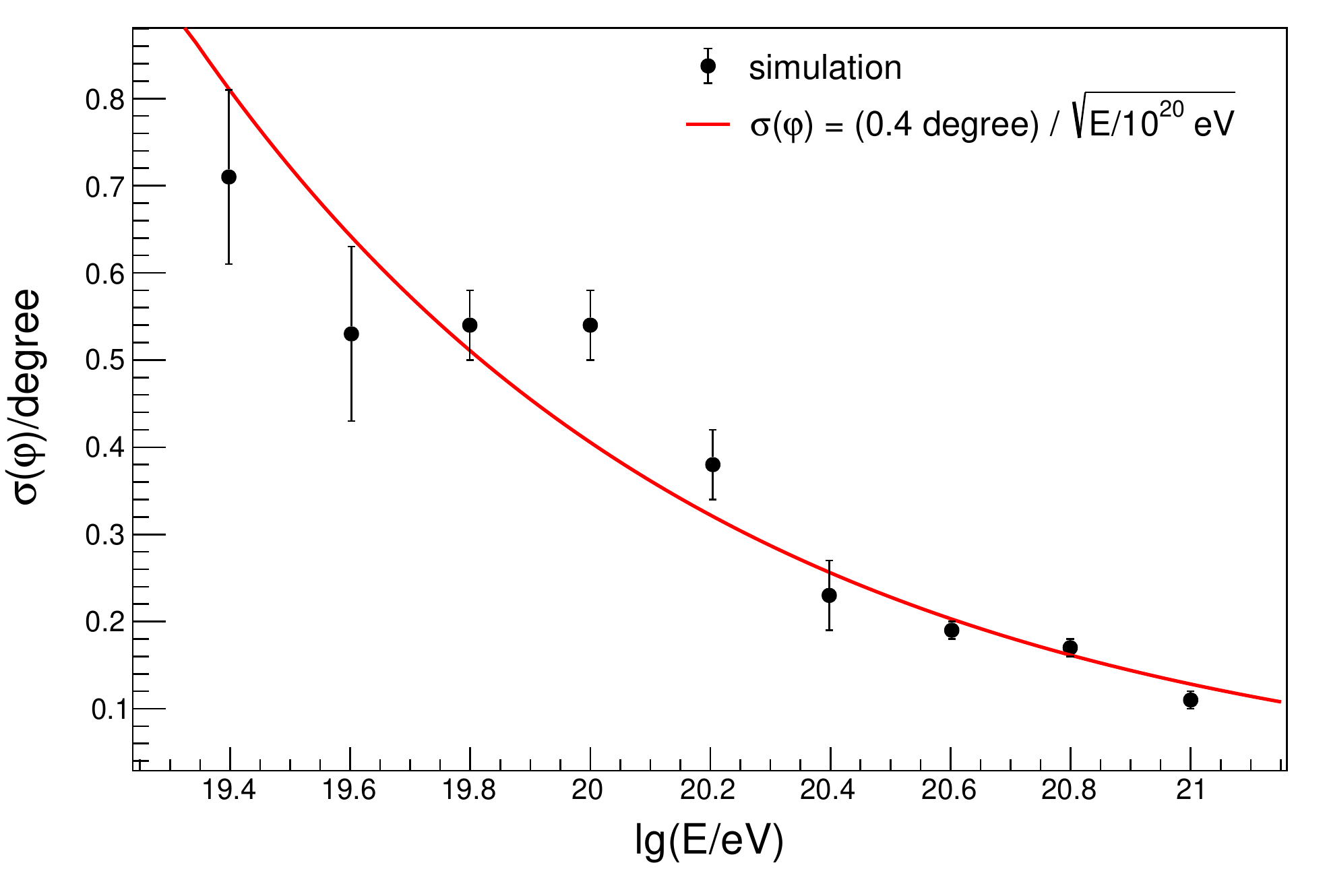}{0.99}
\end{minipage}
    \caption{POEMMA's simulated stereo-reconstructed angular resolution versus
UHECR energy: azimuth ({\it left}) zenith ({\it right}). 
{Adapted from Ref. \cite{Anchordoqui:2019omw}}.
\label{fig14}}
\end{figure*}

POEMMA observations allow the study of different composition models at the highest energies, where composition should become less mixed due to propagation effects (see, e.g.,  \cite{Kotera:2011cp}). The  $X_{\rm max}$ resolution of POEMMA makes it possible to decompose EASs into four groups of nuclear species, gamma-rays, and neutrinos in the highest energy range~\cite{Krizmanic:2013tea}.   
Auger observations show an interesting evolution of the composition with energy consistent with the maximum energy models~\cite{Aab:2017njo}. The current paucity of UHECR data above 40 EeV 
(see Figures  \ref{figSpec} right, ~\ref{figXmax}, and \ref{figNumbCover})
strongly limits definitive tests of different source models with the spectral behavior and composition trends.  Figure~\ref{figXmax} shows the statistical power with which POEMMA will determine the first two moments of the \xmax~ distribution, the mean, $<$\xmax$>$, and the standard deviation, $\sigma(\xmax)$, for energies well beyond the leading observations by Auger. The blue band in Figure~\ref{figXmax} represents the statistical uncertainty assuming the 5 year POEMMA-Stereo UHECR statistics in 0.1 decade energy bands using on a constant composition model that is based on extrapolation to higher energies of Auger data below 40 EeV, where low-statistics measurements of \xmax~ are available.  As is clearly seen, a single POEMMA data point around 100 EeV will have an $<$\xmax$>$ uncertainty that is less than a tenth of the proton-iron separation. With several POEMMA data points above 40 EeV, determination of the UHE composition evolution will be possible. 

In addition, if hot spots in the sky
are observed with more than 20 events, POEMMA can
study a given source composition by the evolution of the
hot spot shape with energy~\cite{Anchordoqui:2019omw,Anjos:2018mgr}.

\begin{figure*}[tb]
\centering
\includegraphics[width=.75\linewidth]{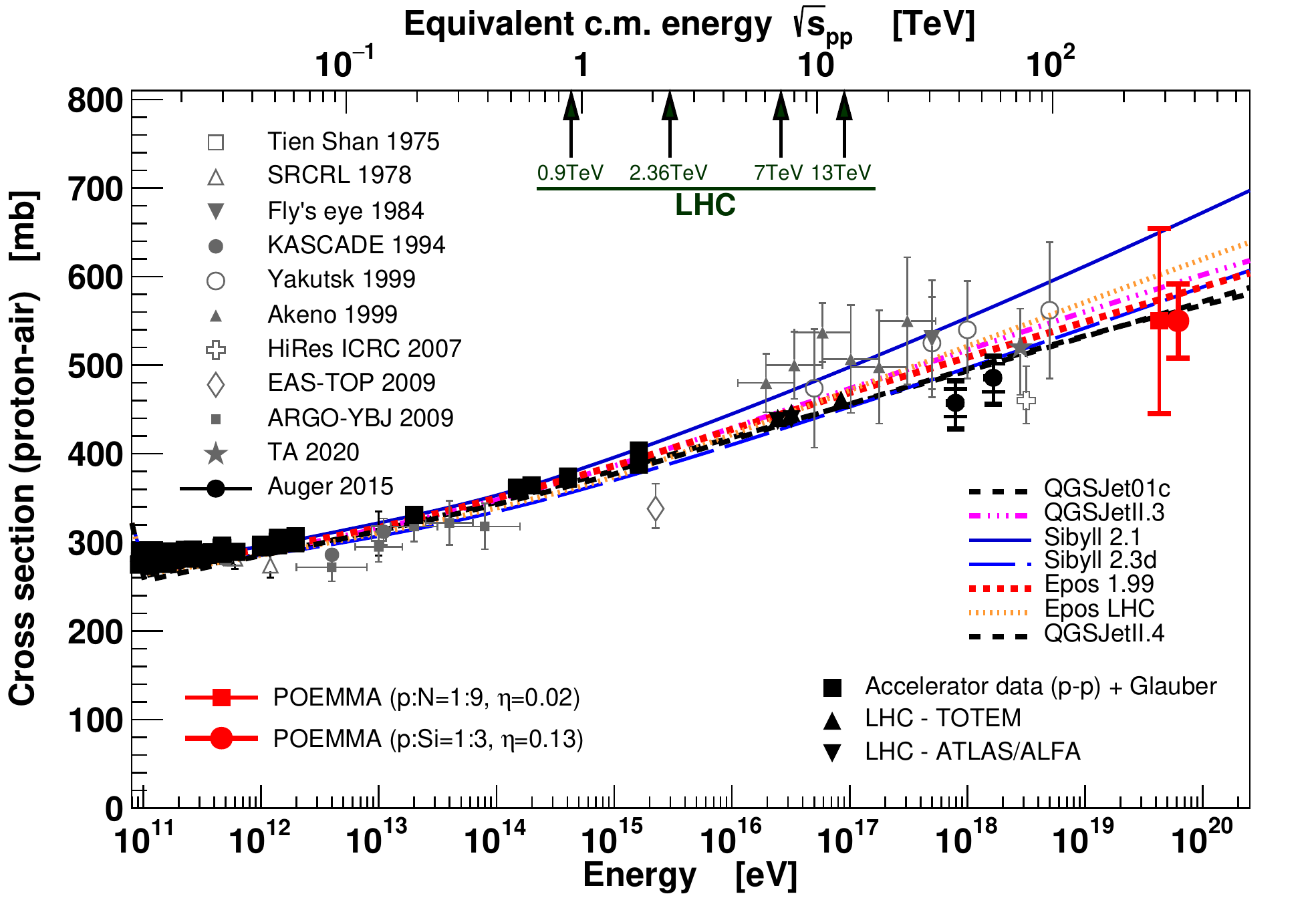}
\caption{ The UHE proton-air cross section as a function of proton energy and the 
projected UHE proton-air cross section measurement with POEMMA for two simplified UHECR composition scenarios (left $p$:N=1:9 and $\eta=0.1\times 0.2 = 0.02$, right $p$:Si=1:3 and $\eta=0.25\times 0.5=0.13$) shown by red markers, including error bars. The two points are displaced for clarity.
Also shown is a compilation of accelerator data converted to a proton-air cross section using the Glauber formalism and measurements
\cite{Nam:1975xk,Siohan:1978zk,Baltrusaitis:1984ka,Mielke:1994un,Knurenko:1999cr,Honda:1992kv,Belov:2007jva,Aglietta:2009zza,Aielli:2009ca},
including the Telescope Array \cite{Abbasi:2020chd} and Pierre Auger Observatory \cite{Ulrich:2015yoo} results. The proton-air cross sections for the QGSJet \cite{Kalmykov:1997te,Ostapchenko:2010vb}, Sibyll \cite{Ahn:2009wx,Engel:2019dsg}, and EPOS \cite{Pierog:2013ria} Monte Carlo programs are also shown. 
}
\label{fig:xs}
\end{figure*}

\begin{figure}[ht]
\begin{minipage}[t]{0.47\textwidth}
  \postscript{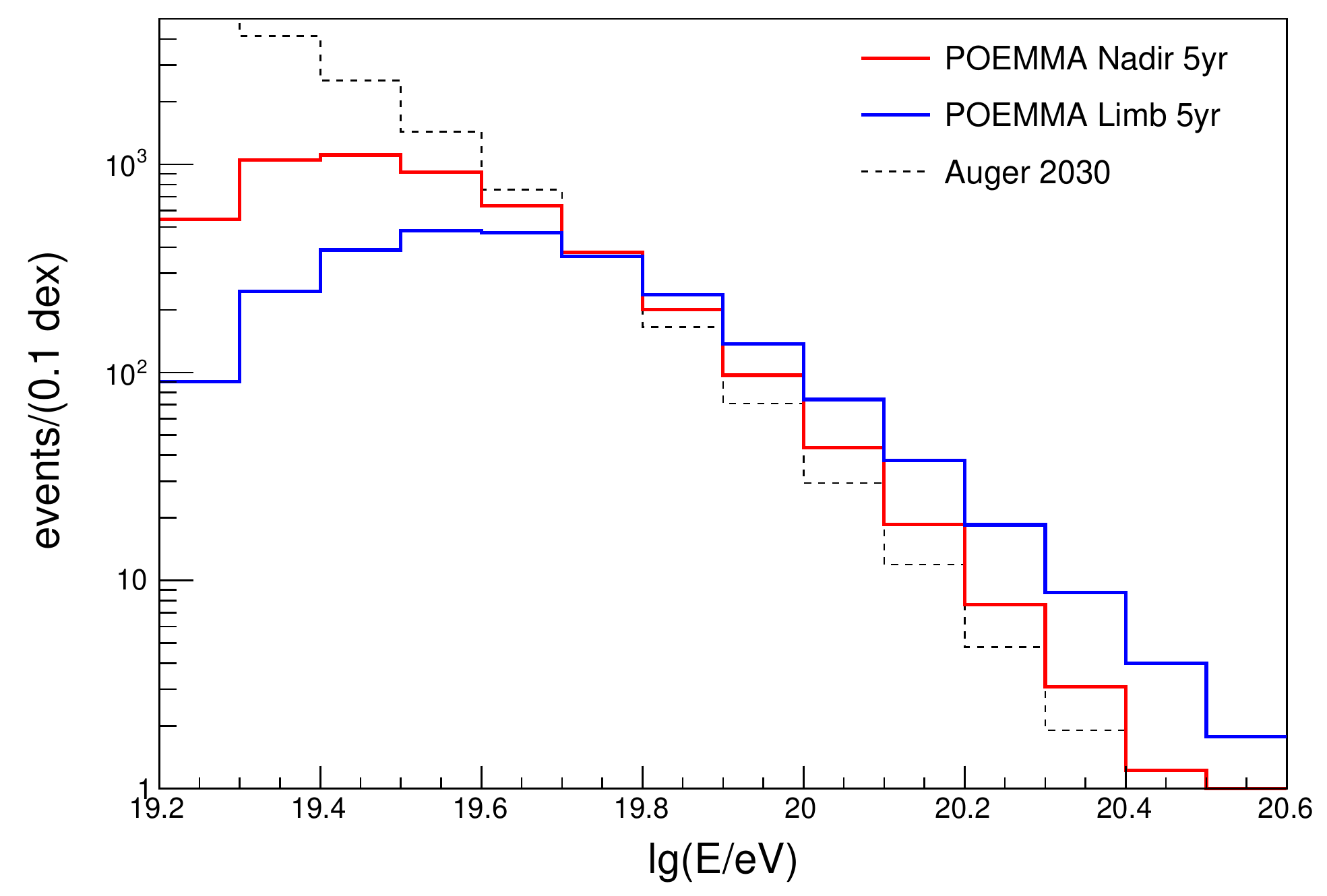}{1.0}
\end{minipage}
\hfill 
\begin{minipage}[t]{0.47\textwidth}
   \postscript{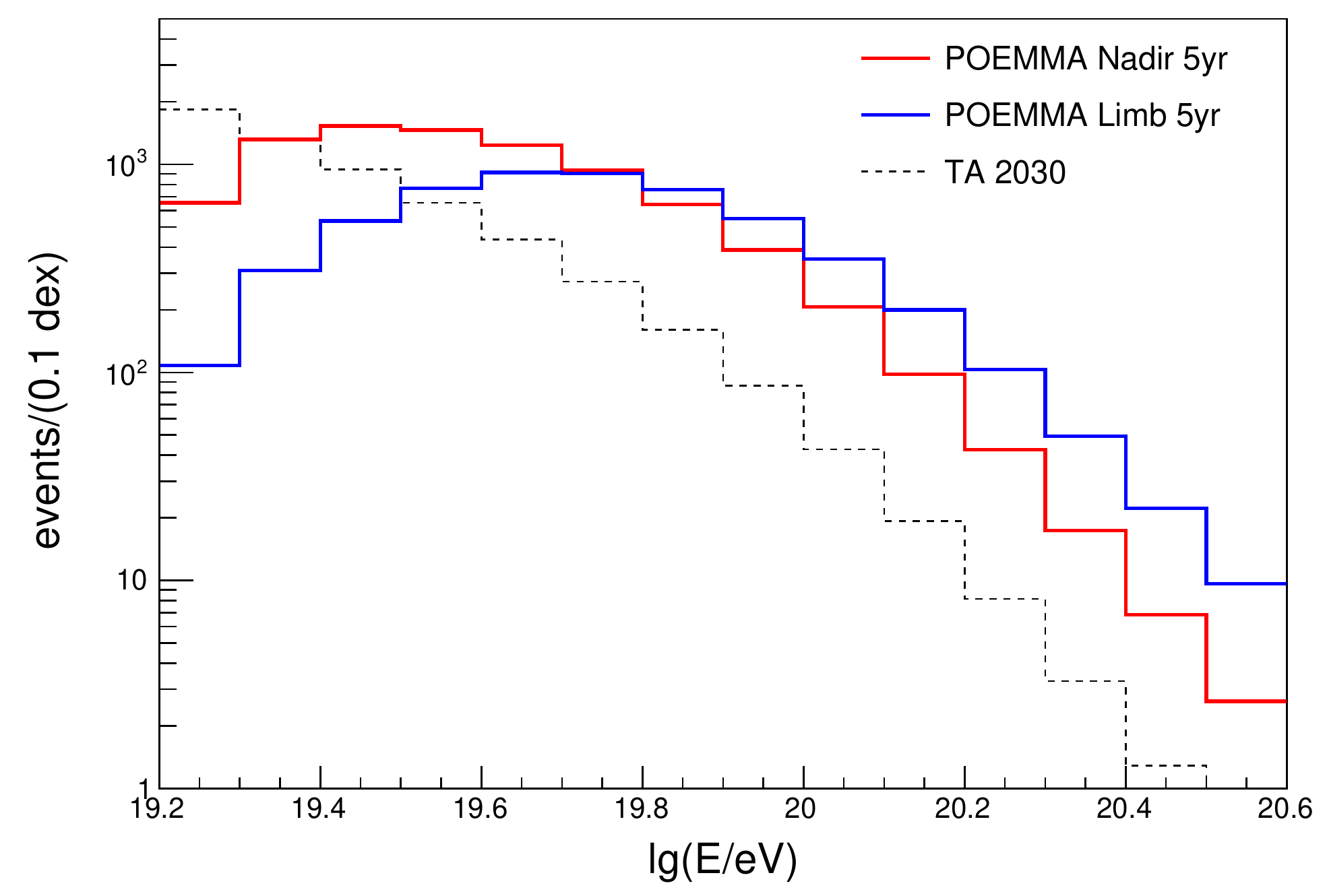}{1.0}
\end{minipage}
  \caption{{\it Left:}  
Number of UHE events detected by POEMMA for five years of observations in POEMMA-Stereo (red) and POEMMA-Limb (blue) operational modes assuming the Auger UHECR energy spectrum. For comparison, the projected event numbers for Auger observations projected to 2030 are indicated by black dashed lines.
  {\it Right:}
  Number of UHE events detected by POEMMA for five years of observations in POEMMA-Stereo (red) and POEMMA-Limb (blue) operational modes assuming the TA UHECR energy spectrum. For comparison, the projected event numbers for TA  observations projected to 2030 are indicated by black dashed lines.}
 \label{figNumbCover}
\end{figure}

The high energy $X_\text{max}$ tail of the distribution $dN/dX_\text{max}$ of events probes the fundamental physics of the proton-air cross section, as outlined in, e.g., \cite{Ulrich:2015yoo}. The distribution falls off according to  $\exp(-X_\text{max}/\Lambda_\eta)$ with $\Lambda_\eta^{-1}\sim \sigma_{p-\text{air}}(E_0>40\ {\rm EeV})$, which corresponds to a cross section at an equivalent center-of-mass energy in nucleon-nucleon collisions of $\sqrt{s_{NN}}=283$ TeV. The $dN/dX_\text{max}$
tail depends on the cosmic ray mass composition. The composition above 40 EeV is modeled here with two simple representative scenarios guided by the cosmic ray composition analysis described in~\cite{Muzio:2019leu}. A conservative UHECR composition scenario with 10\% protons and 90\% nitrogen has a 20\% proton fraction in the high $X_\text{max}$ tail. A less conservative scenario has 25\% protons and 75\% silicon, with a proton fraction of 50\% in the high $X_\text{max}$ tail. Figure \ref{fig:xs} shows the projected cross section measurements with POEMMA for these two simplified models, given a fraction $\eta$ of the $N=1400$ events above 40 EeV, assuming a cosmic ray spectrum parameterized by the Auger Collaboration \cite{Fenu:2017hlc}. An energy resolution of $\Delta E/E=0.2$ and $\sigma(X_\text{max})=35$ g/cm$^2$
are used to estimate the error in the anticipated cross section measurement \cite{Anchordoqui:2019omw}. As Figure \ref{fig:xs} shows, a POEMMA cross-section measurement will correspond to an energy well above the LHC measurements. In the figure, $\sigma_{pp}$ is converted to the proton-air cross section using the Glauber formalism. 

POEMMA will provide a powerful probe of physics in and beyond the Standard Model (BSM) at center-of-mass energies of $\sqrt{s} \approx 100$~TeV, and possibly beyond, well above the reach of particle accelerators. POEMMA's capabilities allow it to probe a unique landscape of BSM models. 
One very compelling search involves extreme-energy photons produced via the decay of super-heavy dark matter (e.g., \cite{Aloisio:2015lva,Alcantara:2019sco}). Possible BSM physics can also be explored by searching for protons, neutrons, and photons from  sources that are apparently too far given their observed energies.  Such observations, related to a spectral recovery past the GZK effect, will test Lorentz invariance with unprecedented precision (for a review, see, e.g., \cite{Stecker:2017gdy}). 

POEMMA is equally sensitive to UHECR sources in both the Northern and Southern Hemispheres. 
POEMMA has full-sky coverage due to its orbit at 525~km altitude and $28.5^\circ$ inclination  and the very large FoVs ($45^\circ$) of the observatories. 
Figure \ref{figSpec} left shows the differential exposure as a function of declination for five years of POEMMA operations in POEMMA-Stereo  mode (purple lines) and  POEMMA-Limb mode (red lines) observations.  The exposures of Auger and TA (including the TAx4 upgrade) are shown as black (TA) and green (Auger) lines respectively assuming operations until 2030.

\begin{figure}[ht]
    \postscript{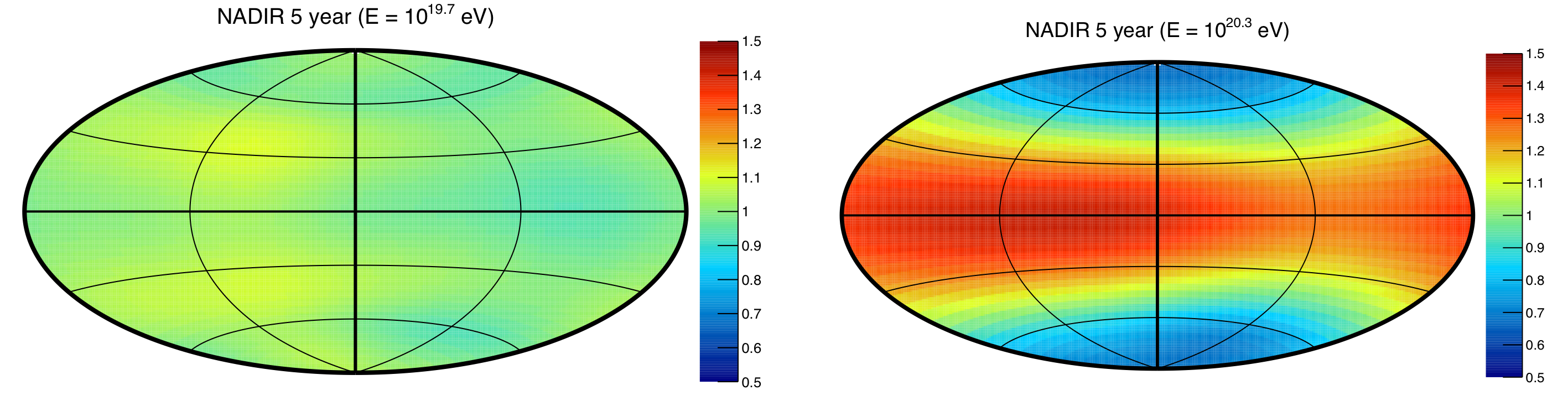}{0.99}
  \caption{POEMMA UHECR sky exposure for POEMMA-Stereo observations in declination versus right ascension.  Color scale denotes the exposure variations in terms of the mean response taking into account the positions of the sun and the moon during a 5-year observation cycle.
{\it Left:} Sky exposure for showers of 10$^{19.7}$ eV. {\it Right:} Sky exposure for showers of 10$^{20.3}$ eV. 
 \label{fig16}}
\end{figure}

The POEMMA-Stereo observation coverage is shown for two EASs energies in Figure \ref{figSpec} left, 10$^{19.7}$ eV (purple dotted line) and 10$^{20}$ eV (purple solid line). These are also displayed as sky exposures in Figure \ref{fig16} in declination versus right ascension.  In Figure \ref{fig16} the color scale denotes the exposure variations in terms of the mean response taking into account the positions of the sun and the moon during a 5-year observation cycle.
The higher exposure for POEMMA-Limb observations is also clear in Figure \ref{figSpec} left,  where red lines for three energies is shown: 10$^{20}$ eV (dotted), 10$^{20.3}$ eV (dashed), and 10$^{21}$ eV (solid). 
 
POEMMA will measure the UHECR source
distribution over the full celestial sphere using a single experimental
framework with a well-defined UHECR acceptance, mitigating the issues
of cross-comparisons inherent to viewing different portions of the sky
with multiple experiments.
The response shown in Figure~\ref{fig16} was
calculated assuming a POEMMA-Stereo configuration aligned along the orbit path. The ability of the space-based POEMMA telescopes to tilt towards the Northern or Southern Hemisphere allows for the sky exposure to be enhanced for a specific hemisphere. Likewise, POEMMA can preferentially point north or south for a sequence of orbital periods to further tailor the UHECR sky
coverage for possible source locations.

\begin{figure*}[ht]
\begin{minipage}[t]{0.51\textwidth}
  \includegraphics[trim = 10mm 13mm 73mm 25mm, clip, width = 1.0\textwidth]{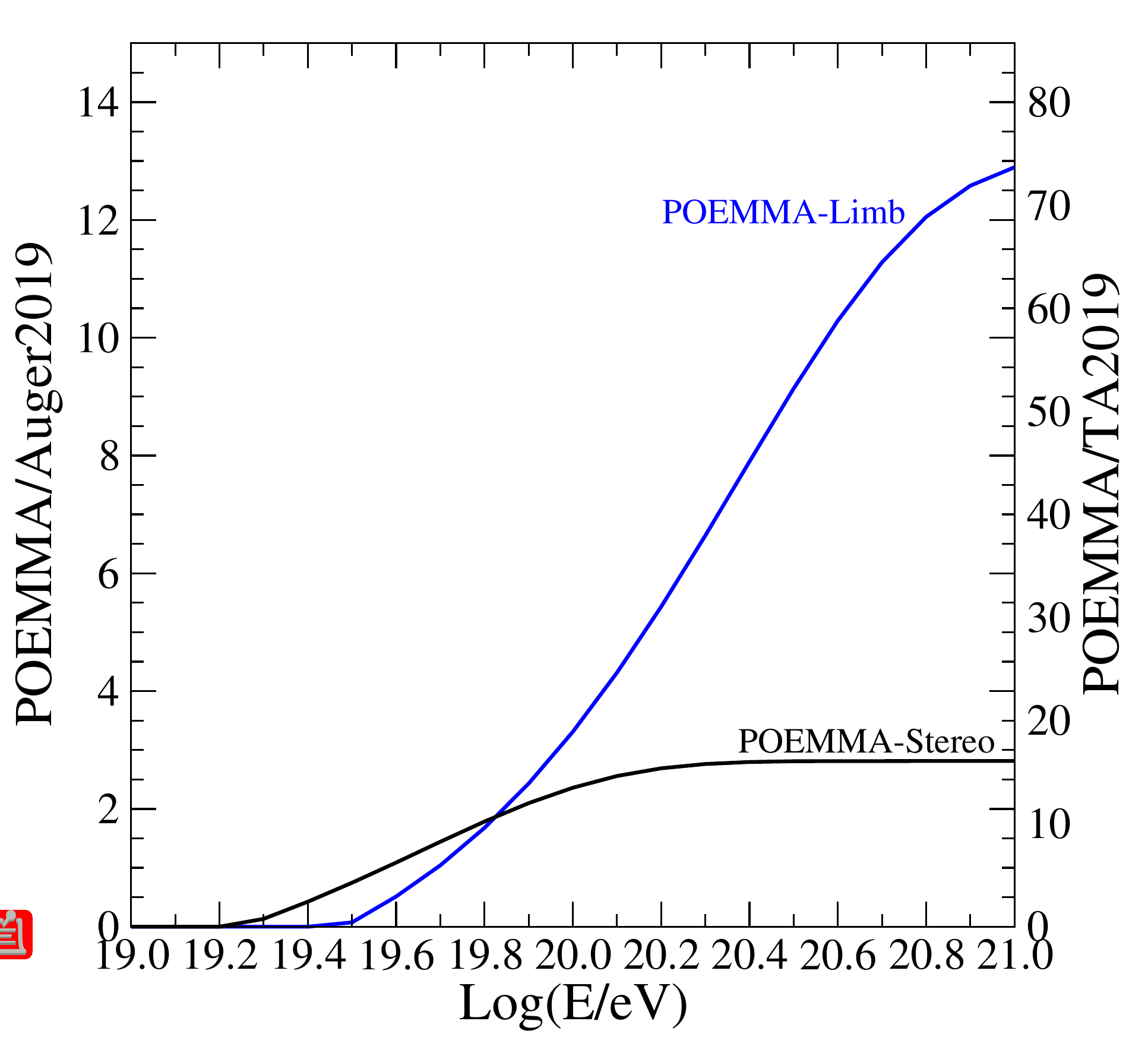}
\end{minipage}
\begin{minipage}[t]{0.48\textwidth}
   \includegraphics[trim = 0mm 8mm 90mm 25mm, clip, width = 1.0\textwidth]{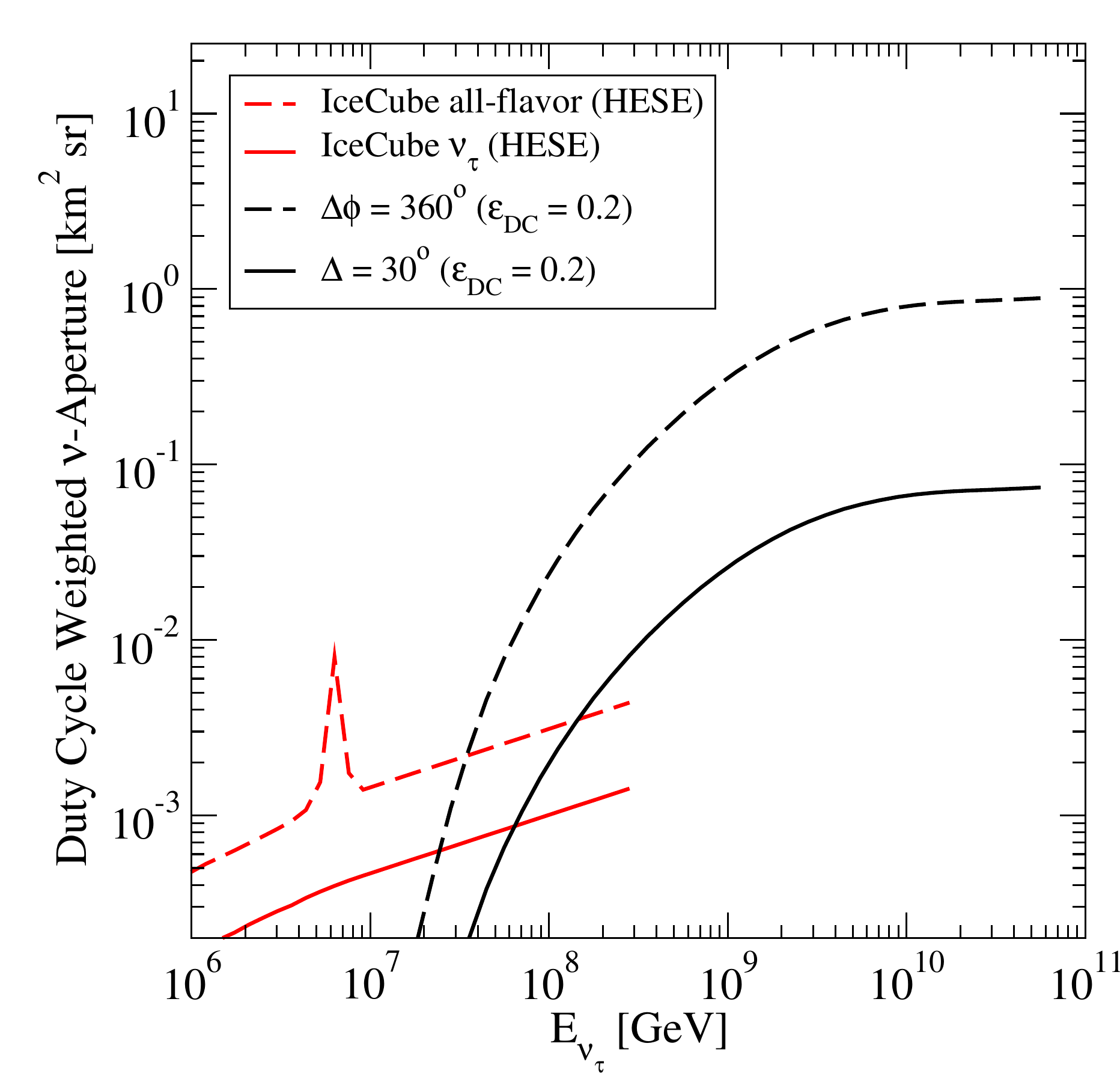}
\end{minipage}
    \caption{{\it Left:} Examples of the 5-year POEMMA stereo UHECR exposure for a satellite  separation of 300 km, assuming a 12\% duty cycle, in units 
Auger exposure~\cite{Verzi2019} and TA exposure~\cite{Ivanov2019,IvanovPriv} reporting at ICRC-2019. The Stereo (Mono) mode has lower (higher) energy threshold, with the mono mode having the higher exposure. {\it Right:} The POEMMA diffuse-flux neutrino aperture as a function of $\nu_\tau$ energy for accepting $\nu_\tau$'s through the up-going \tauon decay EAS. Solid-line is for the current design with a 30$^\circ$ FoV and dashed-line for POEMMA30 ($\times$12), (extrapolating the POEMMA30 sensitivity to 360$^\circ$ FoV in azimuth), and a Duty Cycle, $\varepsilon_{DC}$, of 20\% for both. Also shown is the IceCube all-flavor $\nu$ aperture (dashed line) and $\nu_{\tau}$ (solid line) neutrino aperture for HESE (high-energy starting events). \label{fig15}}
\end{figure*}

Figure~\ref{fig15} left shows an example of a 5-year stereo UHECR exposure in terms of the Auger and TA exposures reported in 2017~\cite{Aab:2017njo}. 
The POEMMA UHECR exposures are calculated from simulations assuming an isotropic flux and an EAS trigger condition based on the modeling of the response of the PDMs in the PFC~\cite{Adams:2013vea}. Simulations of the EAS reconstruction selection criteria lead to an 85\% acceptance for stereo mode and 80\% for the tilted (monocular) configuration for UHECR observations in neutrino mode. The fraction of time POEMMA is viewing the night sky with minimal moonlight is 18\% based on calculations for the POEMMA orbit \cite{Guepin2018}. Previous simulation studies~\cite{Adams:2013vea} have shown that 72\% of events observed from space have the location of shower maximum above clouds based on meteorological cloud height measurements. An additional 5\% reduction is estimated to account for the effects of light pollution from cities and lightning. These lead to an effective duty cycle of 12\% for the UHECR exposure determined after event reconstruction and selection.

When POEMMA is operating in limb observing mode (see Figures~\ref{fig2} right), the two satellites are
separated by $\sim 25~{\rm km}$ and tilted up to  $45^\circ$ away from the nadir to view the limb of the Earth. Consequently, the asymptotic UHECR instantaneous aperture increases by nearly an order of magnitude, albeit with a
higher UHECR energy threshold for reconstructing the observed
events (see Figure \ref{fig15} left). With the smaller satellite separation, the performance is
closer to monocular reconstruction, where the 1 $\mu$s  timing information is needed in order to yield a measurement of the orientation of the EAS in the
atmosphere. While the monocular performance in terms of angular resolution (few
degree near 100~EeV) and composition sensitivity ($X_{\rm max} \sim
100~{\rm g/cm^2}$) is not as accurate as that for the stereo
mode, the energy resolution is still $\sim 20\%$. This energy resolution and the significant increase in statistics (see Figure~\ref{fig3}) allows for the unique study of the spectral shape above 100~EeV and the sky distribution of UHECRs with the highest rigidity.

POEMMA is designed to be able to perform both stereo and monocular reconstruction of the fluorescence signal, with the latter being needed for risk mitigation in the case that one satellite fails to perform properly. In addition, POEMMA will observe $2^\circ$ above the limb to measure UHECR Cherenkov signals~\cite{Neronov:2016iax} as it monitors below the limb for cosmic neutrinos (as discussed  in VI below).

\subsection{Cosmic Neutrino Science}

POEMMA will also be sensitive to the most energetic cosmic neutrinos, from 20~PeV through the EeV scale, thus providing an opportunity to make substantial progress in high-energy astrophysics and fundamental physics.

For astrophysics, POEMMA will measure key components of extreme energy multi-messenger emission, in the form of diffuse and transient cosmic neutrino fluxes.  
(Above $\sim$ 10~PeV, the scattering cross sections of neutrinos and anti-neutrinos on nucleons are virtually identical, so we denote neutrinos and anti-neutrinos simply as neutrinos.) 

A diffuse flux of high-energy neutrinos can be produced by a variety of astrophysical sources, including the unknown UHECRs sources, and in the propagation of UHECRs from extragalactic sources to Earth. 
In particular, at the highest energies, the interactions of UHECRs with the cosmic microwave (GZK effect) and infrared backgrounds lead to a cosmogenic neutrino flux from the decay of pions and neutrons~\cite{Berezinsky:1969qj,Stecker:1978ah}.  The cosmogenic flux depends on the nuclear composition of UHECRs~\cite{Hill:1983xs,Engel:2001hd,Fodor:2003ph,Ave:2004uj,Hooper:2004jc,Anchordoqui:2007fi,Ahlers:2010fw,Kotera:2010yn,Ahlers:2012rz,AlvesBatista:2018zui} providing another way to measure it~\cite{Ahlers:2009rf, Kotera:2010yn}.  Figure \ref{fig15}, right, shows the effective neutrino aperture for the PCC design described here. Given the current data on UHECR composition, the diffuse cosmogenic neutrino flux is too faint for POEMMA to reach with a 30$^\circ$ effective FoV, set by the PCC location at the edge of the focal surface. The POEMMA Collaboration is developing a version of the POEMMA mission with 360$^\circ$ FoV, named POEMMA360, which has an optical design optimized for Cherenkov detection improving the sensitivity to the diffuse neutrino flux.

POEMMA will be uniquely suited for rapid follow-up of ToOs for neutrino observations, because it will orbit the Earth in a period of $95$~mins.~and will be capable of repointing its satellites by $90^{\circ}$ in $500$~s. In combination, these design features will enable POEMMA to access the entire dark sky within the time scale of one orbit. Additionally, an optimal survey strategy will enable POEMMA to achieve quasi-uniform coverage of the full sky on a time scale of a few months for diffuse neutrino flux observations~\cite{Guepin2018}. POEMMA will also have groundbreaking sensitivity to neutrinos at energies beyond $100$~PeV, reaching the level of modeled neutrino fluences for nearby sources in many astrophysical scenarios (see Figures~\ref{fig19a} and \ref{fig19b} and~\cite{Venters:2019xwi}).

Highly energetic cosmic neutrinos are emitted in a number of models of astrophysical transient events, such as gravitational wave events from compact object mergers~\cite[\textit{e.g.,}][]{Kotera:2016dmp,Fang:2017tla}, short and long gamma-ray bursts~\cite[\textit{e.g.,}][]{Murase:2007yt,Kimura:2017kan}, the birth of pulsars and magnetars~\cite[\textit{e.g.,}][]{Fang:2013vla,Fang:2018hjp}, tidal disruption events~\cite[\textit{e.g.,}][]{Lunardini:2016xwi}, blazar flares (\textit{e.g.}, TXS 0506+056~\cite{IceCube:2018cha,Aab:2020xql}), and possibly other high-energy transients.  In models of cosmic neutrino emission, neutrinos are typically produced in the decay of pions, kaons, and secondary muons generated by hadronic interactions in astrophysical sources~\cite{Stecker:1978ah}. Consequently, the expectation for the relative fluxes of each neutrino flavor at production in the cosmic sources, $(\nu_e:\nu_\mu:\nu_\tau)$, is nearly $\left(1 : 2 : 0\right)_{\rm source}$. After neutrino oscillations decohere over astronomical propagation distances, the flavor conversion is properly described by the mean oscillation probability. As a result, cosmic neutrinos should arrive at Earth with maximally mixed flavor ratios, $(1:1:1)_\oplus$, for most extragalactic sources~\cite{Learned:1994wg, Bustamante:2015waa}. POEMMA will observe, in general, one third of the generated neutrino flux via the $\nu_\tau$ flux.

\begin{figure*}[tb]
\centering
\includegraphics[width=0.5\linewidth]{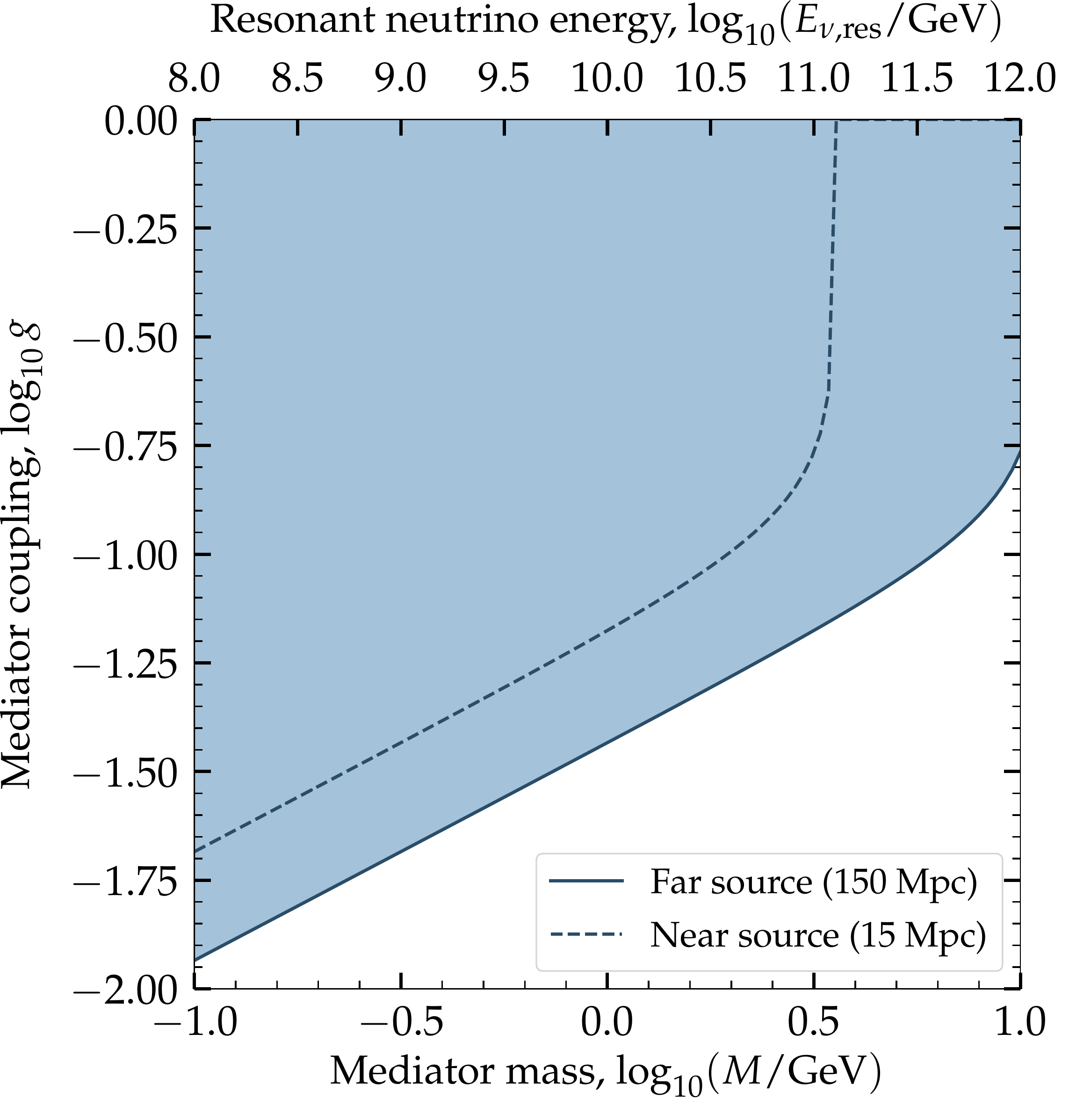}
\caption{Estimated sensitivity of POEMMA to secret neutrino interactions mediated by a new scalar with mass $M$ and coupling strength $g$.  The shaded region can be excluded by observing neutrinos at the resonant energy $E_{\nu,{\rm res}}$ coming from a point source.  GeV-scale mediator masses are presently unconstrained, but could be within reach POEMMA.  For this plot, we fixed the neutrino mass at $m_\nu = 0.05$~eV.
\label{fig:nusi_sens}}
\end{figure*}

Figure\ \ref{fig:nusi_sens} shows that by detecting UHE neutrinos from point sources (see Section\ \ref{section:too_performance}), POEMMA will be able to test {\it secret neutrino interactions}\ \cite{Kolb:1987qy} in a previously unexplored region of parameter space.  Secret neutrino interactions are potential BSM neutrino-neutrino ($\nu\nu$) interactions motivated as solutions to important open issues, including tensions in cosmology\ \cite{Cherry:2014xra, Barenboim:2019tux, Blinov:2019gcj, Escudero:2019gvw} and the origin or neutrino mass\ \cite{Chikashige:1980ui, Gelmini:1980re, Georgi:1981pg, Gelmini:1982rr, Nussinov:1982wu, Blum:2014ewa}.  They occur via a new mediator that couples mainly to neutrinos, whose mass $M$ and coupling strengths $g_{\alpha\beta}$ ($\alpha, \beta = e, \mu, \tau$) are not fixed a priori; here, for simplicity, we assume that the coupling is diagonal and flavor-universal, {\it i.e.}, $g_{\alpha\beta} \equiv g \delta_{\alpha\beta}$.  The $\nu\nu$ cross section becomes resonant at neutrino energy $E_{\nu, {\rm res}} = M^2/(2 m_\nu)$, where $m_\nu$ is the mass of the interacting neutrino.  So far, secret interactions remain unobserved, but they are constrained by a variety of cosmological, astrophysical, and laboratory measurements.  However, constraints on GeV-scale mediator masses are weak or non-existent; see, e.g., \cite{Blinov:2019gcj, Bustamante:2020mep}.  
If there are secret neutrino interactions mediated by GeV-scale masses,  UHE neutrinos emitted by a source could interact with the low-energy relic neutrino background en route to Earth.  If the interactions are strong, they suppress the flux of UHE neutrinos, and none would reach Earth.  This has been studied before in the context of supernova neutrinos and TeV--PeV astrophysical neutrinos\ \cite{Kolb:1987qy, Farzan:2014gza, Ioka:2014kca, Ng:2014pca, Ibe:2014pja, Blum:2014ewa, DiFranzo:2015qea, Kelly:2018tyg, Murase:2019xqi, Shalgar:2019rqe, Bustamante:2020mep}.  Figure\ \ref{fig:nusi_sens} shows the region of parameter space of $M$ and $g$ where the neutrino interaction length is less than 150 Mpc (a far source) or 15 Mpc (a near source).  Inside this region, the flux of emitted UHE neutrinos is suppressed by secret interactions.  Therefore, if POEMMA detects UHE neutrinos from a source located at either of those distances, it would allow us to disfavor a new region of parameter space.  

At the EeV neutrino energy scale, POEMMA may test neutrino emission from
relics from the early universe.  One example involves cosmic neutrinos produced via the decay of highly boosted, strongly coupled moduli---scalar fields---radiated by relic cosmic strings~\cite{Berezinsky:2011cp}, which are quite generic in the string theory landscape~\cite{Conlon:2007gk,AbdusSalam:2007pm}.

A recent BSM possibility was the report by the ANITA
experiment. The observation comprised two intriguing up-going showers with deposited
energies in the range $10^8 \lesssim E/{\rm GeV} \lesssim
10^9$~\cite{Gorham:2016zah,Gorham:2018ydl}.  These anomalous events could originate in the atmospheric decay of an up-going \tauon
produced through the charged-current interaction of a $\nu_\tau$ inside
the Earth. However, the relatively steep arrival angles of these
perplexing events are in  tension with the Standard Model
neutrino-nucleon cross section: the column depth through the Earth for these events is approximately 10 times the neutrino interaction length.
The lack of similar observations from Auger and IceCube implies
that the particle giving rise to ANITA events must
produce an air-shower event rate at least a factor of 40 larger than
that produced by a flux of $\nu_\tau$~\cite{Romero-Wolf:2018zxt}. 
Thus, while the ANITA anomalous events might have a BSM origin~\cite{Cherry:2018rxj,Anchordoqui:2018ucj,Huang:2018als,Dudas:2018npp,Connolly:2018ewv,Fox:2018syq,Collins:2018jpg,Chauhan:2018lnq,Anchordoqui:2018ssd}, recent observations suggest that systematic effects may play a larger role than first anticipated \cite{Gorham:2020zne}. Larger statistics are needed to test this.  POEMMA will have the required FoV to observe anomalous events with high statistics and put a BSM origin to test.  It should be noted that both the PCC and PFC have sensitivity to the Cherenkov signal from BSM upward EAS events, albeit with a higher threshold energy for the PFC. Thus POEMMA will be sensitive to BSM events while in POEMMA-Stereo mode as well as POEMMA-limb observation modes.

\begin{figure*}[ht]
\begin{minipage}[t]{0.47\textwidth}
  \postscript{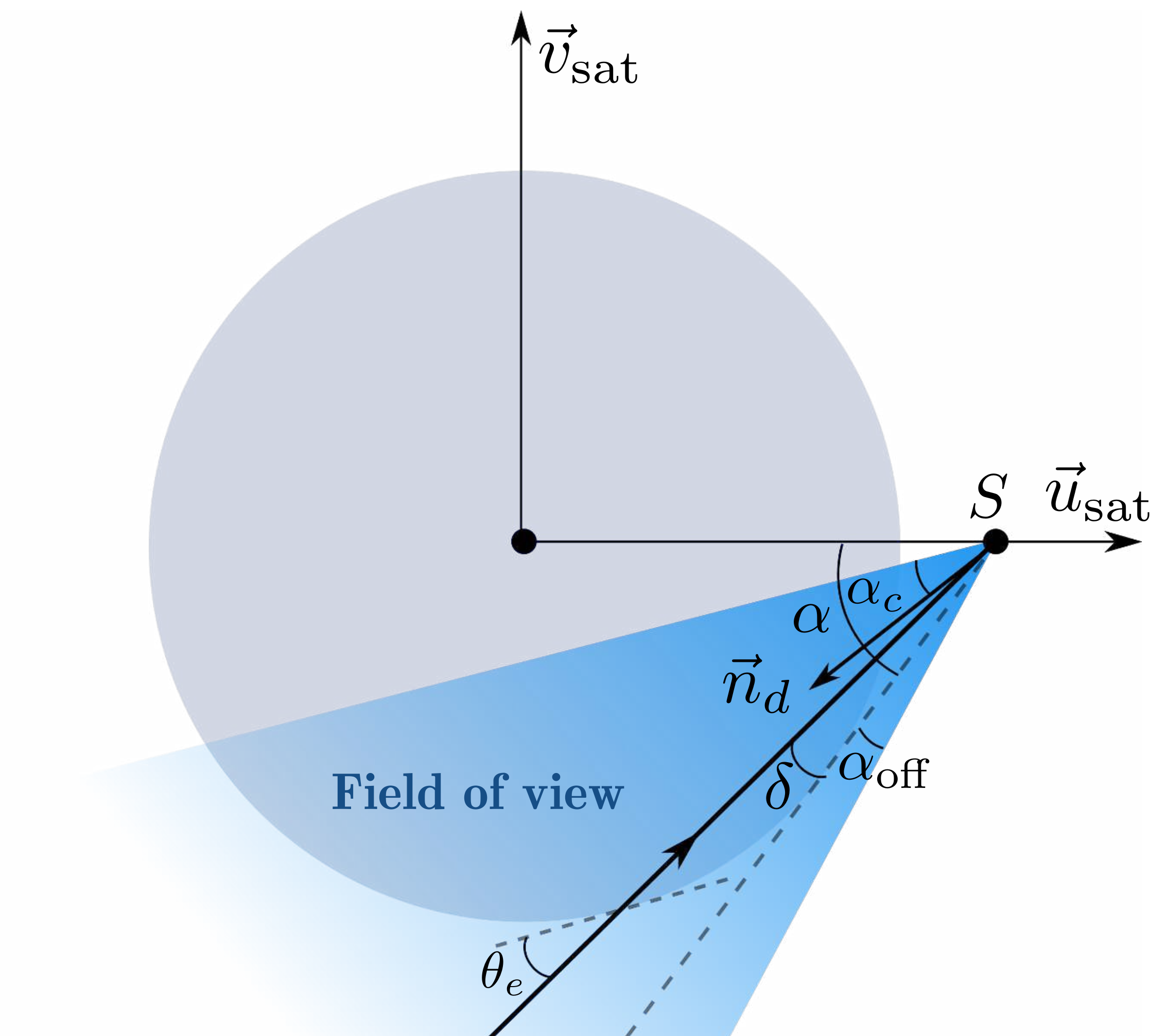}{1}
\end{minipage}
\hfill 
\begin{minipage}[t]{0.52\textwidth}
   \postscript{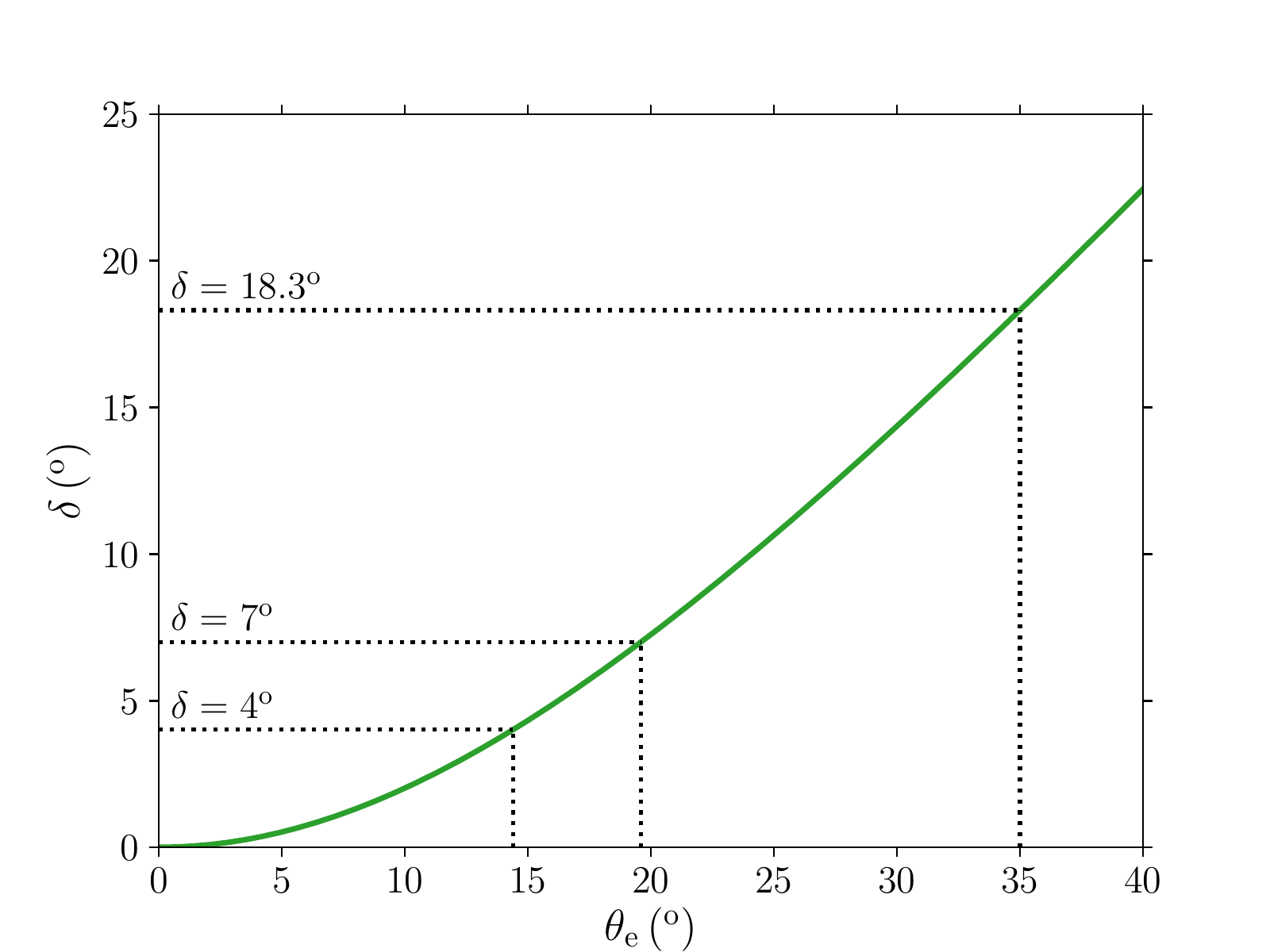}{1}
\end{minipage}
\caption{{\it Left:} Illustration of the geometrical configuration in the orbital plane (satellite position, $\vec{u}_{\rm sat}$, versus satellite velocity $\vec{v}_{\rm sat}$). 
The satellite is located at point S. The arrival direction of an EAS generated by a $\nu_{\tau}$ is characterized by its Earth emergence angle $\theta_e$ and the corresponding angle away from the limb $\delta$ from the point of view of the satellite. The detector has a conical FoV of opening angle $\alpha_c$, with an offset angle 
$\alpha_{\rm off}$ (away from the Earth's limb) and pointing direction $\vec{n_d}$. {\it Right:} Cherenkov viewing angle $\delta$ below the limb versus Earth emergence angle $\theta_e$~\cite{Guepin2018}.
\label{fig17}}
\end{figure*}

\begin{figure*}[tb]
\centering
\includegraphics[width=0.485\linewidth]{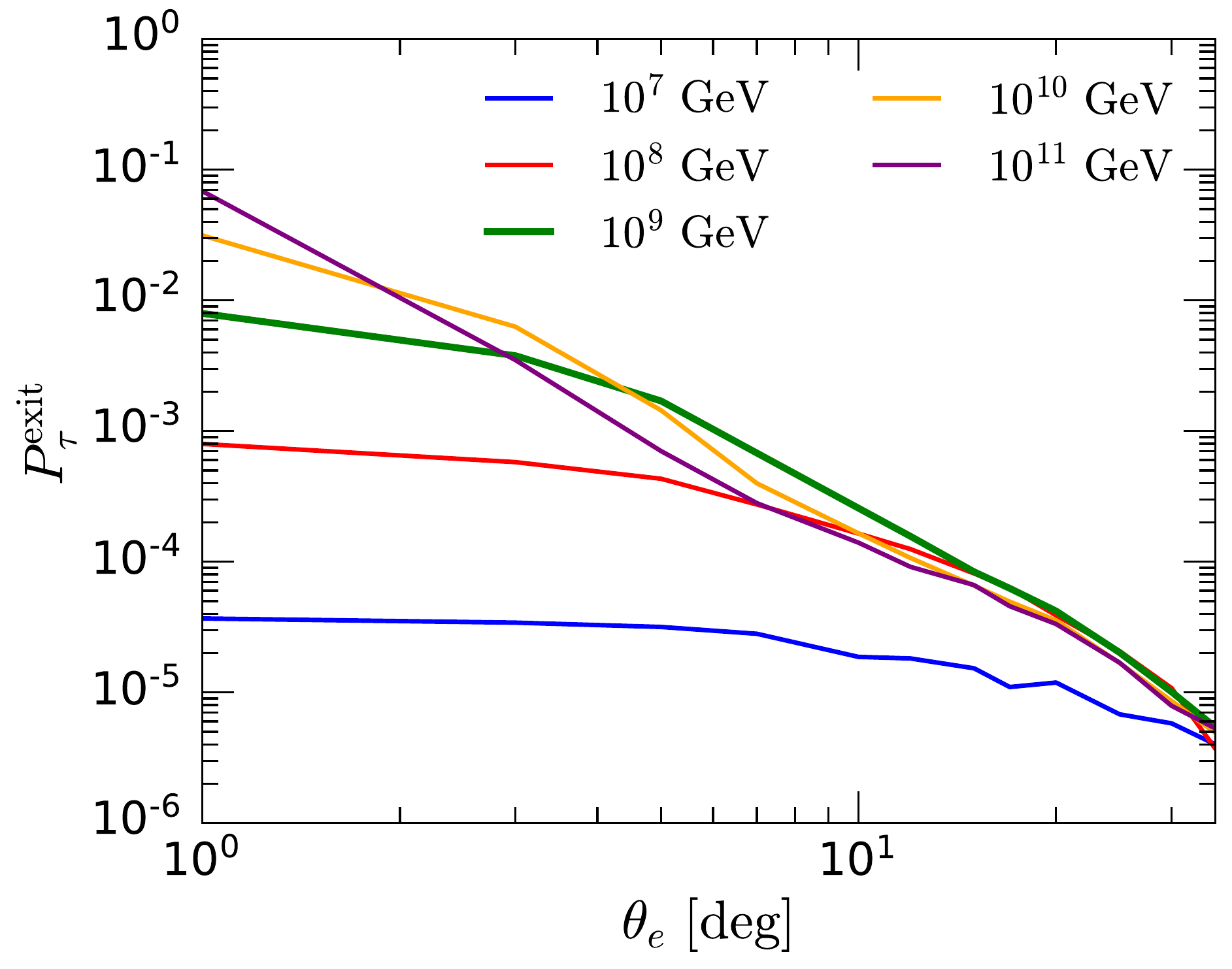}
\includegraphics[width=0.505\linewidth]{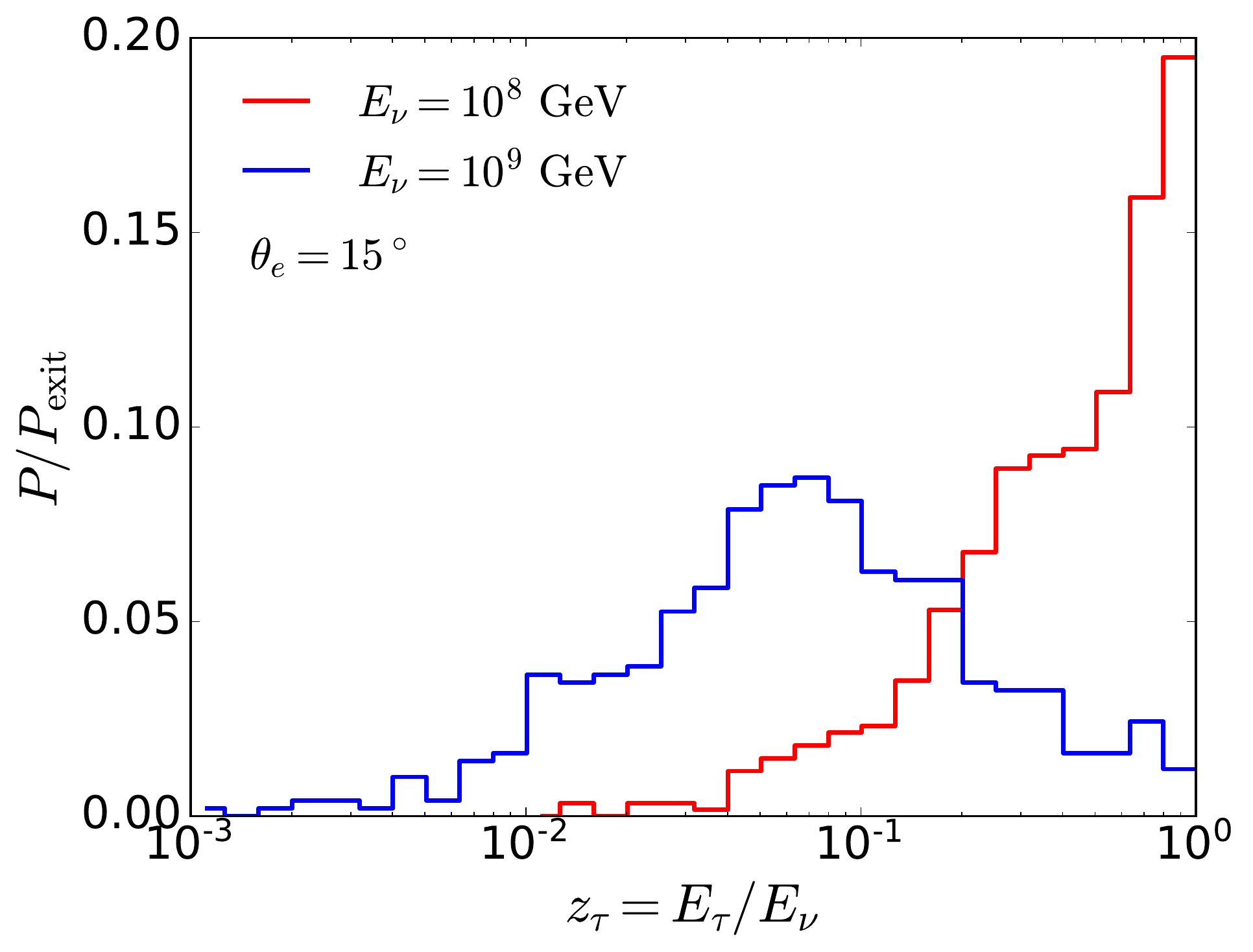}
\caption{{\it Left:} The probability for a \nutau to produce a \tauon which exits the Earth, as a function of Earth emerging angle
relative to horizontal, $\theta_e=1^\circ-35^\circ$, for incident neutrino energies of $10^7$, $10^8$, $10^9$, $10^{10}$ and $10^{11}$ GeV. The ALLM \tauon energy loss model is used, as described in \cite{Reno2019a}. {\it Right:} The \tauon energy distribution as a function of $z=E_\tau/E_\nu$ for \tauons exiting at an Earth emergence angle $\theta_e=15^\circ$ for 
$E_\nu=10^8$ and $10^9$ GeV.
\label{figPexit}}
\end{figure*}

\subsection{POEMMA Performance for Neutrino ToO Observations}
\label{section:too_performance}

POEMMA's sensitivity to the cosmic $\nu_\tau$ flux is based on the observation of Cherenkov emission from EAS caused by the decay of \tauons as they exit the Earth's surface. Observable \tauon decay events for POEMMA start in  directions close to the limb of the Earth located at 67.5$^{\circ}$ from the nadir for POEMMA's 525 km altitude. The geometry of the \tauon emergence angles with respect to the Earth's surface ($\theta_e$) and the corresponding observing angle below the limb for POEMMA ($\delta$) are defined in top panels of Figure~\ref{fig17}.  Assuming Standard Model interactions and observable energies, the Earth emergence angles tend to be below $\theta_e \ls 35^{\circ}$ corresponding to POEMMA observations with  $\delta \ls 18.3^{\circ}$ below the limb. 
(A complete study of the geometry and the corresponding sky coverage of neutrino induced Cherenkov events observable by POEMMA can be found in ~\cite{Guepin2018}.)

The left panel of Figure~\ref{figPexit} shows the 
probability that an incident $\nu_\tau$ emerges from the Earth as a \tauon, as a function of the Earth
emergence angle $\theta_e$. Results are shown for incident $\nu_\tau$ energies of $10^7,\ 10^8,\ 10^9,\ 10^{10}$ and $10^{11}$ GeV for \tauons that emerge from the Earth with angles
$\theta_e=1^\circ-35^\circ$. The right panel shows an example of the emerging \tauon energy distribution for two incident neutrino energies ($10^8,\ 10^9$ GeV) for $\theta_e=15^\circ$.
These results are based on a new calculation developed under the POEMMA probe study~\cite{Reno2019a}.
The new $\nu_\tau$ interaction and \tauon energy loss modeling uses a layered density of the Earth based on the Preliminary Reference Earth Model~\cite{Dziewonski}. 
The $\nu_\tau$ create \tauons in
charged-current interactions with nucleons in the Earth. At the high
energies of interest for POEMMA, the resulting $\tau$ leptons lose energy while propagating through the Earth and have a finite, energy-dependent
probability of escaping the Earth. The 
process is aided by $\nu_\tau$ regeneration~\cite{Halzen:1998be,Reno2019a}:
$\tau$-leptons that decay in the Earth produce lower-energy $\nu_\tau$
that can themselves interact and make new $\tau$-leptons, enhancing the \tauon exit probabilities. 

The left panel of Figure~\ref{figPexit} shows that, because the neutrino cross section increases with energy~\cite{Block:2014kza}, the $\tau$-lepton exit probability is larger at smaller Earth emergence angles $\theta_e$, i.e., for smaller columns depths, since there is less neutrino attenuation.  Larger column depths suppress
the exit probabilities, and the $\tau$-leptons that exit are more likely to be the product of 
regeneration~\cite{Halzen:1998be}. 
The right panel of Figure \ref{figPexit} shows that, even for a fixed column depth ($\theta_e=15^\circ$), regeneration effects are evident.  The energy distribution of the emerging $\tau$-leptons is shifted to a lower energy fraction of the initial neutrino energy for $E_\nu=10^9$~GeV, compared to the case with incident neutrino energy $E_\nu=10^8$~GeV.

\begin{figure*}[ht]
\begin{minipage}[t]{0.49\textwidth}
  \postscript{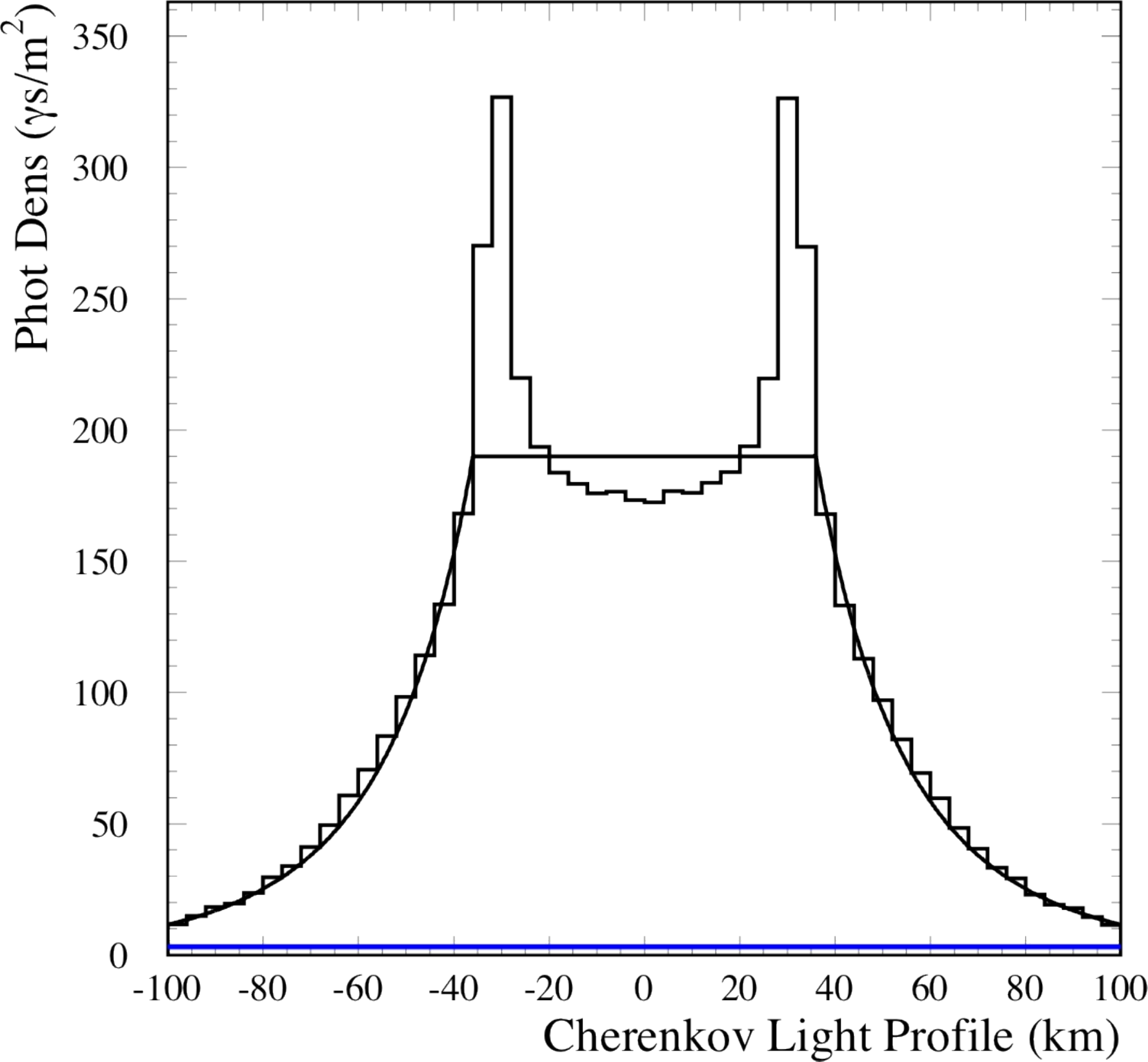}{0.9}
\end{minipage}
\begin{minipage}[t]{0.49\textwidth}
   \postscript{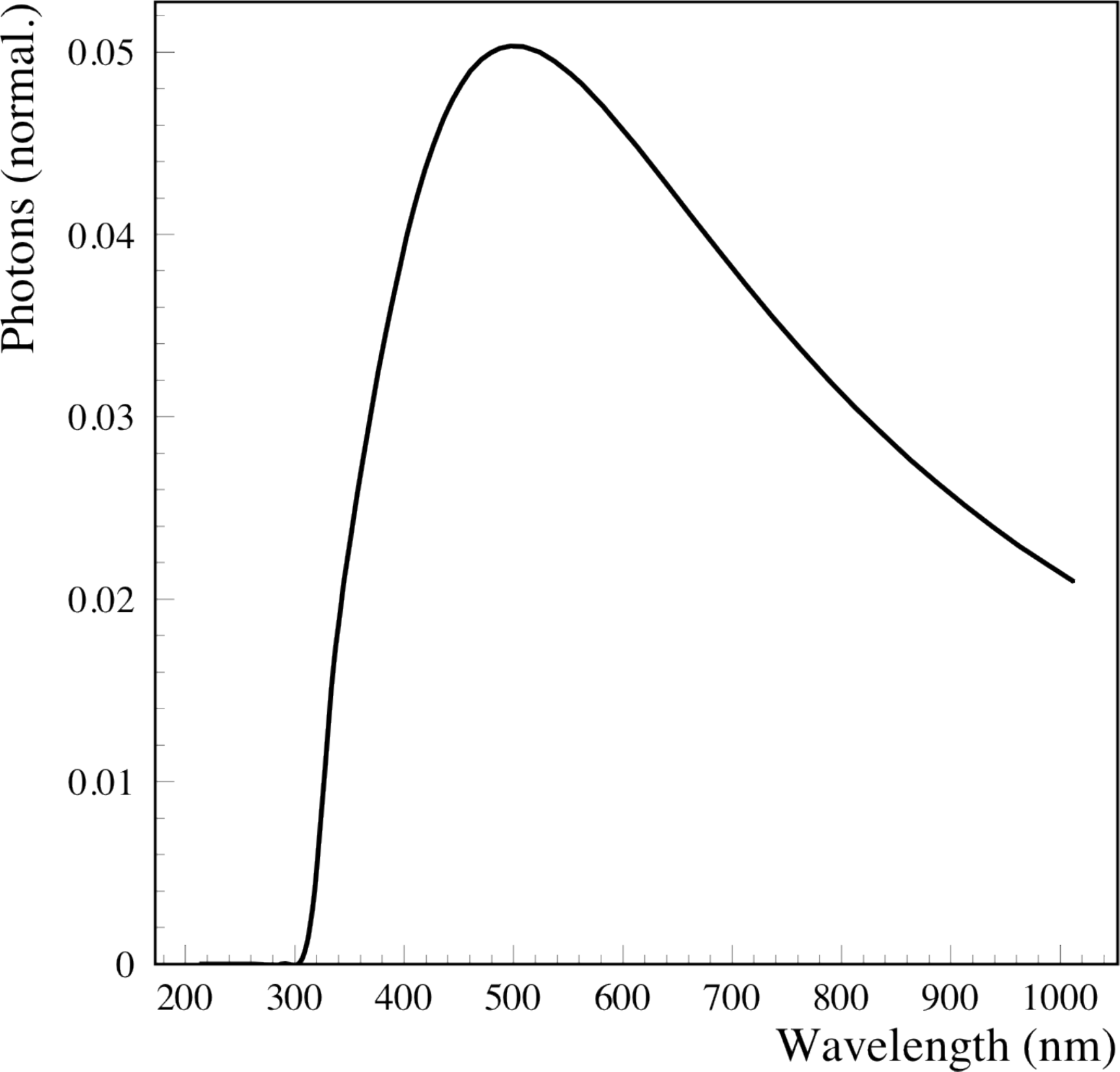}{0.9}
\end{minipage}
\caption{{\it Left:} The spatial profile of the Cherenkov signal
(photons/m$^2$) at 525~km altitude for a 100~PeV upward EAS with
a $15^\circ$ Earth emergence angle. {\it Right:} The simulated Cherenkov spectrum for this EAS observed by a POEMMA telescope, which is well matched to the wavelength response of the PCC.\label{figCprof}}
\end{figure*}

\begin{figure*}[ht]

\begin{minipage}[t]{0.49\textwidth}
\vspace{0.5ex}
   \postscript{Figures/100PeVShowers-CherenkovSpectra-vs-alt-10deg.pdf}{0.9}
\end{minipage}
\begin{minipage}[t]{0.49\textwidth}
\vspace{0.5ex}
 \postscript{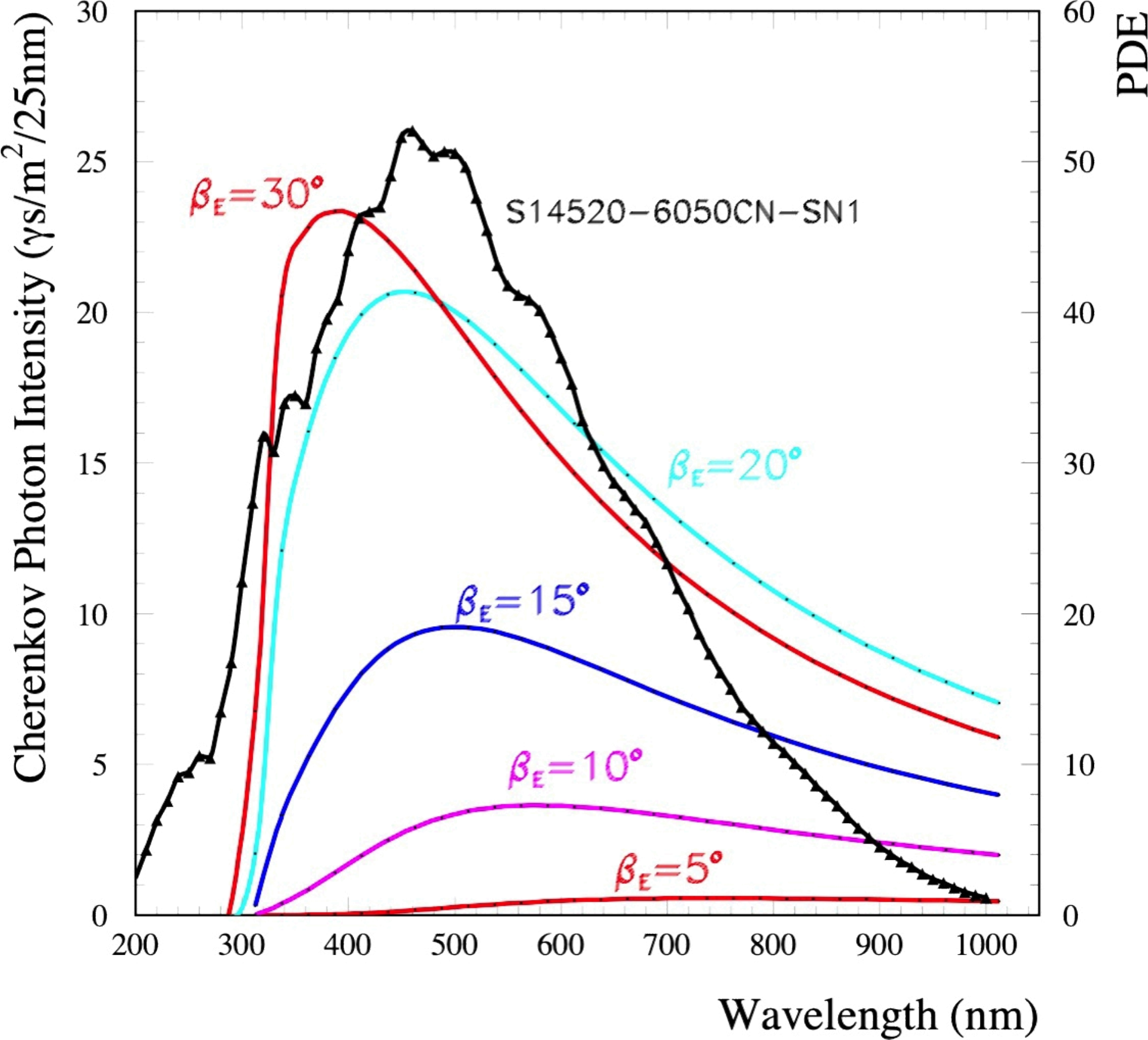}{0.94}
\end{minipage}
\caption{{\it Left:} Cherenkov signal intensity as function of wavelength for 100~PeV upward-moving EASs for $\theta_e = 10^\circ$ Earth emergence angle for a different  EAS starting altitude. {\it Right:} Cherenkov signal intensity as function of wavelength for 100~PeV upward-moving EASs starting at sea level as a function of Earth emergence angle. The measured photon detection efficiency (PDE) of a Hamamatsu S14520 SiPM array \cite{NepomukOtte:2018qjk} is overlaid with the PDE scale given on the right horizontal axis.
\label{figCvar}}
\end{figure*}

\begin{figure*}[ht]
\begin{minipage}[t]{0.48\textwidth}
    \postscript{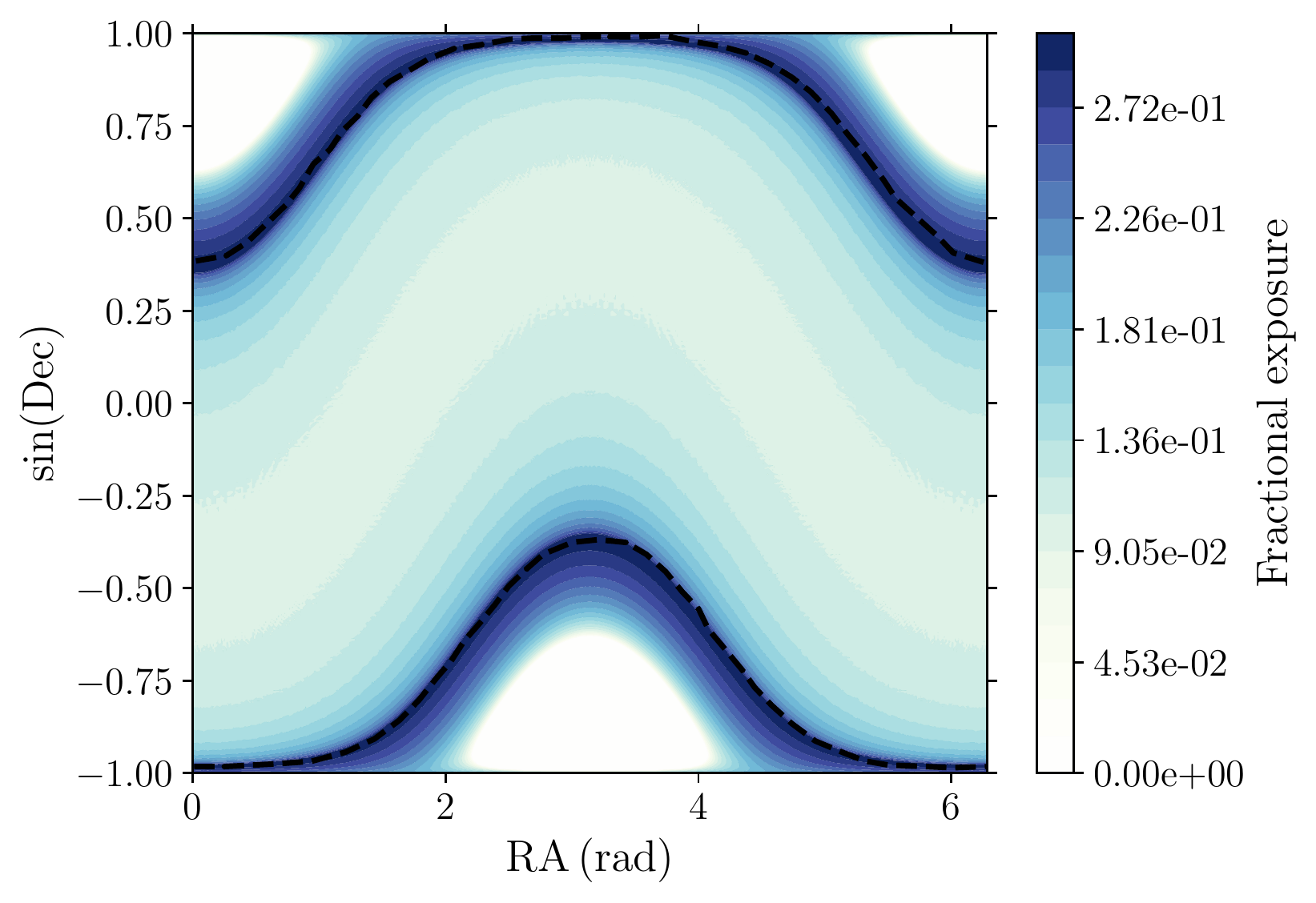}{1}
\end{minipage}
\hfill \begin{minipage}[t]{0.5\textwidth}
  \postscript{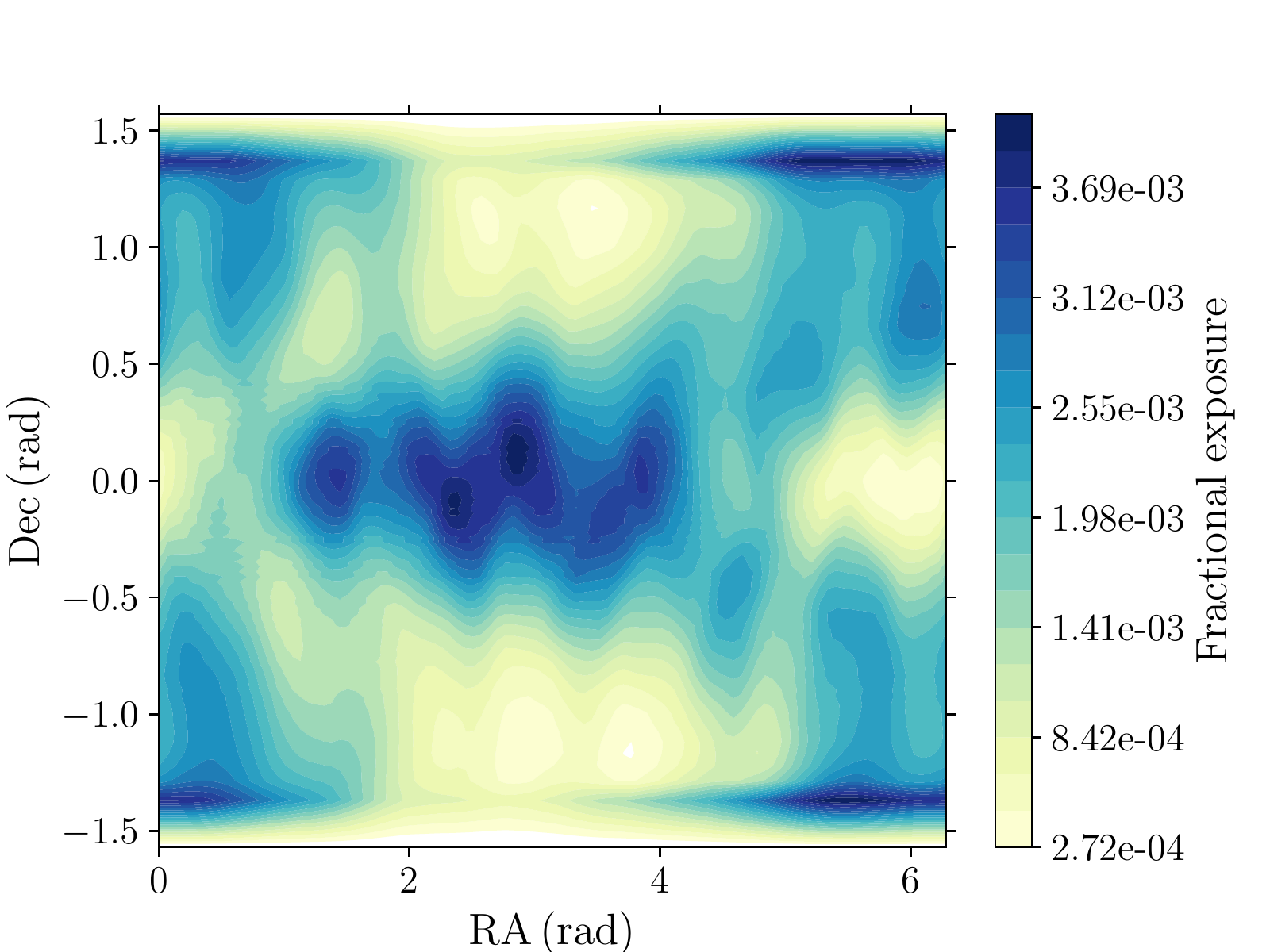}{1}
\end{minipage}
\caption{POEMMA cosmic neutrino sky coverage
{\it Left:} Sky coverage for sources at a given orbital position in the sine of the declination and right ascension, without including the effect of the Sun, at a  given time of the year for viewing angles to $\delta = 18.3^\circ$ below the limb ~\cite{Venters:2019xwi}. {\it Right:} The fractional neutrino sky exposure for one year in declination versus right ascension, assuming a defined variation in the POEMMA limb-pointing directions over the year to achieve full-sky coverage. The calculation takes into account the effects of the sun and moon on the duty cycle for observations~\cite{Guepin2018}. 
\label{fig20}}
\end{figure*}

Once the $\tau$-lepton escapes the Earth, an EAS model is used to
develop the EAS cascade, generate the beamed Cherenkov light, and
attenuate the light using an atmospheric model that includes the
wavelength-dependent attenuation due to aerosols, Rayleigh scattering,
and ozone absorption. The interplay between the $\tau$-lepton
Earth-emergence angle, the $\tau$-lepton energy (and hence decay altitude), the
Cherenkov light generation, and atmospheric absorption determines the
observability of an upward-moving EAS~\cite{Reno2019a}. 

Numerical simulations indicate that the Cherenkov light profile at POEMMA is essentially a flat top with a width of tens of km and a
power-law falloff (as shown in the top left panel of Figure~\ref{figCprof}).  This motivates the satellite separation during neutrino observations to be $\sim 25~{\rm km}$ in order for both satellites to observe a significant portion of the light pool of the same event, allowing the use of time coincidence to lower the detection energy threshold compared to that for a single observation.  POEMMA will observe the upward-moving EAS within a few degrees from the EAS propagation direction (see Figure~\ref{fig4}). The POEMMA optics will focus the Cherenkov signal in a pixel in the POEMMA Cherenkov Camera (PCC) with some spread due to the point-spread-function (PSF) of the optics. Simulations also show that the Cherenkov signal has a temporal width of $\sim 20~{\rm ns}$,  defining the sampling time for the PCC (the SiPM portion of the focal plane). 

The left panel of Figure~\ref{figCvar} illustrates the Cherenkov photon intensity from EASs from \tauon decay as a function of altitude for a fixed EAS energy of $100$~PeV and Earth emergence angle of $\theta_e = 10^\circ$. The right panel of Figure~\ref{figCvar} illustrates the Cherenkov photon intensity from EASs from \tauon decay as a function of Earth-emergence angle for a fixed EAS energy of $100$~PeV and sea level starting altitude. These plots illustrate the large variability in the Cherenkov signal spectra from these upward EASs, which motivated the need for a wide, $200$--$900$ nm wavelength response of the PCC.
The Cherenkov intensity and wavelength dependence are very sensitive to the starting altitude. At the lowest altitudes, molecular and aerosol scattering attenuate
the Cherenkov intensity and push the peak of the spectrum towards longer wavelengths. As the starting altitude increases, the exponential nature of both the aerosol layer and atmosphere itself leads to higher Cherenkov intensities and spectra peaked at lower wavelengths. At even higher altitudes the atmosphere becomes too thin for complete EAS development, leading to a reduction in the Cherenkov intensity for EAS developing at altitudes above $\sim 17$ km.

\begin{figure*}[ht]
\begin{minipage}[t]{0.49\textwidth}
 \postscript{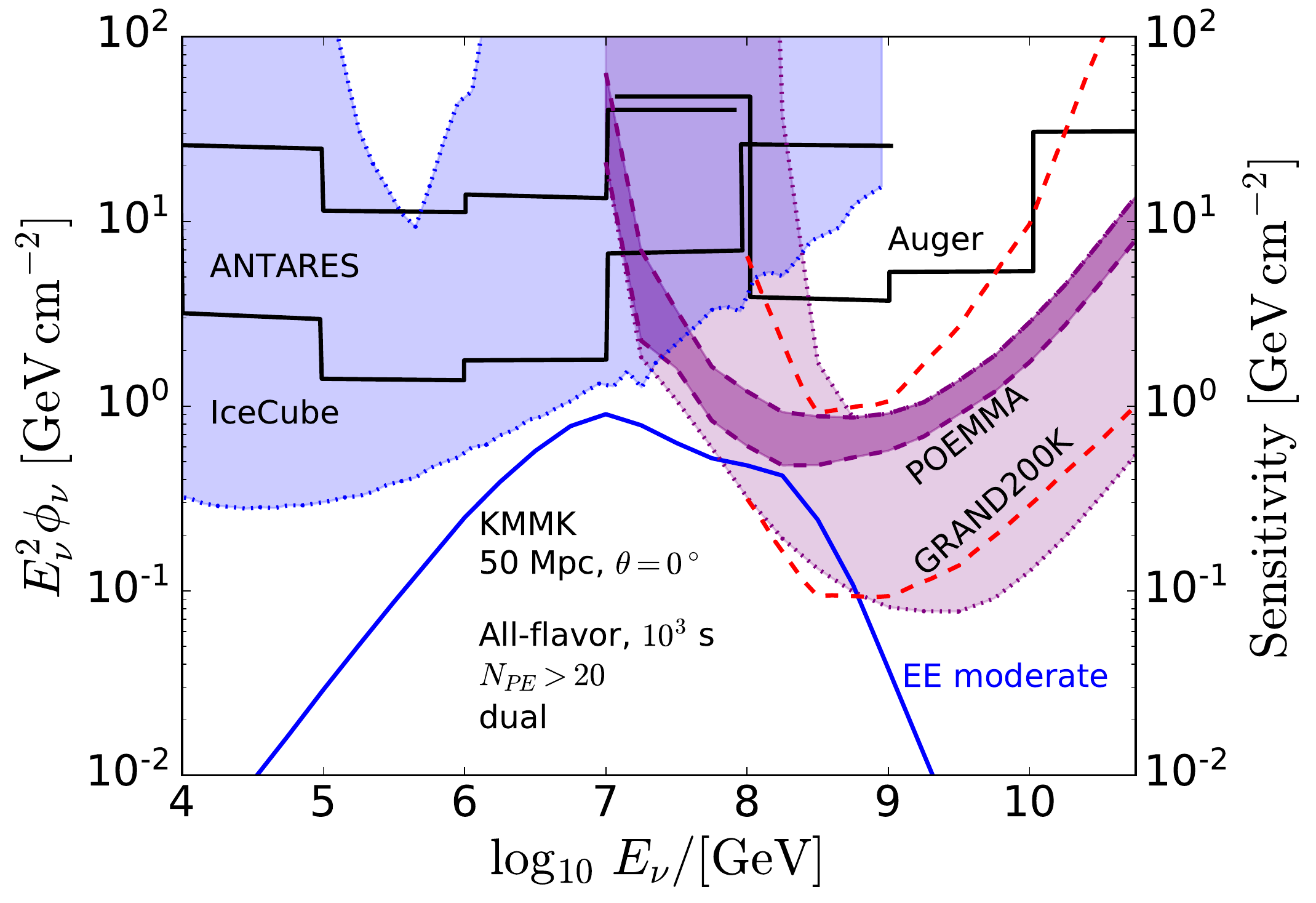}{1}
\end{minipage}
\hfill \begin{minipage}[t]{0.51\textwidth}
  \includegraphics[trim = 27mm 27mm 19mm 40mm, clip, width = 1.0\textwidth]{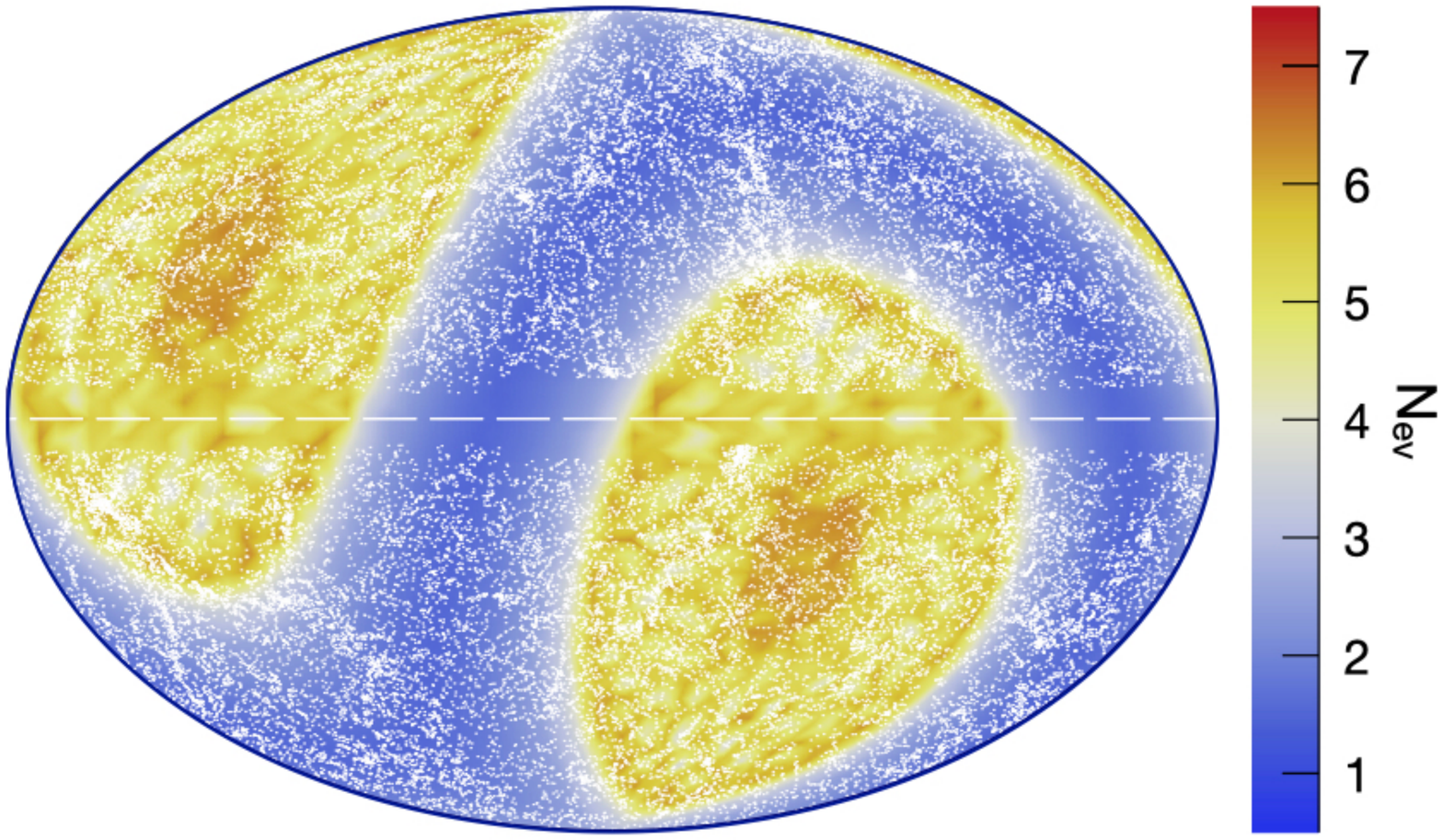}
\end{minipage}
\caption{{\it Left:} POEMMA ToO sensitivity to a short, 1000~s burst shown by the magenta band, where the
dark magenta band corresponds to source locations between the dashed curves in the sky coverage shown in Figure \ref{fig20}. Also shown are all-flavor upper limits from IceCube and Auger (solid histograms) for neutrino searches within $\pm 500$~s around the binary neutron star merger GW170817 ~\cite{ANTARES:2017bia}. The blue bands show the variation of IceCube sensitivity due to celestial source location derived from Ref. \citenum{Aartsen:2016oji} and the red dashed curves represent the projected sensitivity of GRAND200k at zenith angles $90^\circ$ and $94^\circ$ \cite{Alvarez-Muniz:2018bhp},
and models taken from Kimura et al. \cite{Kimura:2017kan} of the all-flavor neutrino fluence from a short gamma-ray burst during the prompt and extended emission (EE) phases, assuming on-axis viewing ($\theta=0^\circ$) and a source at $D=50$~Mpc.  Figure from Ref. \citenum{Venters:2019xwi}.
{\it Right:} Sky plot of the expected number of neutrino events with POEMMA as a function of galactic coordinates for the Kimura et al.~\cite{Kimura:2017kan} short gamma-ray burst with moderate EE model, placing the source at $50$~Mpc. Point sources are galaxies from the 2MRS catalog~\cite{2012ApJS..199...26H}.}
\label{fig19a}
\end{figure*}

As in the air fluorescence case, the brightness of the beamed Cherenkov signal compared to the airglow background sets the energy threshold for observation. The peak of the Cherenkov spectrum spans the $300 \mbox{ nm} < \lambda < 900 \mbox{ nm}$ band due to the dependence of atmospheric absorption on column depth and Earth emergence angle. The dark-sky airglow background is stronger in this wavelength band than for the UV fluorescence case \cite{2003A&A...407.1157H, 2006JGRA..11112307C}.
Simulations show that the time width of the Cherenkov signal for accepted events is $\lesssim 20~{\rm ns}$. Simultaneous viewing of the EAS from \tauon decay with both POEMMA satellites and using a coincidence window of $40~{\rm ns}$ results in a false positive Cherenkov signal of $\sim 0.1\%$, due to the atmospheric air glow background, with sensitivity to $\nu_\tau$ down to $\sim 20~{\rm PeV}$.

A key feature of the POEMMA design is the capability for rapid follow-up of ToOs. Various models of neutrino emission for astrophysical transients predict short and/or long-term neutrino emission associated with the transient event. To search for neutrino bursts with durations of less than a day, POEMMA can reorient its viewing angle in a matter of minutes, slewing up to $90^\circ$ in 500~s. For such short-duration events, the observatories will be operating in the ToO-dual configuration with separate light pools doubling the effective area at higher energies and the optimal sensitivity being achieved for sources that set below the horizon during the event~\cite{Venters:2019xwi}. 
For longer-duration events, POEMMA can monitor the location of a given transient months after its multi-wavelength discovery, and the satellite separation can be reduced to around $25$~km to allow for observations in the ToO-stereo configuration, lowering the energy threshold. The Earth's orbit around the Sun, and the precession of the satellites' orbital plane, allow for full-sky coverage over a few-month timespan, ensuring that long-duration events will come into view regardless of celestial position~\cite{Guepin2018}.

\begin{figure*}[ht]
\begin{minipage}[t]{0.49\textwidth}
\postscript{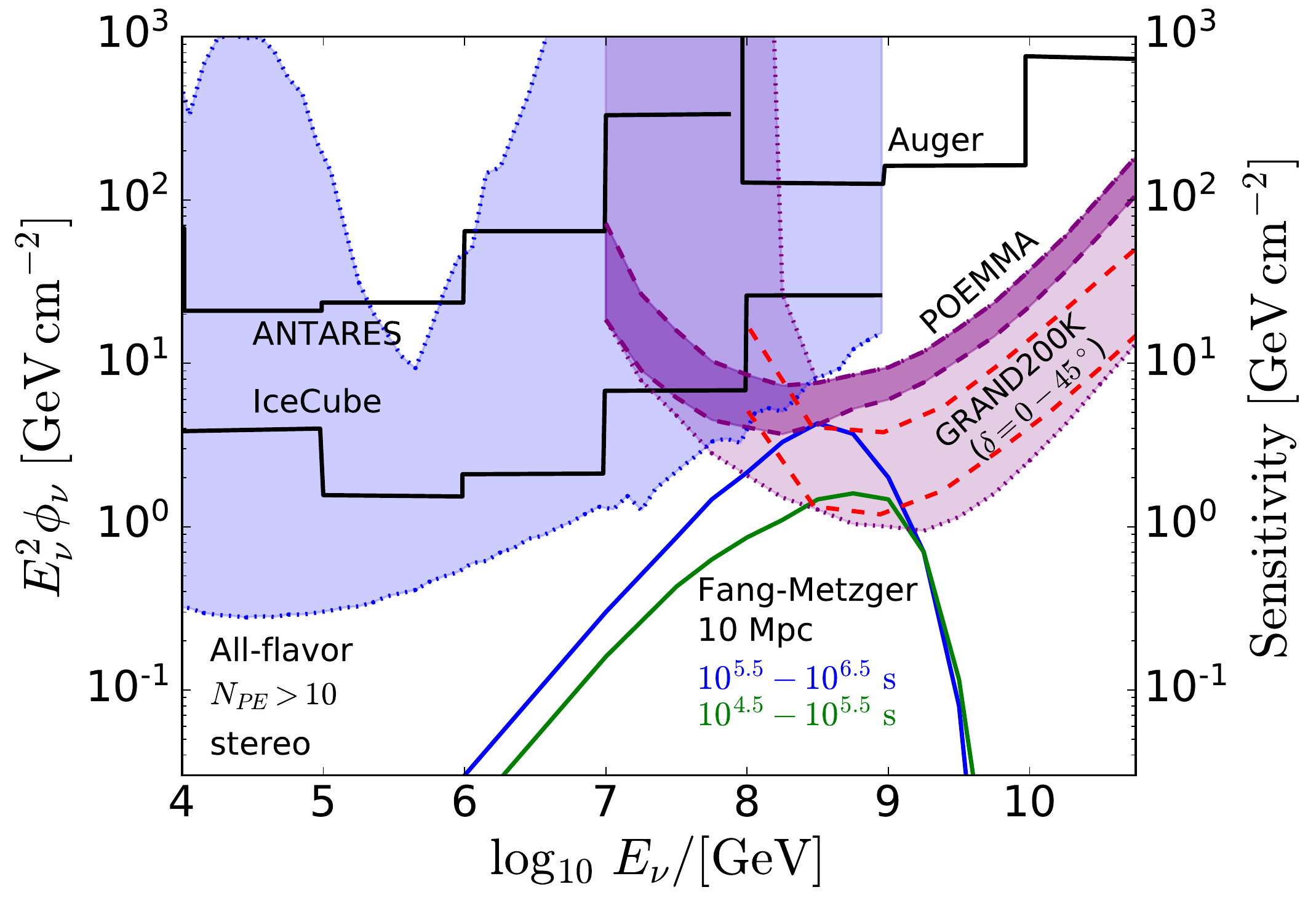}{1}
\end{minipage}
\hfill \begin{minipage}[t]{0.51\textwidth}
  \includegraphics[trim = 27mm 27mm 19mm 40mm, clip, width = 1.0\textwidth]{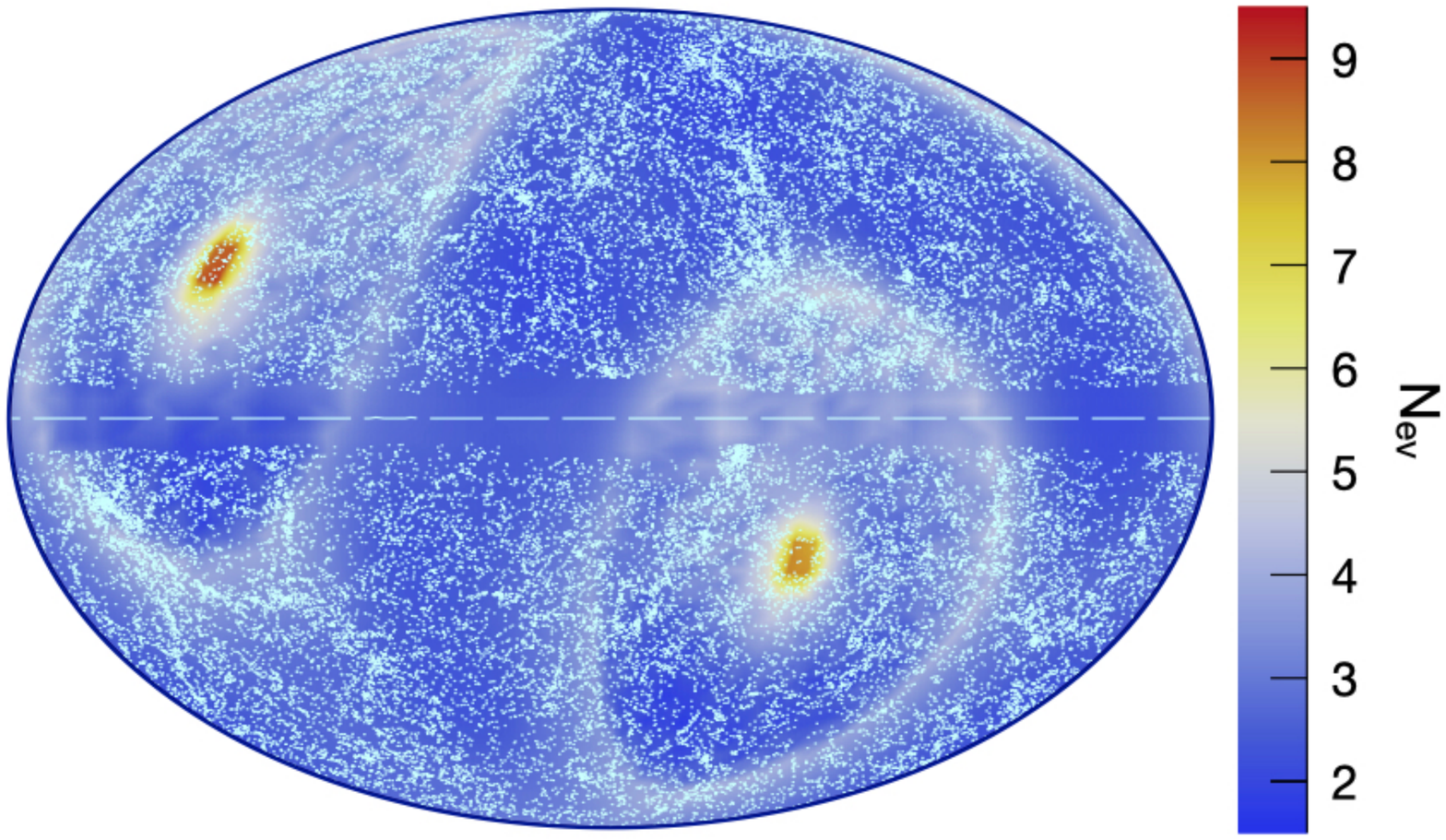}
\end{minipage}
\caption{{\it Left:} POEMMA ToO sensitivities to a long burst shown by the magenta band, where the
dark magenta band corresponds to source locations between the dashed curves in the sky coverage Figure \ref{fig20}. Also shown are the IceCube all-flavor upper limits (solid histogram)from a neutrino search within a 14-day time window around the binary neutron star merger GW170817~\cite{ANTARES:2017bia}. The blue bands show the variation of IceCube sensitivity due to celestial source location derived from Ref. \citenum{Aartsen:2016oji} and the red dashed curves represent the projected sensitivity of GRAND200k at zenith angles $90^\circ$ and $94^\circ$ \cite{Alvarez-Muniz:2018bhp},
and models from Fang \& Metzger~\cite{Fang:2017tla} of the all-flavor neutrino fluence produced $10^{5.5}-10^{6.5}$~s and $10^{4.5}-10^{5.5}$~s after a binary neutron star merger event occurring at a distance of $10$~Mpc. Figure from Ref.  \citenum{Venters:2019xwi}.
{\it Right:} Sky plot of the expected number of neutrino events with POEMMA as a function of galactic coordinates for the Fang \& Metzger~\cite{Fang:2017tla} binary neutron star merger model, placing the source at $10$~Mpc. Point sources are galaxies from the 2MRS catalog~\cite{2012ApJS..199...26H}.}
\label{fig19b}
\end{figure*}

To summarize, the Cherenkov observations of upward-moving $\tau$-lepton induced EAS are based on a neutrino optical Cherenkov Monte Carlo with a POEMMA PCC instrument response model that accounts for the effects of the (dominant) dark-sky air glow background. This Monte Carlo was used to calculate the neutrino sensitivities to the short- and long-burst transient events shown in Figs. \ref{fig19a} and \ref{fig19b}.  Studies of the neutrino energy resolution show that it is a strong function of the Earth-emergence angle of the $\tau$-lepton, which is well-measured ($\lsim 1.5^\circ$) by POEMMA, as well as the energy spectrum of the incident neutrinos. From these studies, the energy resolution can span a fraction of a decade in energy, which is not surprising considering the nature of the processes involved in the chain, i.e. $\nu_\tau \rightarrow$ \taon, \taon energy loss in Earth, \taon decay in the atmosphere, EAS and Cherenkov light generation, atmospheric attenuation of the Cherenkov light, and  POEMMA detection of the Cherenkov light.

POEMMA's exposure for cosmic $\nu_\tau$ sources for one orbital period traces out a band on the celestial sky defined by the inclination of the orbit and
the off-orbit angle for the pointing direction of the telescopes. The left panel of Figure~\ref{fig20} shows the fractional coverage for positions on the sky (given as right ascension and sine of the declination) over the course of a given day of the year assuming detector viewing angles of $\delta = 0^\circ$ to $\delta = 18.3^\circ$ below the limb ~\cite{Venters:2019xwi}. Some sources are located in sky positions that never set below the horizon (shown in white in Figure~\ref{fig20} left) and will only be observed when the Earth's orbit brings that part of the sky into the detectable regions for $\nu_\tau$. The majority of the sky positions in the left panel of Figure~\ref{fig20} are for sources that rise and set at angles close to the orbital plane. A few areas in the sky have one order of magnitude more fractional exposure because sources in this region stay just below the horizon during a portion of POEMMA's orbit: these are shown in the darker blue colors in the left panel of Figure~\ref{fig20}.

The right panel of Figure~\ref{fig20} shows the one-year $\nu_\tau$ sky
coverage, based on a specific set of defined repoints for each orbital
period, demonstrating the ability to cover the full sky yearly ~\cite{Guepin2018}.   The figure illustrates the unique capability of POEMMA to adjust its observing strategy to benefit from the flexibility of space-based observations. This makes it possible to follow 
cosmic neutrino sources over the full sky. An in-depth analysis of the best cosmic neutrino target for POEMMA should optimize repoints during a given observational period giving priority to the most likely high-energy neutrino sources to be followed-up as ToOs for different time windows after each transient.

\begin{table*}[ht]
\begin{minipage}[t]{0.5\textwidth}
  \centering
  \vspace{30pt}
  \includegraphics[trim = 50mm 20mm 45mm 20mm, clip, width = 1.0\textwidth]{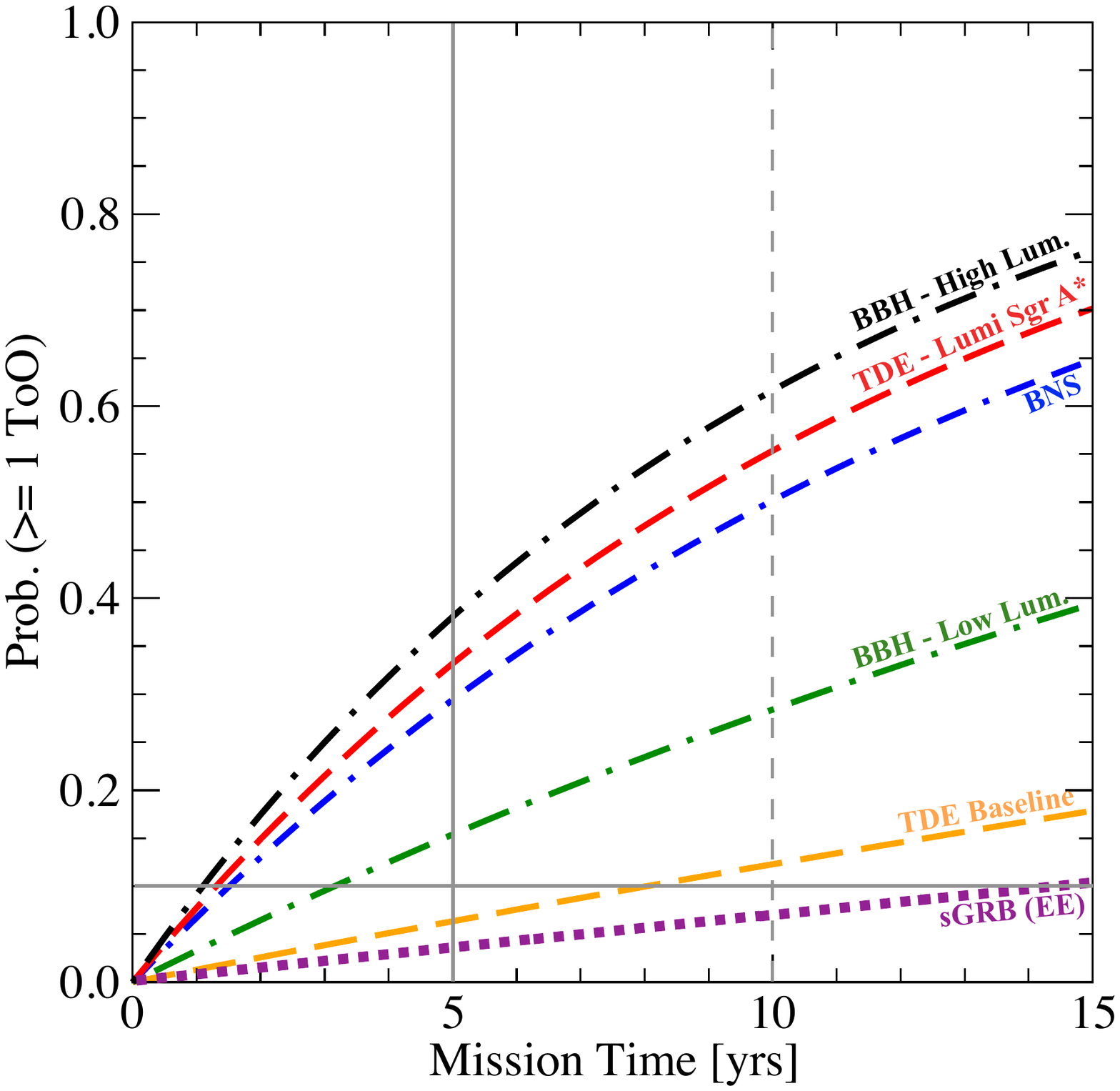}
\end{minipage}
\begin{minipage}[t]{0.5\textwidth}
\centering
\vspace{0pt}
\begin{tabular}{|C{0.73in}|C{0.55in}|C{0.55in}|C{0.55in}|} 
 \hline
 \hline
 \footnotesize{Source Class} & \footnotesize{$\nu$ Horizon Distance} & \footnotesize{Mission Time for $10$\% Prob.} & \footnotesize{Model Reference} \\ [0.5ex]
 \hline
 \footnotesize{TDE $M_{\rm SMBH} = 5 \times 10^6 M_{\odot}$ Lumi Scaling} & \footnotesize{$128$ Mpc} & \footnotesize{1.5~yrs.} & \footnotesize{\cite{Lunardini:2016xwi}} \\
 \hline
 \footnotesize{TDE Base Scenario} & \footnotesize{$69$ Mpc} & \footnotesize{8~yrs.} & \footnotesize{\cite{Lunardini:2016xwi}} \\
 \hline
 \footnotesize{BH-BH merger Low Fluence} & \footnotesize{$43$~Mpc} & \footnotesize{3~yrs.} & \footnotesize{\cite{Kotera:2016dmp}} \\
 \hline
 \footnotesize{BH-BH merger High Fluence} & \footnotesize{$137$~Mpc} & \footnotesize{1~yr.} & \footnotesize{\cite{Kotera:2016dmp}} \\
 \hline
 \footnotesize{NS-NS merger} & \footnotesize{$16$~Mpc} & \footnotesize{1.5~yrs.} & \footnotesize{\cite{Fang:2017tla}} \\
 \hline
 \footnotesize{sGRB Moderate Extended Emission} & \footnotesize{$90$~Mpc} & \footnotesize{14.5~yrs.} & \footnotesize{\cite{Kimura:2017kan}} \\
 \hline
 \hline
\end{tabular}
\end{minipage}
\caption{{\it Left:} Poisson probability of POEMMA detecting at least one ToO versus mission time for several modeled source classes. {\it Right:} Promising source classes for detecting at least one ToO event with POEMMA based on a Poisson probability of $\gtrsim 10$\%. Also included are the horizon distance for detecting one neutrino per ToO event, the mission time for $10$\% chance of detecting $\geq 1$ ToO event, and the model reference. \label{figPTab}}
\end{table*}

Figures~\ref{fig19a} and \ref{fig19b} provide performance plots for POEMMA ToO observations for both short- and long-duration astrophysical transient events. The left panel of Figure~\ref{fig19a} shows the sensitivity for short bursts of $\sim$1000~s, assuming that POEMMA is in the ToO-dual configuration and that sources are in the observable part of the sky. The lighter blue band shows the range of sensitivities achievable by POEMMA depending on the source location on the sky. The dark blue band corresponds to sky locations between the dashed curves in the sky coverage plot in Figure~\ref{fig20} left. For comparison, Figure~\ref{fig19a} left includes histograms denoting the IceCube and Auger neutrino fluence upper limits (scaled to 3 flavors)
from neutrino searches within $\pm 500$~s around the binary neutron star merger GW170817~\cite{ANTARES:2017bia}.  
Projected sensitivities of GRAND200k~\cite{Alvarez-Muniz:2018bhp} for zenith angles $90^\circ$ and $94^\circ$ are shown with the dashed red lines. Also plotted in Figure~\ref{fig19a} left are models taken from Kimura et al. \cite{Kimura:2017kan} of the all-flavor neutrino fluence from a short gamma-ray burst during the prompt and extended emission (EE) phases, assuming on-axis viewing ($\theta=0^\circ$) and a source at $D=50$~Mpc. The right panel of Figure~\ref{fig19a} shows the corresponding sky plot of the expected number of neutrino events for this model~\cite{Venters:2019xwi}. In this scenario, POEMMA will be able to detect at least one neutrino in every region of the sky, provided that the source location is in an observable region at the time of the event. In the scenario in which corroborating evidence from multi-messenger observations is not available, POEMMA will exclude models, including background-only models, predicting less than $\sim 0.3$ neutrinos at the level of $\sim 5\sigma$ in $\sim 50$\% of the sky.

The left panel of Figure~\ref{fig19b} shows the sensitivity for long bursts, assuming that POEMMA is in the ToO-stereo configuration.
For comparison, the solid histogram denotes the IceCube neutrino fluence upper limits (scaled to 3 flavors)
from a neutrino search within a 14-day time window around the binary neutron star merger GW170817~\cite{ANTARES:2017bia}.
GRAND200k's projected declination averaged sensitivity for $0^\circ <|\delta|<45^\circ$ 
is indicated by the dashed red lines~\cite{Alvarez-Muniz:2018bhp}. Also shown are models from Fang \& Metzger~\cite{Fang:2017tla} of the all-flavor neutrino fluence produced $10^{5.5}-10^{6.5}$~s and $10^{4.5}-10^{5.5}$~s after the birth of a millisecond magnetar produced by a binary neutron star merger event occurring at a distance of $D = 10$~Mpc. The right panel of Figure~\ref{fig19b} shows the corresponding sky plot of the expected number of neutrino events for this model~\cite{Venters:2019xwi}. The sky plot includes the reduction in exposure at each location due to the sun and the moon (see Figure~\ref{fig20}). In this scenario, POEMMA will exclude models predicting less than $\sim 0.3$ neutrinos at the level of $\sim 5\sigma$ in every region of the sky. The left panels in Figures~\ref{fig19a} and \ref{fig19b} illustrate that the neutrino spectra from the both long-burst and short-burst neutrino transients can have peak flux at energies above 10 PeV where POEMMA is sensitive while the neutrino flux at lower energies would not be currently detectable with ground-based instruments with IceCube-scale sensitivity, regardless of the location of the transient in the sky.

In order to identify candidate sources classes that are most likely to result in ToOs for POEMMA, the occurrence of transient events was modeled as a Poisson process with a ToO rate calculated from the cosmological event rate for the source class and the horizon distance for detecting one event based on an assumed model~\cite{Venters:2019xwi}. The left panel of Table~\ref{figPTab} shows the Poisson probability of POEMMA detecting at least one ToO event as a function of mission time for a variety of candidate source classes. The most promising source classes are those for which POEMMA has a better than $10$\% chance of detecting at least one ToO within the proposed mission lifetime of $3$--$5$ years. As listed in the right panel of Table~\ref{figPTab}, these source classes include jetted tidal disruption events~\cite{Lunardini:2016xwi}, binary black hole mergers~\cite{Kotera:2016dmp}, and binary neutron star mergers~\cite{Fang:2017tla}. Short gamma-ray bursts with an extended emission phase~\cite{Kimura:2017kan} might also result in a detectable ToO for POEMMA, though on a timescale of $\sim 15$ years. Table~\ref{figPTab} right also provides the neutrino horizon distance for each modeled source class.

\subsection{POEMMA Diffuse Neutrino Performance}

\begin{figure*}[ht]
\begin{minipage}[t]{0.49\textwidth}
  \includegraphics[trim = 72mm 40mm 65mm 40mm, clip, width = 1.0\textwidth]{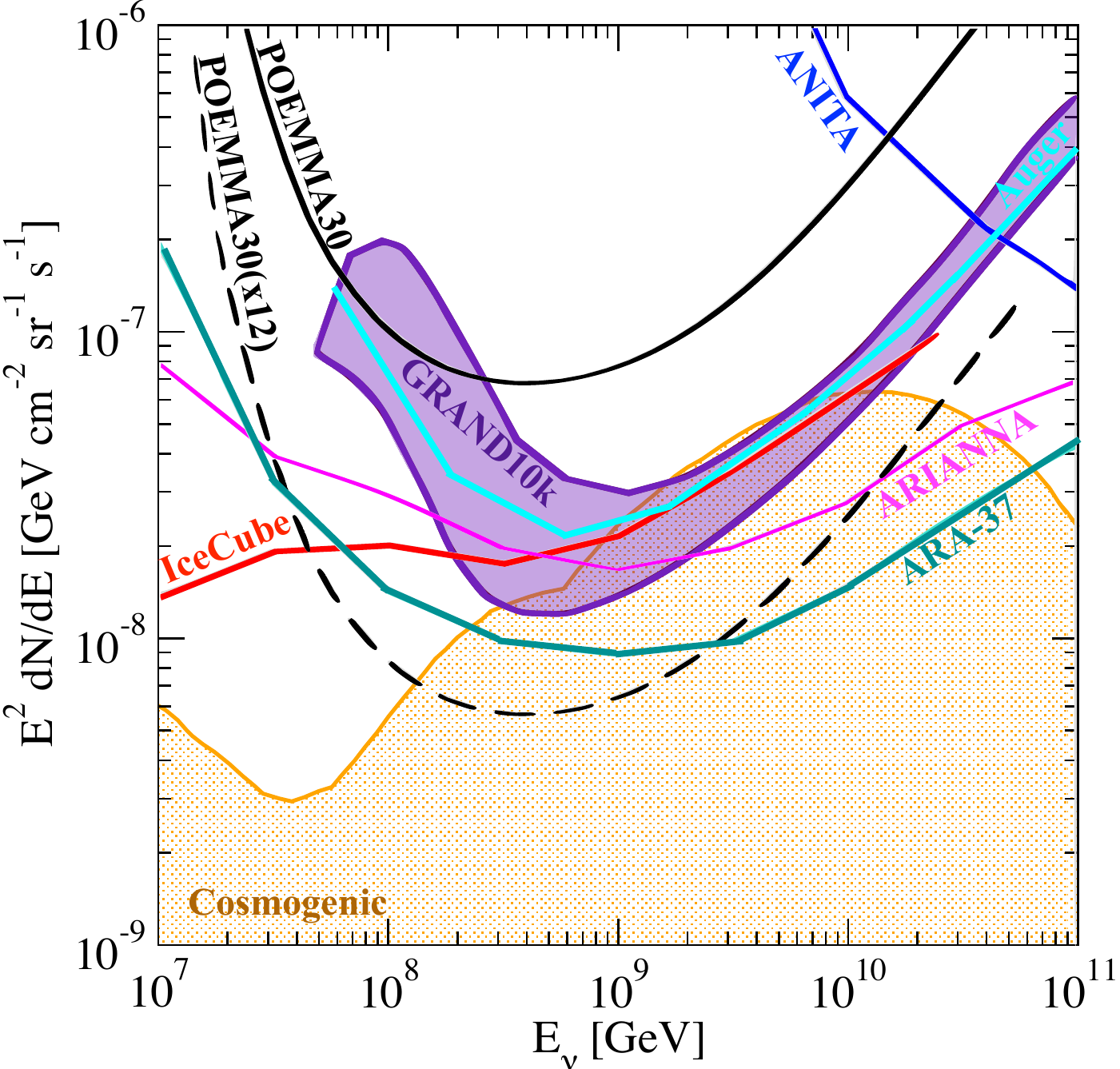}
\end{minipage}
\hfill 
\begin{minipage}[t]{0.49\textwidth}
  \includegraphics[trim = 68mm 40mm 69mm 40mm, clip, width = 1.0\textwidth]{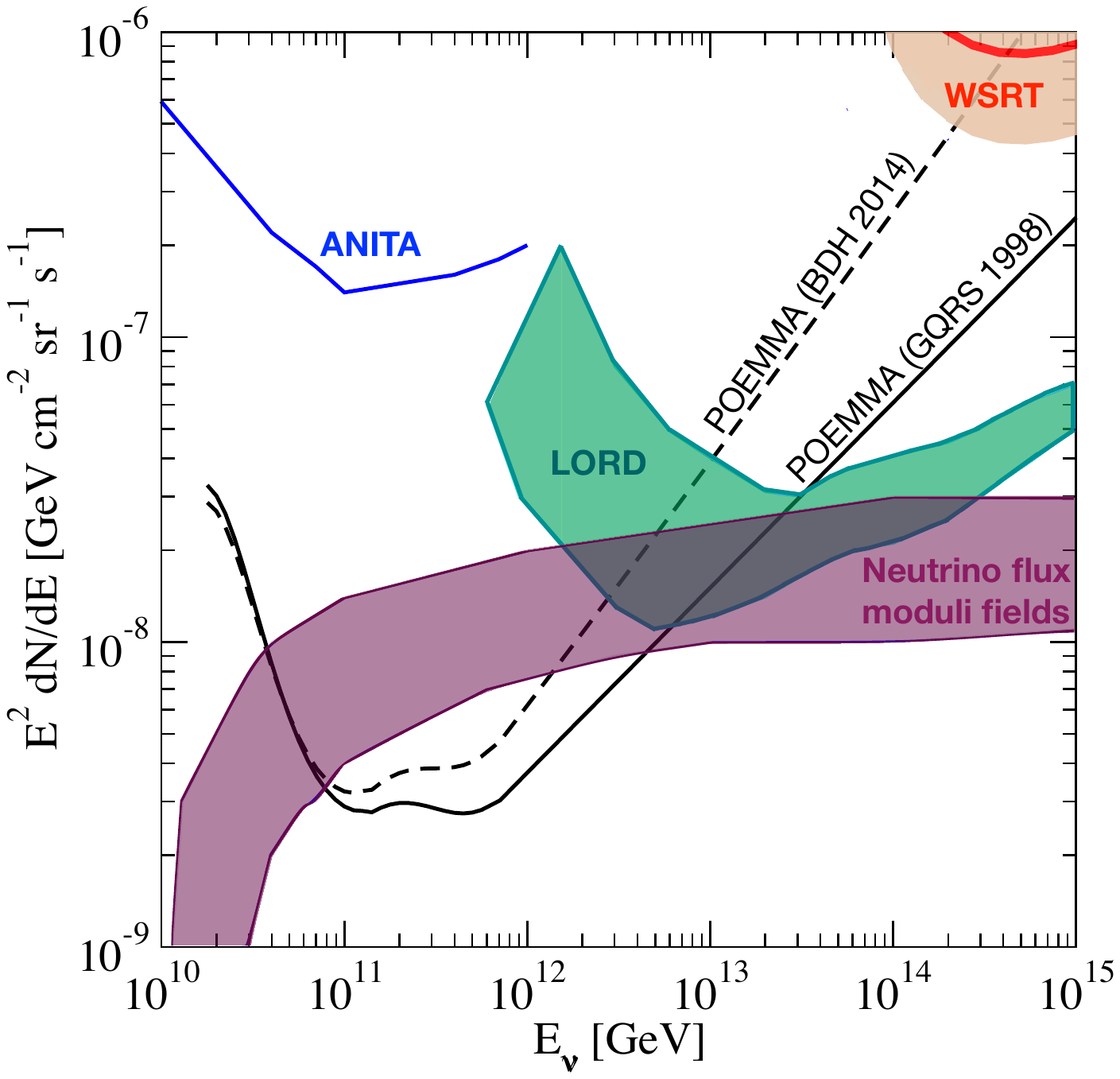}
\end{minipage}
    \caption{POEMMA 5-year sensitivity to EAS showers resulting from neutrino interactions in the Earth or in the atmosphere. {\it Left:} The black solid curve shows the POEMMA sensitivity to $\tau$-induced EAS showers arising from $\nu_{\tau}$ interactions in the Earth and detected via optical Cherenkov emission (scaled to three flavors)~\cite{Reno2019a}. The black dashed curve is the sensitivity for POEMMA30($\times$12) (extrapolating the POEMMA30 sensitivity to 360$^\circ$ FoV in azimuth). The $90$\% CL upper limits from Auger (scaled for sliding decade-wide neutrino energy bins and for three flavors)~\cite{Aab:2019auo}, IceCube~\cite{Aartsen:2018vtx}, and ANITA I-IV~\cite{Gorham:2019} are shown along with $3$-yr. sensitivity projections for ARIANNA~\cite{Barwick:2014pca}, ARA-37~\cite{2012APh....35..457A}, and GRAND10k~\cite{Fang:2017mhl}. For comparison,  the combined allowed ranges from \cite{vanVliet:2019nse} and \cite{Heinze:2019jou} for the all-flavor cosmogenic neutrino flux arising from UHECR interactions with the cosmic microwave background. {\it Right:} Black curves show the POEMMA sensitivity to EAS showers arising from neutrino interactions in the atmosphere and detected via fluorescence from charged-current and neutral-current interactions from all three neutrino flavors. The solid (dashed) curve is calculated using cross sections from Ref.~\cite{Gandhi:1998ri} (Ref.~\cite{Block:2014kza}). For comparison, predictions for strongly coupled string moduli (maroon band)~\cite{Berezinsky:2011cp} are also shown along with upper limits from ANITA I-IV (blue line) \cite{Gorham:2019} and Westerbork Synthesis Radio Telescope (WSRT; red line with tan band) \cite{Scholten:2009} and a projected sensitivity for LORD  (green band) \cite{Ryabov:2016fac}. 
    \label{figX}}
\end{figure*}

Figure \ref{fig15} right shows the diffuse-flux neutrino aperture for POEMMA as a function of $\nu_\tau$ energy assuming a  20\% duty cycle with telescopes in the neutrino stereo configuration. The current POEMMA design has an effective neutrino aperture that surpasses that for IceCube HESE (high-energy starting events) \cite{Aartsen:2018vtx} above 70 PeV for diffuse neutrino flux searches. (Note 100\% duty cycle is assumed for IceCube.) The results are shown for two cases, one with an azimuth angle span of 30$^\circ$, which corresponds to the current conceptual design for POEMMA, and with 360$^\circ$ azimuth viewing. The left panel of Figure~\ref{figX} shows the corresponding 5-year sensitivity curves for both azimuthal spans. The benefit of increasing the azimuth range of the limb observations is clear and will drive the development for the eventual POEMMA360 mission.

While in POEMMA-Stereo mode, POEMMA will monitor $\sim 10^{13}$ metric tons of atmosphere, allowing for detection of fluorescence from EASs produced by UHE neutrino interactions that occur deep in the atmosphere. Simulations of isotropic neutrino events at energies beyond 30~EeV and starting at a slant depth of $>$ 2000 g/cm$^2$ demonstrate that POEMMA will have high sensitivity to fluorescence from neutrinos at these energies~ \cite{Anchordoqui:2019omw}. The right panel of Figure~\ref{figX} provides all-flavor sensitivity curves for POEMMA to UHE neutrinos using the fluorescence technique for cross sections taken from \cite{Gandhi:1998ri} and \cite{Block:2014kza}. For comparison, Figure~\ref{figX} right includes current upper limits from ANITA I-IV \cite{Gorham:2019} and WSRT \cite{Scholten:2009}, sensitivity projections for the future Lunar Orbital Detector (LORD)~\cite{Ryabov:2016fac}, and a range of modeled neutrino fluxes resulting from models of strongly coupled moduli in a string theory background with $G \mu \sim 10^{-20}$, where $G$ is Newton's constant and $\mu$ the string tension \cite{Berezinsky:2011cp}. 
The POEMMA sensitivity curves were calculated assuming observations in the POEMMA-Stereo configuration with a 12\% duty cycle and 5 years of observation time.  The energy resolution of atmospheric neutrino-induced EAS observed via fluorescence measurements is $<20\%$ for the EAS measurements, similar to that for UHECRs. However, the neutrino energy resolution is determined from the fraction of neutrino energy going into the EAS which depends on the CC and NC neutrino cross-sections and y-dependence (see Ref. \cite{Anchordoqui:2019omw}).

Figure~\ref{figX} right shows that POEMMA will probe a significant part of the parameter space, providing a method of searching for strongly coupled moduli, which complements searches based on gravitational effects of strings, like
structure formation, cosmic microwave background, gravitational radiation, and gravitational lensing. The strongest current bound from lensing effects is estimated to be $G\mu \lesssim 10^{-7}$~\cite{Christiansen:2010zi}, while millisecond pulsar observations lead to
$G\mu \lesssim 4 \times 10^{-9}$~\cite{vanHaasteren:2011ni}. Next-generation gravitational-wave detectors are expected to probe $G\mu \sim 10^{-12}$~\cite{Damour:2004kw,Olmez:2010bi}. Thus,
POEMMA will attain sensitivity to a region of the parameter space more
than 10 orders of magnitude below current limits and $\sim 8$ orders of
magnitude smaller than next-generation gravitational-wave detectors.

\subsection{Atmospheric Science}
\label{subsec:AtmScience}

Through its combination of light collecting power, large field of view, fine pixelization, relatively fast timing, absolute photometric calibration, and vantage point from LEO, the POEMMA instrument is capable of transformative measurements of atmospheric transient luminous events (TLEs).  
The focus will be global surveys that span regions of known strong electrical storms, over land and ocean in the Northern and Southern Hemispheres. POEMMA's unprecedented sensitivity to TLEs observed from space also brings high potential for new discoveries. 

\begin{figure}[b]
    \postscript{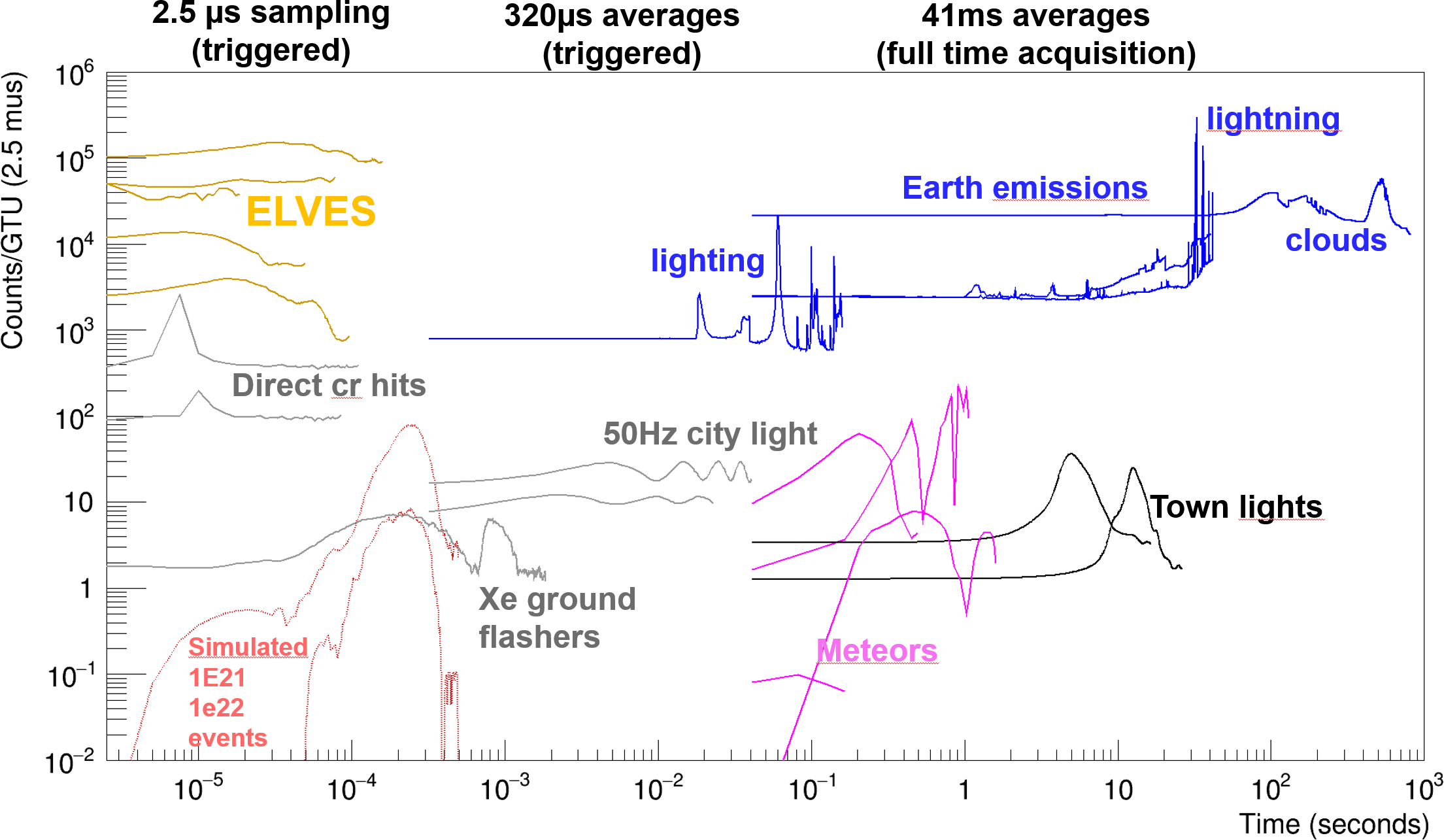}{0.99}
  \caption{Temporal profile of various signals observed by Mini-EUSO. The fast sampling rate (2.5 $\mu s$) allows for studies of UHECRs, elves and other fast phenomena. The averaged sampling can be used to characterize slower events such as lightning, meteors, UV emissions of artificial and natural origin. All curves refer to experimental data from ISS altitude except the Monte Carlo simulated light curves of UHECRs. From Ref. \cite{Bacholle:2020emk}.
 \label{mini-euso-events}}
\end{figure}

The largest space instruments flown so far that target UV emission from the Earth's atmosphere are the Tracking Ultraviolet Setup (TUS) experiment on the Lomonosov Satellite~\cite{Adams} and Mini-EUSO~\cite{Capel:2018} on the International Space Station. For reference, light curves from the various phenomena observed by Mini-EUSO are shown in Figure \ref{mini-euso-events}.

 Transient Luminous Events (TLEs) that have been identified in the upper atmosphere include sprites, elves, and TEBs (Terrestrial Electron Beams). 
 Sprites (sometimes denoted as an acronym for Stratospheric/mesospheric Perturbations Resulting from Intense
Thunderstorm Electrification) are the result of high energy cloud-to-ground lightning. They are  caused by atmospheric gravity waves that grow in the electric fields produced by lightning. Sprites occur in the mesosphere at 50 to 90~km altitude. Their appearance varies and their color ranges from blue to red with a duration between 10 and 100~{\rm ms}.  

Elves, also referred to as ELVES (Emission of Light and Very Low Frequency perturbations due to Electromagnetic Pulse Sources),  are expanding rings of light that are generated in the base of the ionosphere above strong lightning.  A rapid change of current in the lightning process can generate an EMP of field strength sufficient to accelerate electrons in the ionosphere to sufficient energy to excite N2 molecules. As the excited molecules relax, they emit light. The UV component of this light is produced by the same fast atomic transitions that produce the well-known UV fluorescence signature of extensive air showers that are generated in the troposphere by high energy cosmic rays.  Elves can reach many hundreds of km in diameter. Their brightness, size, exact shape, expansion rate and timing structure depend on properties of the lightning below. Their typical time scale ranges from a fraction of a ms to several ms. The ability of a cosmic ray fluorescence detector to measure more than 1000 elves in detail over one particular $3\cdot10^6\text{\,km}^2$ region of the Earth has been demonstrated using the Pierre Auger Observatory~\cite{Aab:2020, Merenda:EOS:2020}.

The set of TLEs described above can be easily identified by POEMMA given their morphology and time evolution. For example, the fast timing and light gathering power of POEMMA is well-suited to measure elves in detail. Elves carry signatures of the fundamental properties of high-current lightning. High-current lightning is produced by extremely strong convective thunderstorms. These destructive storms are becoming more frequent and severe as the climate warms \cite{Pachuri:2014,Diffenbauch:2013,Brooks:2013}. Studies of the properties of the strong convective storms occurring today offer a proxy for studying extreme storms of the future.

POEMMA will conduct a global survey of elves and their timing structure (single vs. double peaks vs. multiple peaks). In combination with the lightning locations and currents as measured in coincidence by ground-based lightning detectors, this combined data set would translate into a global survey that identifies cloud-to-ground vs intra-cloud lightning. Traditional radio-based lightning sensors on the ground are challenged to make this distinction on their own because the EMPs from the direct and mirrored charge current distributions of a lightning stroke will reach the antenna at the same time. 

\begin{figure}[hb]
    \postscript{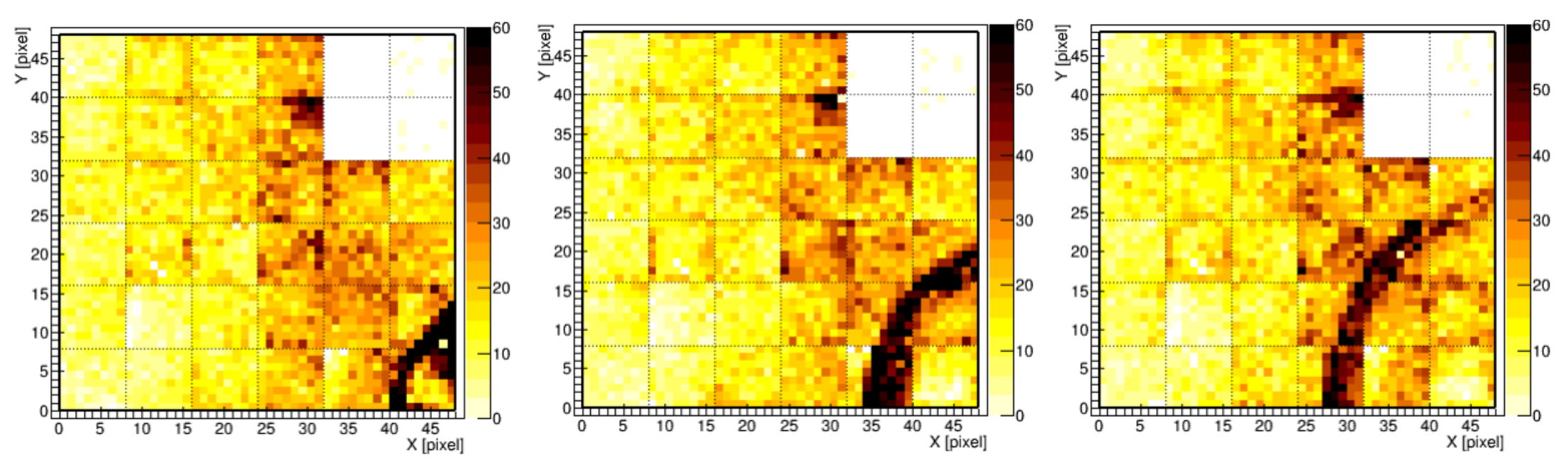}{0.99}
  \caption{The time evolution of an elve as observed by the photo detection module of the Mini-EUSO instrument looking down from the International Space Station. The three frames shown here each have a duration of 2.5$\mu$s and are separated by 7.5$\mu$s. Each pixel views $\simeq 4.5 \times 4.5$ km as projected down to the typical 90 km altitude of an elve. The POEMMA fluorescence camera will feature 36 of these photo detection modules. (The pixels in the blank square regions were temporarily operating at reduced sensitivity due to a previous bright light that triggered the high voltage power supply safety mechanism.)
 \label{mini-euso-elve}}
\end{figure}

An example of an elve that was recorded by looking down from LEO is shown in Figure \ref{mini-euso-elve}. This elve was observed from the International Space Station by Mini-EUSO. The spatial sampling is $\simeq 4.5~\text{km}$ at the nominal 90~km height of the base of the ionosphere and the temporal sampling is $2.5~\mu\text{s}$. With greater statistics of elves acquired with the resolution of POEMMA, we will be able identify all the possible lightning discharges that can induce elves on a global scale and recover valuable information about strong electric storms.

TEBs  are injected into the magnetosphere by terrestrial gamma-ray flashes (TGFs)~\cite{Dwyer:2008}. These electrons are guided by the Earth's magnetic field toward the conjugate point in the opposite
hemisphere where most are expected to re-enter the atmosphere, creating a nitrogen fluorescence signal similar to that from and EAS and lasting 1-10 ms. If the geomagnetic field is stronger at the conjugate point, some of these electrons will mirror and return to the other hemisphere where they will re-enter the atmosphere. A fluorescence signal from TEBs entering the atmosphere has not yet been detected. POEMMA will be capable of detecting these reentrant TEBs.

\subsection{Meteors and Macros}

\begin{figure*}[ht]
\begin{minipage}[t]{0.53\textwidth}
    \postscript{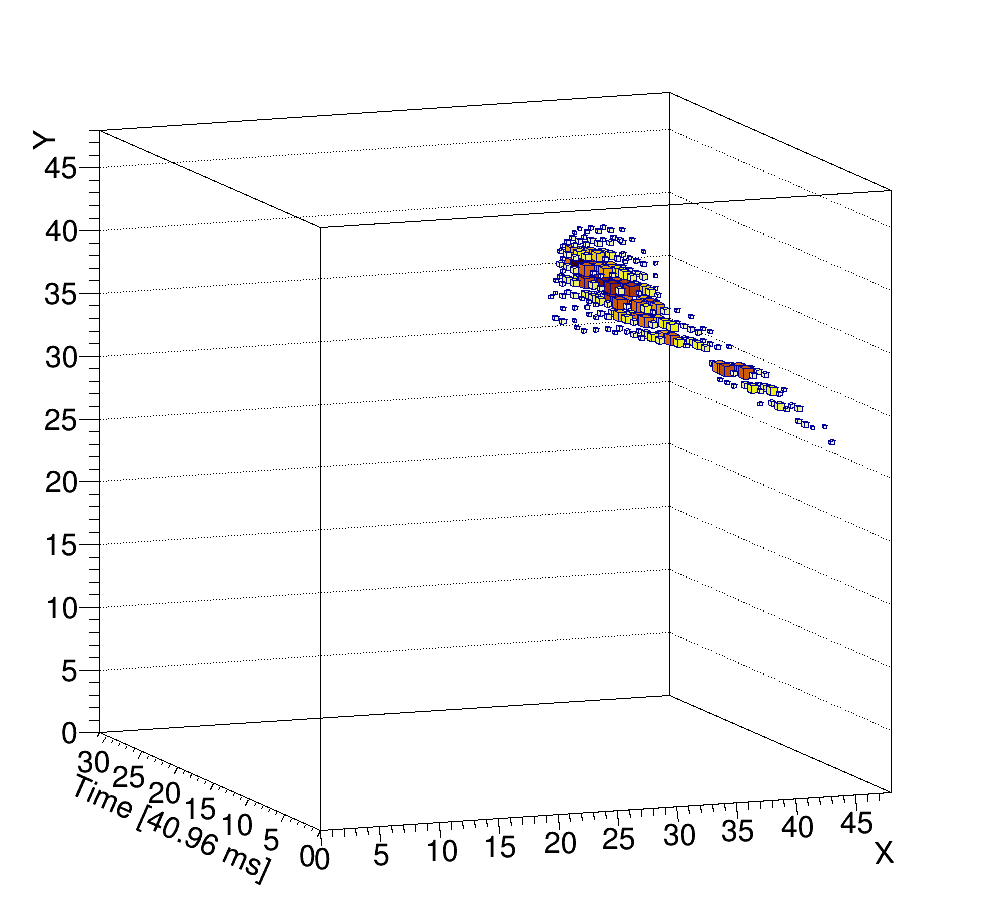}{1}
\end{minipage}
\hfill \begin{minipage}[t]{0.47\textwidth}
  \includegraphics[width = 1.0\textwidth]{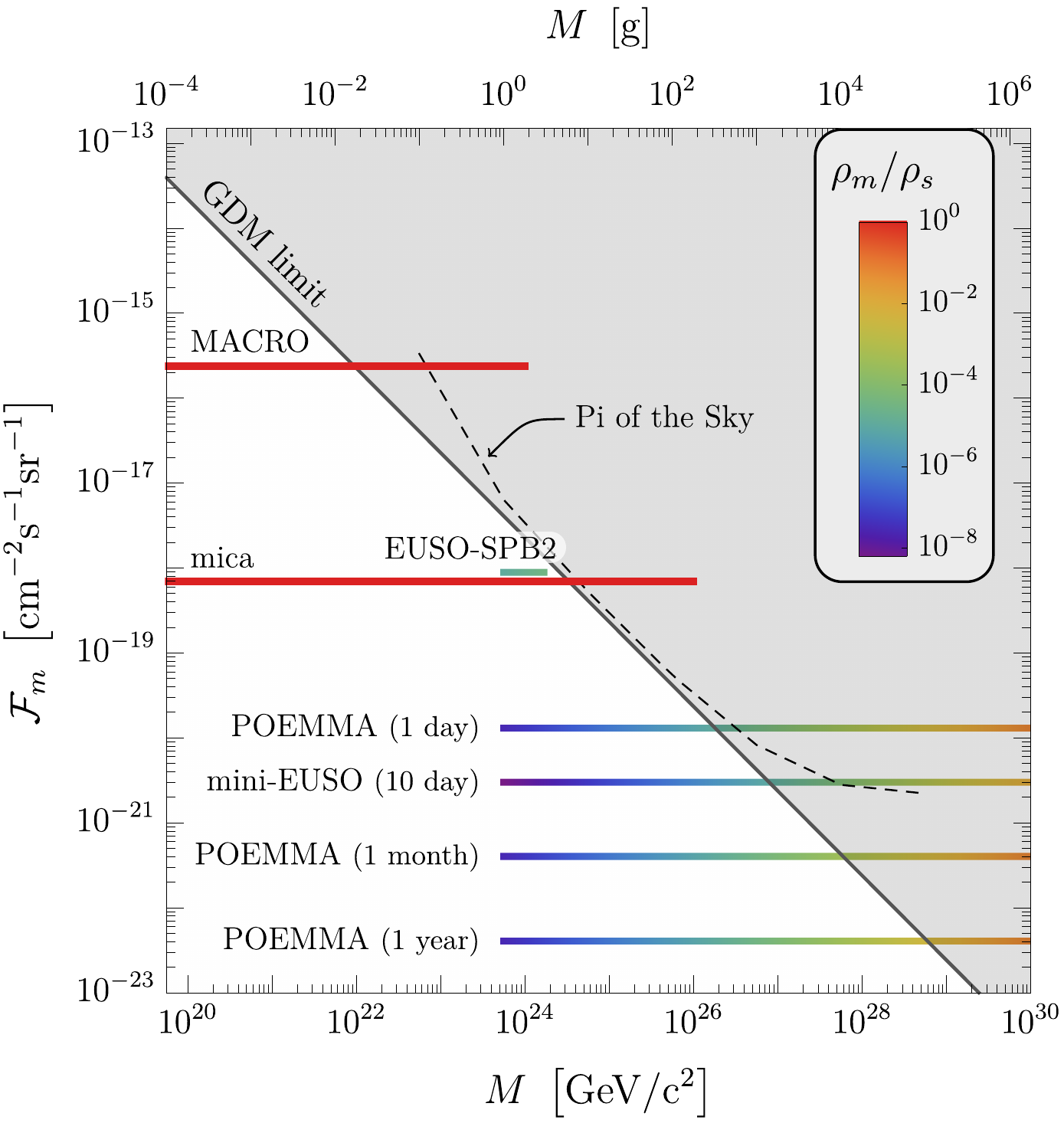}
\end{minipage}
\caption{{\it Left:} A meteor track as observed by Mini-EUSO at a 40.96 ms sampling rate. The  X and Y axis represent the pixels of the focal surface plotted versus time. The color scale and the size of boxes correspond to the number of counts deposited in a pixel.
{\it Right:} The projected POEMMA 90\% confidence level upper limit on the macro
flux ${\cal F}_m$ (as a function of the macro mass $M$)
resulting from null detection over different time spans of acquired
data compared to EUSO-SPB2~\cite{Paul:2021bhh}, Mini-EUSO, and other
experiments~\cite{Price:1986ky,Price:1988ge,Ambrosio:1999gj,Piotrowski:2020ftp}. The
Galactic dark matter (GDM) limit is indicated for comparison. 
\label{meteor_nuclearites}}
\end{figure*}

Meteor and fireball observations are key to the understanding of both
the inventory and physical characterization of small solar system
bodies orbiting in the vicinity of Earth. On the basis of studies performed for the JEM-EUSO (Joint Experiment Missions for Extreme
Universe Space Observatory)
program~\cite{Adams:2014vgr,Abdellaoui:2017},  we estimate that POEMMA will be able to
detect meteors to an apparent magnitude $\mathbf m = +10$. This is a
superior sensitivity compared to that of Mini-EUSO, which can detect
meteors to $\mathbf m = +6$. In the left panel of
Figure~\ref{meteor_nuclearites} we show the temporal and spatial development of a meteor as
observed from space by Mini-EUSO: the high sampling rate and resulting
averages enable the precise determination of the meteor's
characteristics like velocity and point of origin. The higher
performance of POEMMA may result in the detection of a larger number
of events and a more detailed determination of their parameters.
Thus, 
POEMMA will record a statistically significant flux of meteors,
including both sporadic ones, and events produced by different meteor
showers.  Being unaffected by adverse weather conditions, and thereby 
having a higher duty cycle and higher chance to detect rare events,
POEMMA can also be a very important facility to search for
interstellar meteors. It can also detect very bright meteors due to
high dynamic range. Thanks to POEMMA's stereoscopic vision, it can
make a 3D reconstruction of the meteor's trajectory. 

Moreover, the observing strategy developed to detect meteors may also
be applied for the detection of macroscopic dark matter (or macro for
short)~\cite{Jacobs:2014yca}. These are composite objects which are able to
evade extant bounds on dark matter while accommodating astronomical
and astrophysical observations. The most appealing macro candidate, originally
proposed by Witten~\cite{Witten:1984rs} and worked out in several
subsequent works
(e.g.,~\cite{Farhi:1984qu,DeRujula:1984axn,Alcock:1988re}), is a
``macroscopically'' sized nugget of strange quark matter (or nuclearite), wherein
the dark matter would have nuclear density and
may be (nearly) a result of entirely Standard Model physics. A related
compelling cadidate is a dark quark nugget (or dark nuclearite)~\cite{Bai:2018dxf}. Because
of the exciting possibility that macros emerge from essentially the
same Standard Model physics as ordinary nuclei, we adopt $\rho_s \sim
3.6 \times 10^{14}~{\rm g/cm^3}$~\cite{Chin:1979yb} as a reference
density of particular interest. However, depending on the confinement
scale and the magnitude of the dark baryon asymmetry, the dark quark nugget's density may span several orders of magnitude~\cite{Bai:2018dxf}. In the spirit
of~\cite{Sidhu:2018auv}, herein we consider macro densities in the generous range $10^6 < \rho_m/{\rm
  (g/cm^3)} < 10^{15}$.  Most
macros are expected to move at higher velocities than meteoroids, and
to exhibit a wider range of possible trajectories, including objects
moving upward after crossing the Earth. This allows a clean
discrimination of the macro
signals~\cite{Adams:2014vgr}. In the right panel of
Figure~\ref{meteor_nuclearites} we show the macro-detection sensitivity
of POEMMA, Mini-EUSO, and EUSO-SPB2~\cite{Paul:2021bhh}, together with existing
limits~\cite{Price:1986ky,Price:1988ge,Ambrosio:1999gj,Piotrowski:2020ftp}. The sensitivities shown in the figure are estimated following
the procedure outlined in~\cite{Paul:2021bhh}. POEMMA specific details were presented elsewhere~\cite{Anchordoqui:2021xhu}. With a null result, after 1 day/month/year of integrated data
(which would correspond to 5 days/months/years acquisition with a duty cycle
of 20\%), POEMMA would be able to provide an upper limit to macro flux, which
is $\sim$2/3/4 orders of magnitude more restrictive than those obtained by existing experiments.

The sensitivities shown in  Figure~\ref{meteor_nuclearites}  assume a strategy in which POEMMA acquires data with a time resolution of 1.024 ms (1024 GTUs), the trigger requires a signal excess in one pixel above 3$\sigma$ during 1 time frame, and this condition is repeated for more than 4 time frames in adjacent pixels. These conditions are currently being adopted in the Mini-EUSO search for meteors.  

It is important to underline that the estimated sensitivity might be conservative. Indeed it has been obtained by using the same approach as JEM-EUSO and Mini-EUSO, which are based on instruments with monocular vision that are nadir pointing. In this sense POEMMA, with its stereoscopic system, would allow a reduction in the threshold on the mass and it would relax the tight quality cuts used in JEM-EUSO to distinguish a macro from a meteor.

\begin{figure}
\begin{center}
\includegraphics [width=1\textwidth]{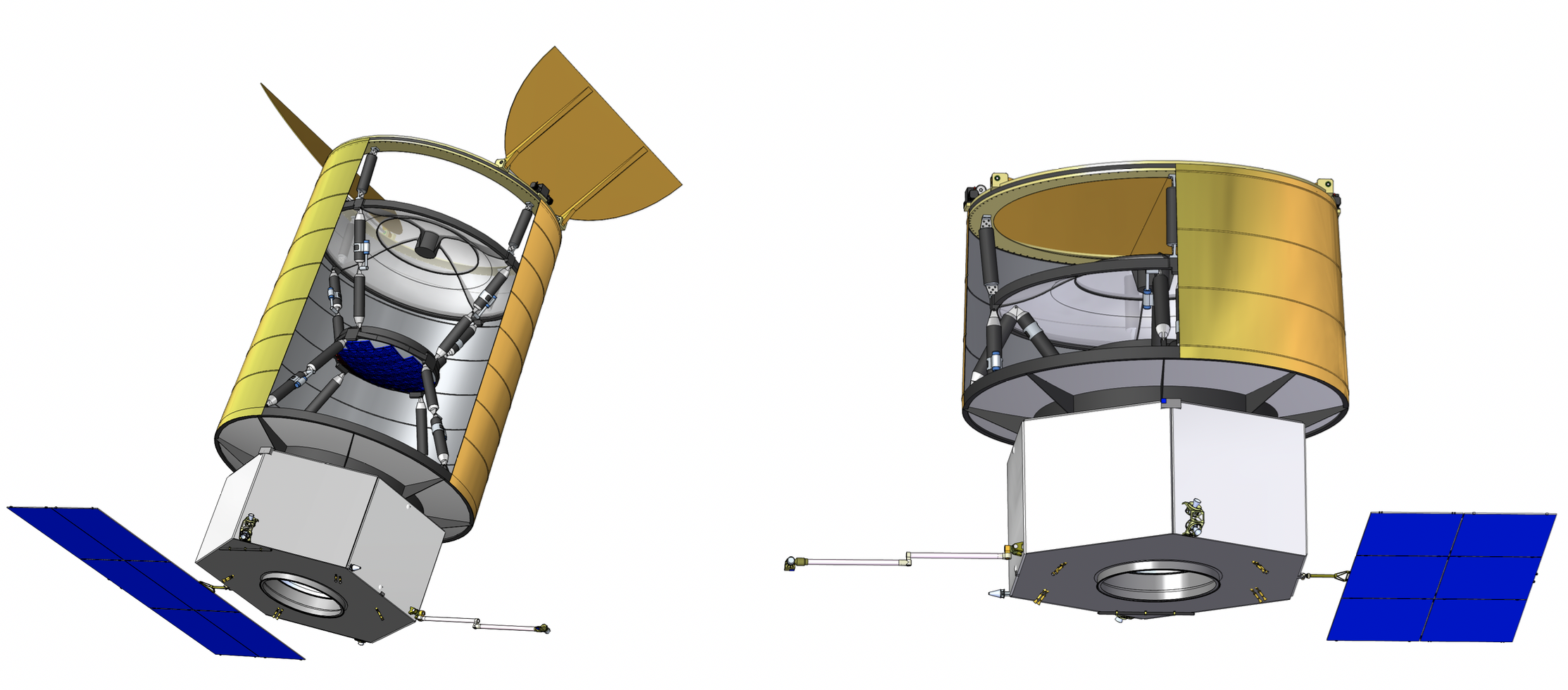}
\caption{POEMMA telescope (instrument and spacecraft) deployed with open shutter doors (left) and in stowed position for launch (right). Cutaways in the light shield display the internal structure of corrector plate and focal surface in the middle of the payload (blue). Spacecraft bus shown with solar panel (blue) and communications antenna deployed in both images. {Adapted from Ref. \cite{Olinto:2019mjh}}.}
\label{fig6}
\end{center}
\end{figure}

\section{POEMMA Observatory}

The design of the POEMMA observatory and mission evolved from previous work on the OWL~\cite{Stecker:2004wt,Krizmanic:2011hs} and JEM-EUSO~\cite{Adams:2013vea} designs, the CHANT concept~\cite{Neronov:2016zou}, and the sub-orbital payloads EUSO-SPB1~\cite{Wiencke:2017cfi} and EUSO-SPB2~\cite{Adams:2017fjh}. The final POEMMA telescope (instrument and spacecraft) and mission concepts were refined during two intense weeks at the Integrated Design Center (IDC) of the Goddard Space Flight Center (GSFC) (see Figure \ref{fig1}). The POEMMA instrument was developed at the IDC Instrument Design Laboratory (IDL) from July 31 to August 1, 2017.
The POEMMA spacecraft and mission concepts were developed at the IDC Mission Design Laboratory (MDL) from October 30 to November 3, 2017.

The POEMMA observatory is comprised of two telescopes. Each telescope has an instrument and a spacecraft bus. Here we detail the instrument subsystems and how they relate to the scientific performance of the POEMMA telescopes. We also describe the major subsystems of each of the identical spacecraft buses and how their performance supports the science observations. The mission concept is discussed in \S 4.

\subsection{POEMMA Instrument}

 The POEMMA instrument is  designed to provide significant advances in the exposure to UHECRs and cosmic neutrinos over the entire sky to potentially discover their origin and study their physics and astrophysics. The instrument detects these extremely energetic particles by electronically recording the signals generated by EASs in the night side of the Earth's atmosphere. 
EASs develop at speeds close to the speed of light with particle longitudinal distribution profiles reaching tens of kilometers at ultra-high energies (see Figure \ref{fig4} left). The central element of the POEMMA instrument  is a high sensitivity optical system that measures and locates two types of emission from these EASs: the faint isotropic emission due to the fluorescence of atmospheric nitrogen excited by air-shower particles (with emission in the $300 \lesssim \lambda/{\rm nm} \lesssim 500$ range, Figure \ref{fig4} right), and the bright, forward-collimated Cherenkov emission from EASs directed at the POEMMA observatory (with the light distribution and emission spectrum as in Figure \ref{figCprof}).  POEMMA optics and cameras are designed to optimize the wavelength coverage, time sampling, and pixel sizes to best reconstruct ultrahigh energy EASs using both the fluorescence and Cherenkov signals.

Each POEMMA instrument features a large diameter optical system (mirror and corrector plate, see Figure \ref{fig1} and \ref{fig7}) to collect the weak fluorescence and Cherenkov signals. Photons are focused onto the segmented focal surface (Figure \ref{fig10}) and subsequently digitized as short videos of fluorescence traces (Figure \ref{fig9} right) and as Cherenkov light pulses. These two types of signals are expected in different parts of the focal plane. Cherenkov signals are observed below the limb for cosmic neutrinos and above the limb for UHECRs. These Cherenkov photons  are focused in a region near the edge of the focal surface where the POEMMA Cherenkov Camera (PCC) is located. Over the remainder of the focal surface,  fluorescence photons are recorded by the POEMMA Fluorescence Camera (PFC). Each camera design is optimized for the unique optical signals, which it records to meet the science objectives.  The instrument system architecture is based on a large number of identical, highly parallel electronic channels designed to meet the high standards of a NASA Class B mission. Aerospace grade components have been identified for key elements within the instrument to ensure reliability for the mission. 

The POEMMA instruments are designed for deployment after launch. A stowed configuration (see Figure \ref{fig6} right) enables two identical satellites to be launched together on a single Atlas V rocket (see Figure \ref{fig12}). Space qualified mechanisms extend each instrument after launch to their deployed position to begin observations. The instrument design includes light-tight shutter doors that must open and close during each orbit. The shutter doors and other critical subsystems in the instrument and spacecraft incorporate redundancy to minimize operational and mission risks.

Key elements of the POEMMA instruments and their functionality are described below. 

\subsubsection{POEMMA Optics}

\begin{figure*}[ht]
\begin{minipage}[t]{0.49\textwidth}
  \postscript{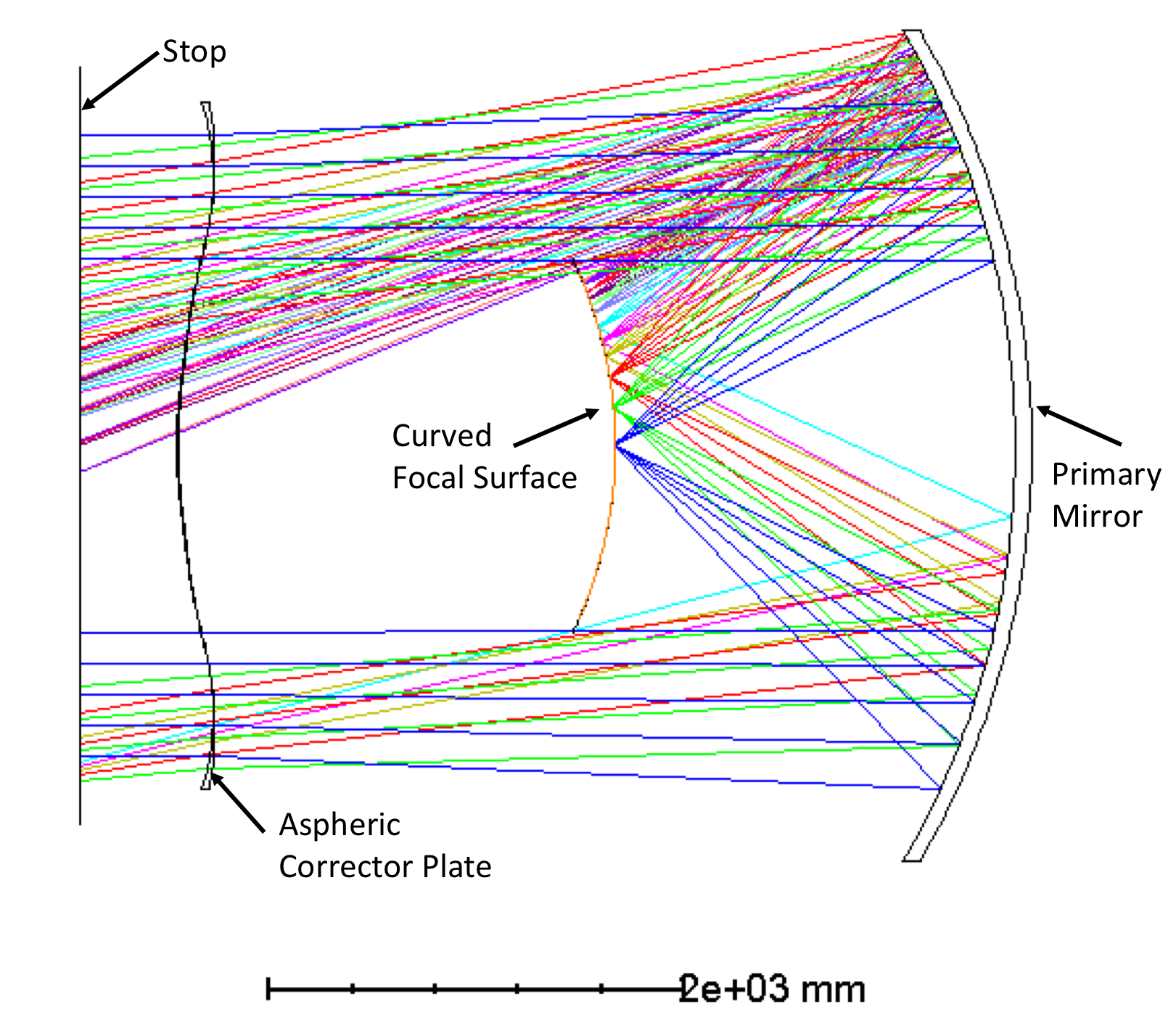}{0.9}
\end{minipage}
\hfill \begin{minipage}[t]{0.49\textwidth}
   \postscript{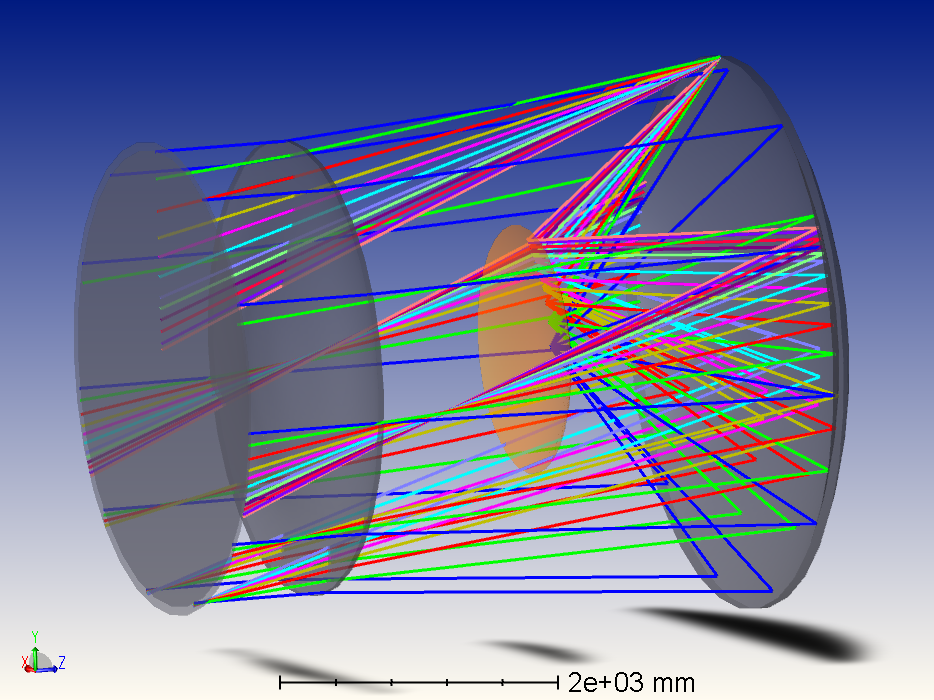}{0.99}
\end{minipage}
    \caption{{\it Left:} Cross sectional layout of POEMMA optics. {\it Right:} Model of POEMMA optics.
     \label{fig7}}
\end{figure*}

\begin{figure*}[ht]
\begin{minipage}[t]{0.52\textwidth}
  \postscript{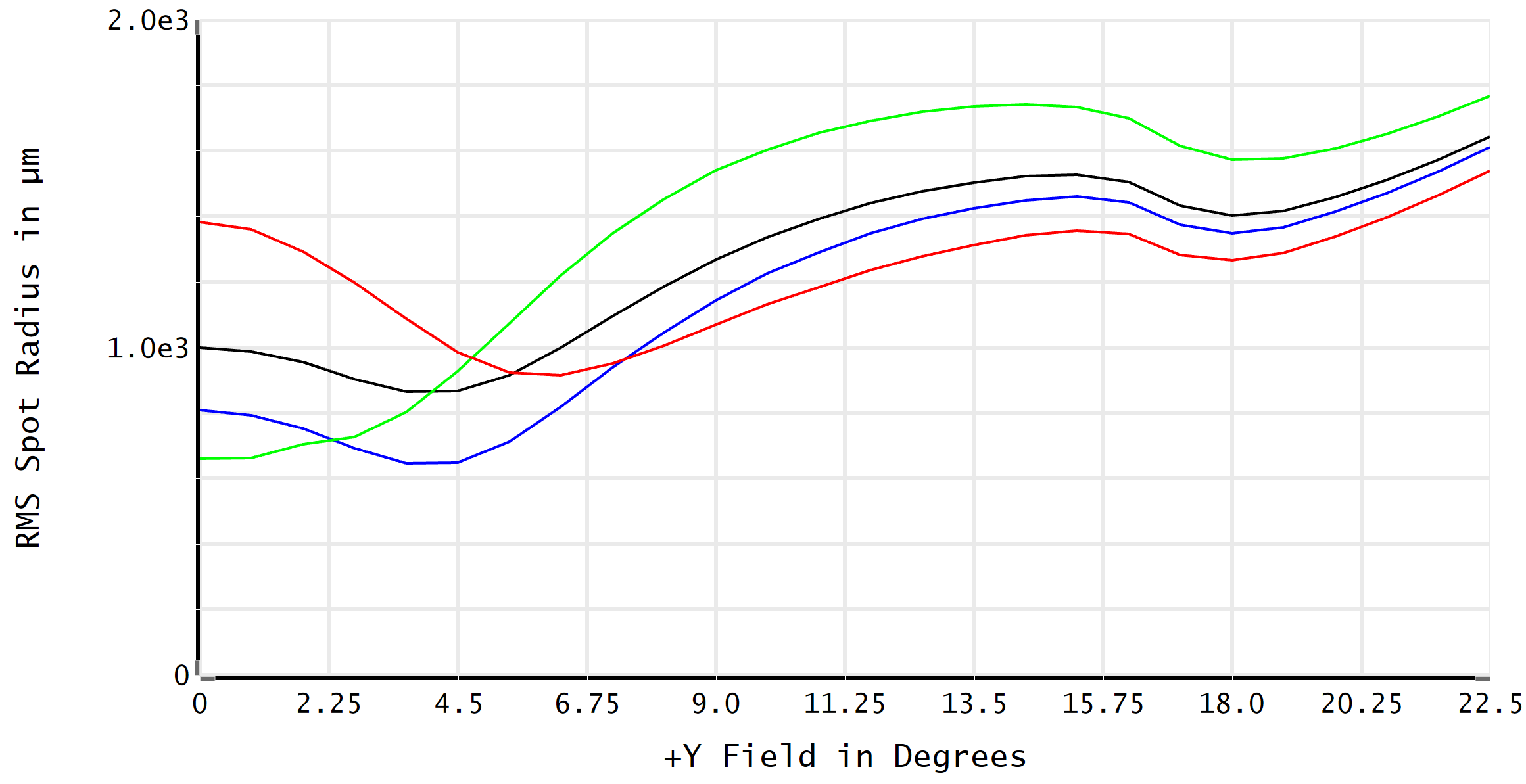}{1}
\end{minipage}
\hfill \begin{minipage}[t]{0.46\textwidth}
   \postscript{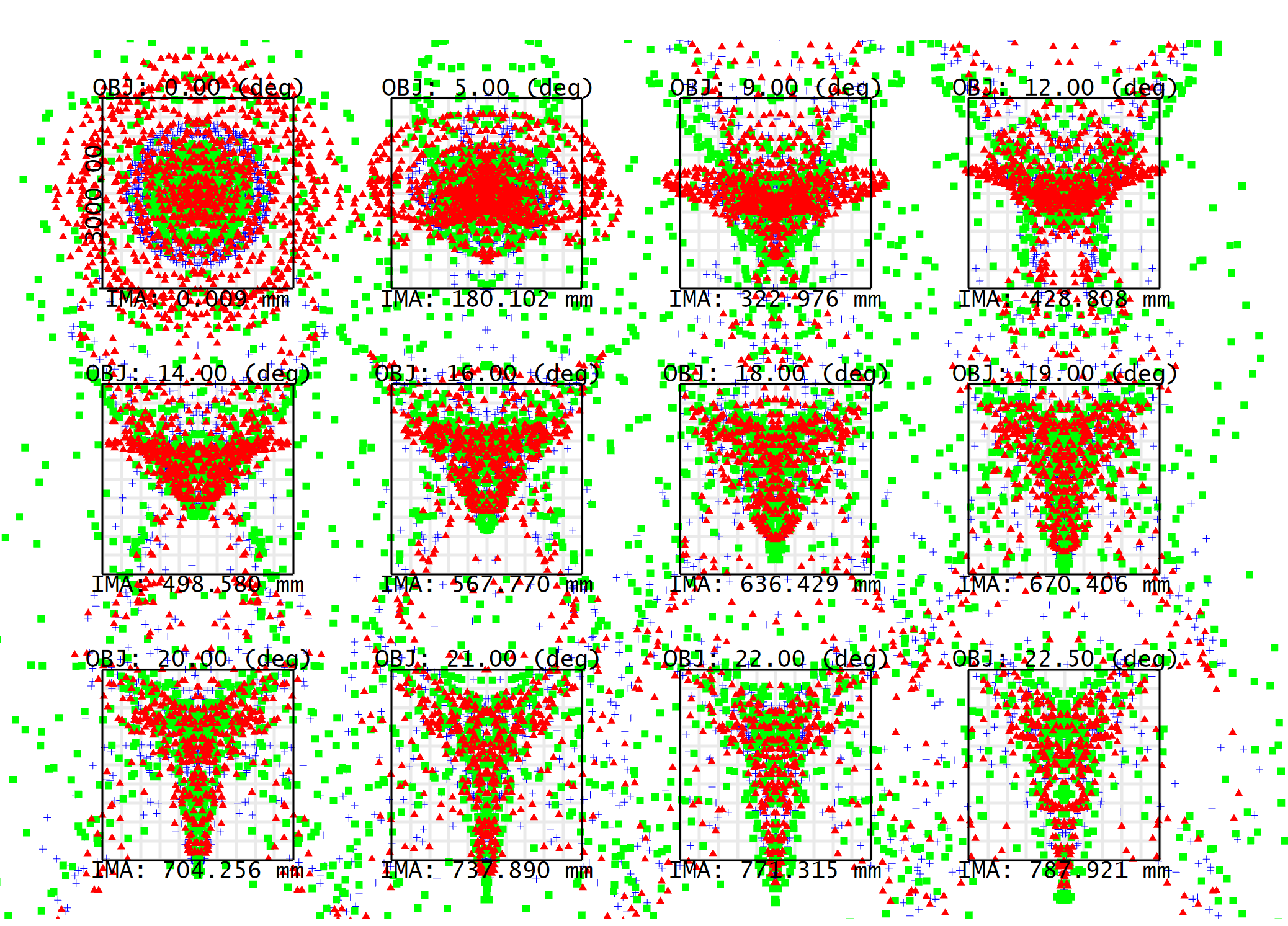}{0.8}
\end{minipage}
    \caption{{\it Left:} RMS of the spot radius as a function of incident angle, plotted for polychromatic (black) 0.36 (blue) 0.33 (green) and 0.39 (red) microns.  {\it Right:} Spot diagram, over 45 degree FoV, plotted against 3 mm pixel size areas; dots for 0.36 (blue) 0.33 (green) and 0.39 (red) microns. 
    \label{fig8}}
\end{figure*}

\begin{figure*}[ht]
\begin{minipage}[t]{0.47\textwidth}
    \postscript{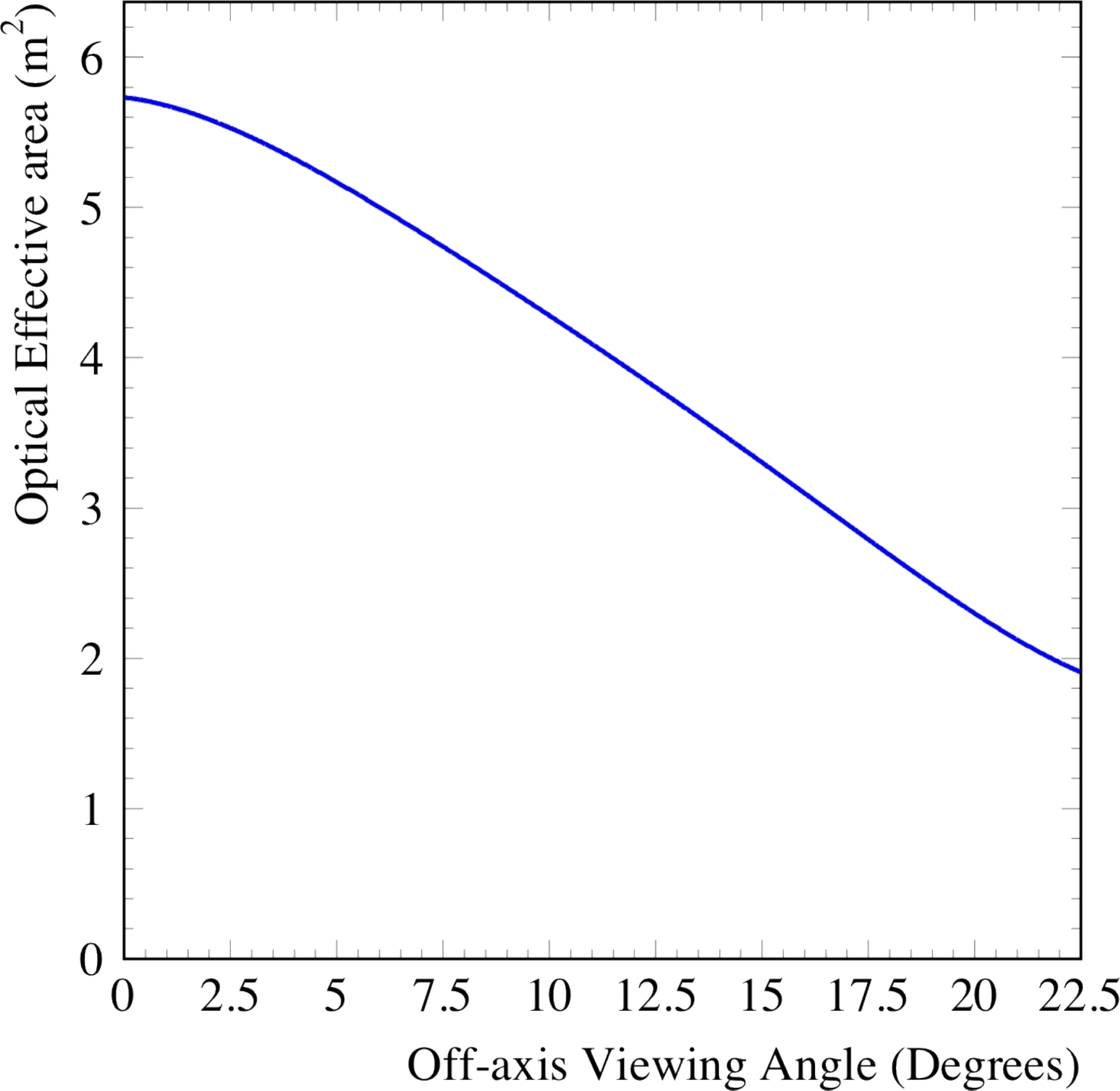}{1}
\end{minipage}
\hfill \begin{minipage}[t]{0.52\textwidth}
  \postscript{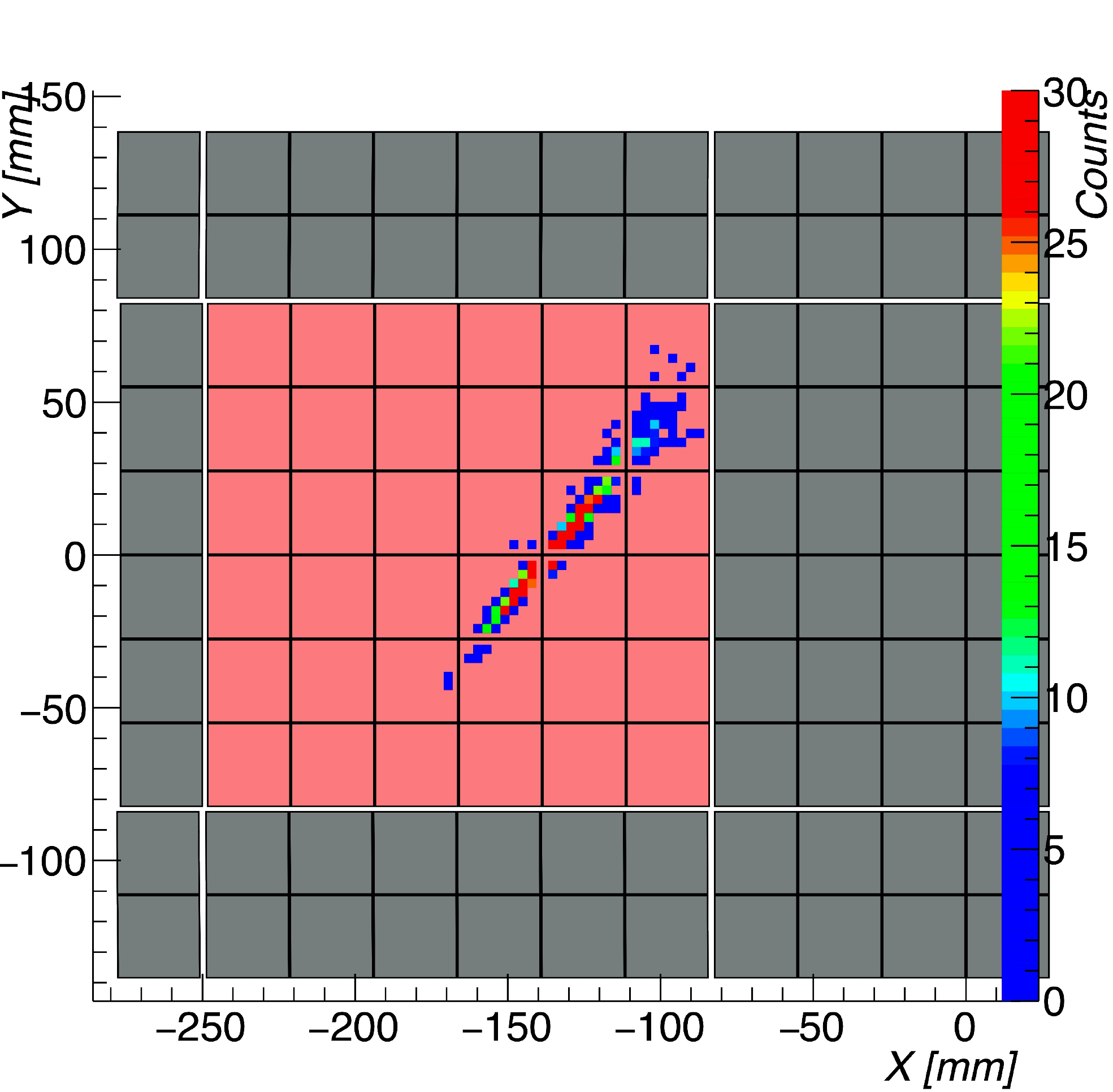}{1}
\end{minipage}
\caption{{\it Left:} The POEMMA optical {\it effective} area as a function of viewing angle.   {\it Right:} EAS fluorescence image on one POEMMA Photo Detector Module (PDM) for a 100 EeV shower. 
\label{fig9}}
\end{figure*}

The POEMMA telescope optical design is optimized for the space-based measurements of UHECRs and cosmic neutrinos. Each POEMMA instrument is based on a Schmidt telescope design (see Figure \ref{fig7}) with a large nearly spherical primary mirror (4 m diameter). The optics includes a thin refractive aspheric aberration corrector (3.3 m diameter, 8.55 m$^2$ area) at the center of curvature of the primary mirror, and a convex spherical focal surface (1.61 m diameter). The corrector lens optimizes the optical aperture producing  a very large optical collecting area and a massive field-of-view (45$^\circ$ full FoV). Taking into account the obscuration of the stop and the focal surface, the optical collecting area is 6.39 m$^2$.  Once the optical transmission is also considered, the optical {\it effective} area is 5.71 m$^2$ on-axis and is shown in Figure \ref{fig9} for other viewing angles. 
This optical design requires an offset between the corrector and the stop, and a slightly aspheric primary mirror (only ~0.8 mm of departure from the base sphere) to achieve good image quality over the entire field-of-view (FoV).
The volume constraints of the launch vehicle were also considered in the optical design.
The 4-m diameter of the monolithic POEMMA primary mirror is set to fit within the dual payload adapter (DPA) the Atlas V launch vehicle. 

To accurately resolve the UHECR EAS development needed to measure the EAS evolution in fluorescence measurements,  a spatial resolution of $\sim$ 1~km from a 525~km orbit is required. This leads to pixel angular resolution of $\sim 0.1^\circ$ or smaller.  Given the 3 mm linear pixel size of the focal surface photosensors, the optics design achieves an instantaneous FoV (iFoV) of 0.084$^\circ$ per pixel and results in a system with an effective focal length (EFL) of 2.04 m, and an effective pupil diameter (EPD) of 3.3 m.  Because the EAS image, and therefore the detectors, are located between the optical aperture, i.e., entrance pupil, and the 4-m diameter primary mirror, the focal plane obscures a portion of the collecting area. However the optical aperture is oversized to ensure that the 6.39m$^2$ optical collecting area is achieved. 

The point-spread-function (PSF) of the POEMMA optics has a RMS diameter that is no more than the 3~mm spatial pixel size of the photodetectors in the focal plane (see Figure \ref{fig8}) over the entire FoV. Note that the UHECR and neutrino EAS imaging requirements are nearly a factor $10^4$ away
from the diffraction limit, implying optical tolerances closer to that for a microwave dish than an astronomical telescope. The optics specifications are detailed in Table \ref{tab-2} in the Appendix.  (A cross section layout is shown in Figure \ref{fig7} left, spot diagrams on the 3 mm pixel scale are shown in Figure \ref{fig8} right, the RMS spot size versus viewing angle in Figure \ref{fig8} left, and the effective area as a function of viewing angle is in Figure \ref{fig9} left.)   Table~\ref{OpticsTable} details the specification of the optical design of the POEMMA Schmidt telescopes.

\subsubsection{POEMMA Focal Surface}

\begin{figure}[tbp]
    \postscript{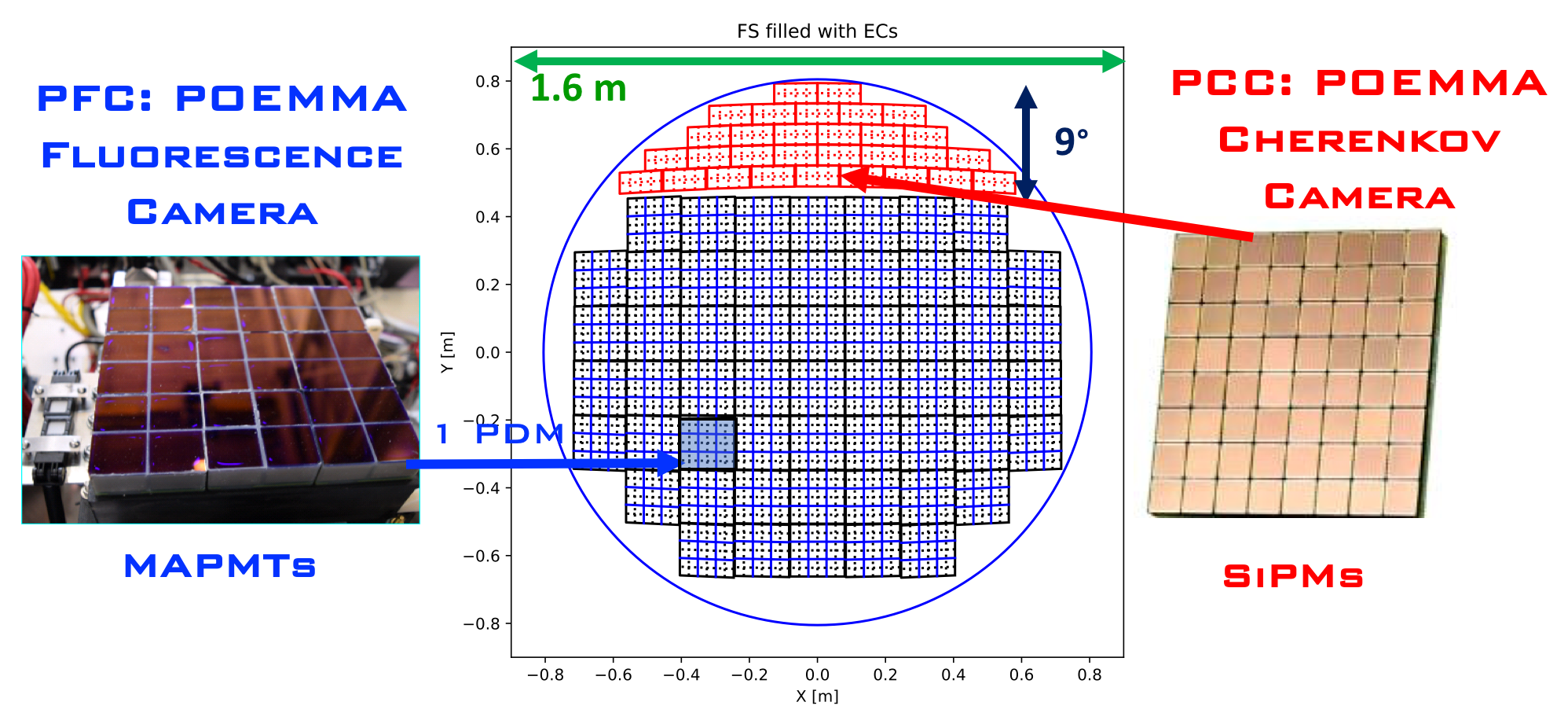}{0.99}
    \caption{Layout of the photon sensors on the focal surface. The red tiles represent the Cherenkov light sensors (each tile is an FSU composed of 8 SiPMs) and the blue tiles correspond to the fluorescence light sensors (each tile is a PDM composed of 36 MAPMTs). {Adapted from Ref. \cite{Olinto:2019mjh}}.
    \label{fig10}}
\end{figure}

The focal surface (FS) of each POEMMA telescope forms a convex spherical surface matched to the design of the optical system (Figure \ref{fig7}). The radius of curvature of the FS is  2.07 m. The 45$^\circ$ FoV of the optical design leads to a FS diameter of 1.61 m. The FS is divided into two sections: the POEMMA Fluorescence Camera (PFC), optimized for EAS fluorescence signals, and the POEMMA Cherenkov Camera (PCC), optimized for EAS Cherenkov signals. 

The PFC includes the center of the FoV and covers 80\% of the focal surface area. The PFC records the EAS video in 1$\mu$s frames (or Gate Time Units, GTU) in the $300 \lesssim \lambda/{\rm nm} \lesssim 500$ wavelength band using multi-anode photomultiplier tubes (MAPMTs) each covered with a near-UV transparent BG3 filter. The MAPMTs are baselined to be Hamamatsu (R11265-203-M64) units with ultra-bialkali photocathodes that provide high quantum efficiency.  Each MAPMT has 64 pixels in an $8 \times 8$ array. Each MAPMT pixel measures 3 mm on a side and the entire MAPMT measures $30 \times 30$ mm$^2$. The MAPMTs are grouped $2 \times 2$ into Elementary Cells (ECs), and ECs are grouped $3 \times 3$ into Photon Detector Modules (PDMs), as originally developed for the JEM-EUSO mission \cite{Adams:2013vea}. The PFC is composed of 55 PDMs, as shown in Figure \ref{fig10}, for a total of 1,980 MAPMTs containing a total of 126,720 PFC pixels.

Figure \ref{fig9} left shows the results of an ESAF simulation \cite{Berat:2009va} of an EAS on the PDM of a POEMMA focal surface. The signals recorded by the PFC measures the shower evolution and maximum peak, $S_{\rm max}$, which is proportional to the energy of the UHECR. The stereo reconstruction of the EAS images from both POEMMA observatories precisely determines the geometry of the EAS and thus the column depth where $S_{\rm max}$ occurs, denoted as $X_{\rm max}$ in units of ${\rm g/cm^2}$ (see Figure \ref{fig4}).  $X_{\rm max}$ is sensitive to the type of particle initiating the EAS because different particle species have different cross sections. By accurately measuring $S_{\rm max}$ and $X_{\rm max}$, POEMMA can determine the energy and the nature of the UHECRs, e.g., proton, heavier nucleus, neutrino, photon, or even non-standard model particles, and how the UHECR composition changes above $\sim$20 EeV.

The POEMMA Cherenkov Camera occupies the upper 9$^\circ$ of the FoV and $\sim10\%$ of the focal plane area (see Figure \ref{fig10}) and is populated with silicon photomultipliers (SiPMs). SiPMs are relatively new yet mature photodetectors. They are currently used or planned to be used in several high-energy physics and astroparticle physics experiments. Their compactness and high spectral sensitivity make them particularly interesting for POEMMA. 

Most of the Cherenkov light arriving at the telescope has wavelengths $>300$\,nm, mainly due to ozone absorption, with a  tail extending into the IR. The photon detection efficiency of the latest SiPMs peaks above 50\% at $400\,$nm  \cite{NepomukOtte:2018qjk,Otte:2016aaw}. We have characterized many SiPMs to detect Cherenkov light from air-showers  with instruments on the ground and in space \cite{NepomukOtte:2018qjk,Otte:2016aaw,Otte:2018uxj}. Using these measurements, we find that SiPMs have an almost optimal spectral response for our application \cite{Reno2019a, Krizmanic:2020shl}. 

Features identified as nuisances in early SiPMs, which could potentially hamper their use in demanding experiments like POEMMA, were positively addressed over the last several years. For example, the temperature dependence of the gain and PDE was high and considered a problem. But because state-of-the-art SiPMs now operate at 10\% to 20\% overvoltage, small temperature-dependent variations of the breakdown voltage (typically 0.1\% per K) do not cause significant changes in gain or PDE \cite{Otte:2015ggr}. The remaining relative temperature dependence of the PDE is about 1\% per K and comparable to the relative temperature dependence of the classical bialkali photomultiplier's quantum efficiency like the one used in the fluorescence camera. 

Another classical nuisance of SiPMs used to be optical crosstalk, the correlated and simultaneous firing of SiPM cells. Current designs have trenches between cells to prevent light from transversely propagating through the SiPM, thus largely suppressing optical crosstalk. Even at high operating voltages such as those  we plan to use for POEMMA to achieve a close to 100\% breakdown probability, optical crosstalk amounts to only a few percent \cite{NepomukOtte:2018qjk}. The most significant impact optical crosstalk has on the operation of POEMMA is an increased threshold due to the higher probability of seeing larger fluctuations at the trigger level \cite{Vinogradov:2011vr}.  With the few percent optical cross talk we measure, we find in simulations that the trigger threshold increases insignificantly.
Similarly, the impact of optical crosstalk on the energy reconstruction is insignificant compared to the photon flux's statistical fluctuations and the large photon flux from the night sky background.
 The large night sky background intensity is also more than ten times larger than the intrinsic dark count rate of SiPMs.

In the simulations we presented above, we modeled the SiPM with the characteristics measured in our laboratories \cite{NepomukOtte:2018qjk,Otte:2018uxj}.

In the PCC, the Cherenkov signals are recorded with a sampling rate of 100\,MS/s. The SiPMs in the PCC are to be assembled in arrays of $8 \times 8$ pixels with a total area of $(31 \times 31)$\,mm$^2$. The SiPM arrays are, furthermore, grouped $4\times 2$ into Focal Surface Units (FSUs). A total of 30 FSUs forms the full PCC with 15,360 pixels (Figure \ref{fig10}). 

 As shown in Figure~\ref{fig10}, the PCC measures Cherenkov signals up to 9$^\circ$ away from the edge of the telescope's FoV and in a $\sim30^\circ$-wide band.   This angular span of the PCC is amply sufficient for viewing and following ToO neutrino sources since the transient source angular error is $\lesssim 2.5^\circ$, e.g. the largest effective Cherenkov emission angle from the brightest EAS determined from simulations \cite{Venters:2019xwi}. The PCC contains 15,360 $3\times3$ mm$^2$ pixels, with each pixel signal digitized using 100 MHz sampling.  The front-end electronics for the PDMs and FSUs will be located directly behind the focal surface. These signals are transported to the data acquisition electronics located underneath the primary mirror and in the satellite bus.

The PCC also records the Cherenkov light flashes from EASs produced by UHECRs viewed above the limb of the Earth. By pointing the PCC such that it spans $2^\circ$ above the Earth's limb, the rate of UHECR Cherenkov detections will be measured. Thus, the effect of a possible background to observing \tauon decays  induced by $\nu_\tau$ interactions in the Earth will be determined. 

The baseline FSU design developed in the IDL included a silicon pad detector attached behind each SiPM array. The purpose of these pad detectors are to identify and subsequently veto SiPM background signals from charged-particle radiation in the LEO environment, especially from galactic cosmic rays. Monte Carlo simulations were performed to simulate the flux of cosmic rays penetrating into POEMMA and striking the SiPM array. These simulations also recorded whether the cosmic rays striking a SiPM also struck any of the silicon pad detectors. These calculations determined the background rate from all the cosmic ray hits on the SiPMs. It was determined that the background count rate in the SiPMs from cosmic rays was far less than that from the dark-sky air glow background \cite{Krizmanic:2020shl}. Events triggered by the night sky air glow background must be suppressed by setting of a high count-rate threshold. Because of this high threshold, the cosmic ray background makes a negligible contribution. Thus, silicon pad detectors will not be included in future designs of the FSU. It should be noted that the high photo-electron threshold imposed by the dark-sky  air glow background constraints also suppresses noise triggers from the SiPM dark count rate.

\subsubsection{Atmospheric Monitoring System}

The Atmospheric Monitoring System (AMS) uses two infra-red (IR) cameras on each POEMMA telescope. These cameras will observe clouds in and around the FoV of the main instrument \cite{Adams2015a,Adams2015b,Allen2018}. The IR images collected through an 8 $\mu$m filter on one camera and through a 11 $\mu$m filter on the other camera will be analyzed to identify the presence of clouds and to estimate cloud top heights \cite{Anzalone2019}.

The FOV of each IR camera is 60$^\circ$x45$^\circ$, with a corresponding pixel FoV of 0.094$^\circ$.  Each picture has a resolution of 640 by 480 pixels with 16 bits per pixel with an uncompressed picture size of 615 kB. Each IR camera will take one image per minute, an interval based on the instrument FoV and the 7.6 km/s speed of the satellites. 
 
 The IR cameras will be housed in a small  enclosure mounted at the center of the POEMMA optics corrector lens assembly, which is a blind spot caused by the focal surface, and behind the shutter (see Figure \ref{fig1}). Power and data will be delivered through cabling connected to the main electronics box. The camera system will use internal controllers to maintain a stable temperature within 1$^\circ$C of its operating range of -40$^\circ$C to 55$^\circ$C. The IR camera assembly will include an integrated calibration target that can be moved in front of the camera lenses.
\subsubsection{Calibration Systems}
\label{subsec:CalibrationSystems}

The POEMMA on-board calibration system will feature pulsed light emitting diodes (LEDs) that periodically illuminate the focal surface of the instrument. The data recorded will be used to monitor its relative photometric calibration throughout the mission following the methods proposed for the JEM-EUSO mission \cite{Adams2015a}. The LED system on each satellite will include 3 LED modules and one control unit. The LED modules will be mounted on the bottom side of the corrector plate holding frame. The control unit will be mounted behind the mirror. Each LED module contains seven LEDs (340, 360, 390, 500 , 650, 800 and 950 nm). These are mounted behind an opal diffuser together with a digital-to-analog converter and analog drivers. During data taking operations, a subset of LEDs, for example 360 nm and 500 nm, will be flashed individually, typically once per orbit. LEDs will also be flashed during portions of the orbit when science data acquisition operations have been suspended because the moonlight background is too high. Occasionally, detailed information on all focal surface channels will be acquired by executing a dedicated full instrument calibration run. During this run the LEDs will be flashed sequentially by the controller strobes through a range of wavelengths, light pulse amplitudes and temporal widths to span the dynamic range of the instrument with an emphasis on the properties of the expected signals from astroparticle and atmospheric science targets of interest, from the 10's of nanoseconds long Cherenkov signal from EASs to the meteor signal observed to the second time scale.

\subsubsection{Data System}

The electronics and real-time software for the two sections of the focal surface (the PFC and the PCC) function independently so each can be optimized for their specific task. A single Data Processor (DP) will read out the data from both systems, store it locally and then transmit it to the spacecraft bus for transmission to the ground.

The search for UHECRs will be based on the PFC detection of fluorescence from EASs using the MAPMTS in the PDMs. An EAS will show up as a bright, persistent spot that will typically move across part of the focal surface depending on the viewing angle (see Figure \ref{fig9} right). Each PDM will contain one SPACIROC ASIC for each MAPMT, and one PDM FPGA to search for signal persistence. The ASIC counts photoelectrons for each MAPMT anode individually during a 1$\mu$s GTU. The FPGA will search those counts for a persistent signal in excess of the background. The threshold for the brightness and duration of the persistent signal will be adjusted to yield a rate of no more than 7 Hz in total across the whole MAPMT system. Taking into account that the scene viewed by each pixel changes every 100 ms due to the orbital motion of the spacecraft, 
the relative threshold adjustment for the PFC to account for changes in brightness level 
should be monitored, and changed if necessary, at a rate that is fast enough to respond to these changes, e.g. at rate larger than 10 Hz. We note that during the EUSO-SPB1 mission  thresholds were adjusted at the MAPMT level at a rate of 0.3 MHz.
When persistence is detected, 
all the data will be passed to an FPGA readout board that will implement a search for contiguous tracks. The threshold for defining a track will be adjusted so that the total rate is no more than 0.1 Hz. When a track is detected the DP will begin the readout process. 

For atmospheric TLEs, including elves, a longer readout time extending to several ms with a specialized trigger will be implemented, similar to that used in Mini-EUSO \cite{Capel:2018}. For the measurement of slower phenomena, such as meteors, a data sampling at ms scale is needed and recommended.

The search for cosmic neutrinos will be based on the PCC detection of Cherenkov light from EASs initiated by \tauon decay. The Cherenkov signal is much shorter ($10$--$100$ ns) compared to the fluorescence signal of an EAS and requires a different readout concept. Instead of counting individual photoelectron signals, the SiPM signals are continuously scanned and the readout is triggered if a signal above a preset threshold is sensed. The threshold will be set high enough to reduce the background rate,  due to the dark-sky airglow and other sources, to an acceptable level. Once triggered, the signals of the SiPMs within a region in the PCC where the trigger occurred are digitized. The Cherenkov signals are digitized with a sampling rate of 100\,Ms/s to limit their contamination by signals from background dark-sky airglow photons. 

Upon receipt of a trigger in either the PFC or the PCC, the DP will read out data and store it locally, eventually transferring it to the spacecraft bus that handles transmission to the ground stations. The DP will also handle housekeeping functions, interfaces to various auxiliary systems (e.g., atmospheric monitoring system), and exchange of commands and data with the spacecraft bus.  The total data rate is estimated to be $<1$ GB/day.

\subsubsection{Mechanical Structure} 

The POEMMA instrument structural concept was reviewed by Marshall Space Flight Center(MSFC) engineering staff. The most challenging parts of the mechanical system are the deployment of the large optical elements, corrector plate and focal surface, light-shield, and the mechanical shutter. Structural analysis of the conceptual design shows a first fundamental frequency after deployment at 7.9 Hz, 60\% above the IDL goal of 5 Hz, which is appropriate for such a large structure. An analysis of the stowed configuration during launch could not be completed due to the lack of design details of the DAP and launch vehicle at this stage of the design. However, Engineering solutions are available to meet the 15 Hz guideline through additional design effort during the instrument preliminary design phase.

The deployment of the focal surface and corrector plate rely on one-time operations of actuators that drive the folded struts to reach full extension and then lock them into place for the duration of the mission. No further adjustment of the optical system is required due to the sizable tolerances afforded by the coarse EAS imaging requirements (which is $10^4$ away from the diffraction limit). 

The operation of the shutter doors will be critical over the duration of the mission. The shutter must be closed when the spacecraft is approaching the day side of the orbit to prevent the exposure of the focal plane to bright light. For a five year mission, this is $\sim$55,400 cycles. The shutter design  is a motor driven gear system with redundant motors and gear boxes for each half of the shutter. The two shutter doors are composed of light-weight honey comb structures. An overlapping seam at the center prevents light leaks when the doors are closed. The shutter will protect the optical system during daytime portions of the orbit and, together with other mitigation strategies, will reduce the fluence of atomic oxygen impinging on the corrector plate during the mission. The large area of the shutter surfaces will also function as part of the thermal sub-system in maintaining a controlled environment. 

\subsubsection{Light shield} 

To prevent stray light from entering the POEMMA instrument, provide micrometeoroid protection and for thermal control, a collapsible shroud will surround the optics (Figure \ref{fig6}). The concept for this shroud was developed by L'Garde, Inc. \cite{Garde:2019}.The shroud will consist of a set of three nested cylinders. Each cylinder will have Beta cloth on the outer surface and flat-finish black Kapton on the inner surface. Multi-layer insulation can be added between the inner and outer surfaces, if needed for thermal control. When POEMMA is stowed for launch, the nest of cylinders will be collapsed and secured. The bottom of the outer cylinder will be attached to the periphery of the mirror and the top of the inner cylinder will be attached to the frame of the shutter door. The top of the outer cylinder, the bottom of the inner cylinder and both the top and bottom of the middle cylinder will have locator pins that engage holes in the periphery of the mirror and in the door frame respectively to secure them during launch. As POEMMA deploys, these pins will disengage. Cables (stowed in tubes at launch) will deploy and pull the middle cylinder into position. Rolling Kapton light baffles between the cylinders will insure that the shroud is light tight both when stowed and when deployed.

\subsubsection{Thermal System}

Thermal modeling of the POEMMA instrument in the IDL determined that passive cooling can provide sufficient thermal control of the instrument. In particular, the requirements are driven by the temperature range ($0$--$20$ $^\circ$C), gradient ($\pm 5$ $^\circ$C), and stability requirements for the SiPM modules in focal surface (FS). The large surface area of the FS radiates heat to the interior volume of the instrument. The black interior of the light shield (used to minimize stray optical light in the telescope) and the white interior of the shutter doors absorbs the heat and the high emmittances of the light shield and doors radiate the heat into space. Constant conductance heat pipes and the aluminum mechanical structure minimize the temperature gradient across the focal plane. In this design, no radiators on the spacecraft are required.  Heaters (300 W) are required to keep the instrument in the desired temperature range, especially in the worst cold case and mechanical thermostat control can keep the temperature in the range of $\pm 5$ $3^\circ$C. The thermal modeling analysis was performed assuming 30\% margins.

\subsection{Spacecraft bus}

\begin{figure*}[ht]
\begin{minipage}[t]{0.52\textwidth}
    \postscript{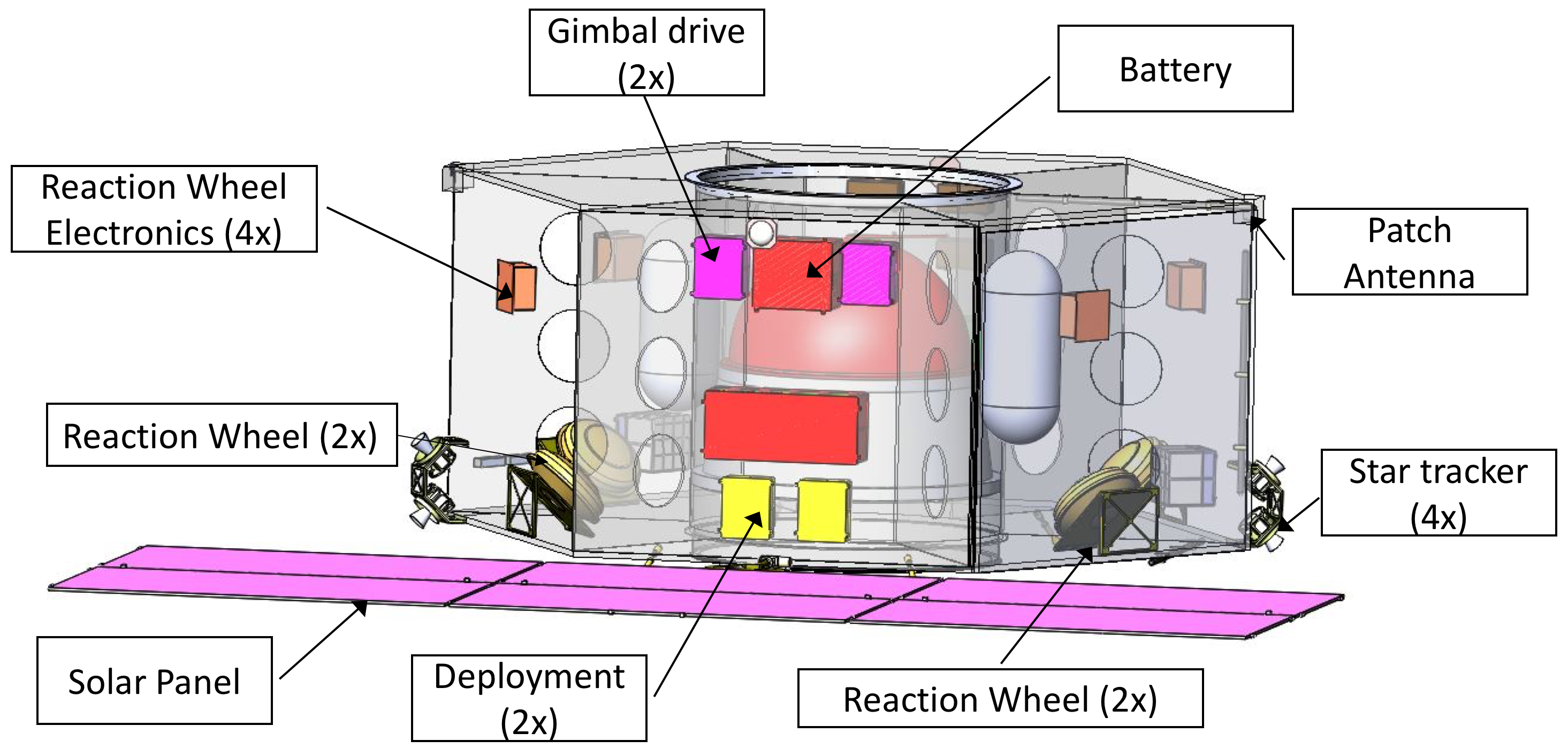}{1}
\end{minipage}
\hfill \begin{minipage}[t]{0.46\textwidth}
  \postscript{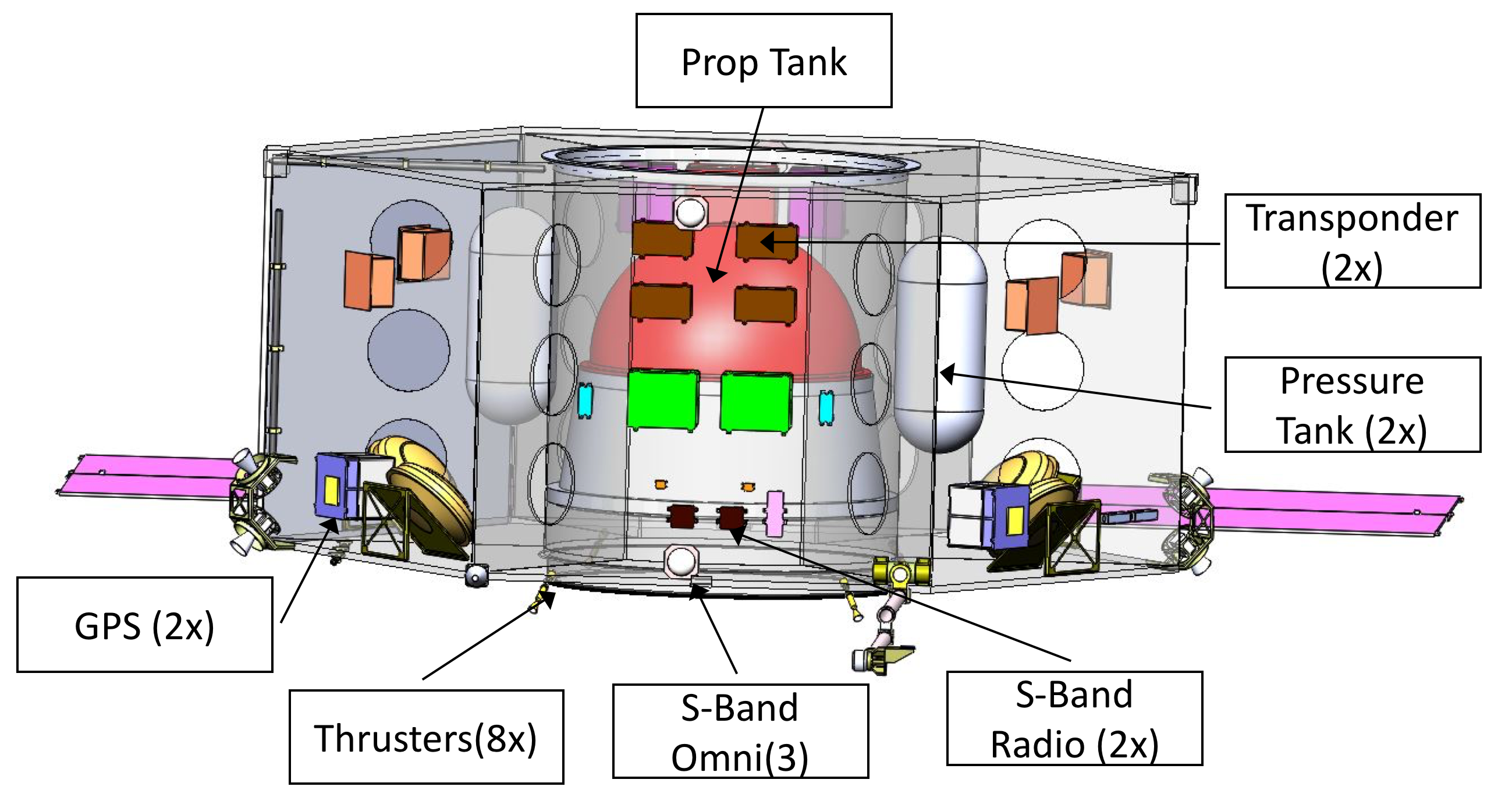}{1}
\end{minipage}
\caption{Spacecraft bus MDL design with some components highlighted.  \label{figY}}
\end{figure*}

The spacecraft bus (Figure \ref{figY}) is a hexagonal cylinder 1.55 m tall with an outside diameter of 3.37 m weighing 1,073 kg (including a 25\% contingency). Located behind the POEMMA mirror (see Figure \ref{fig6}), it provides basic services to the telescope including on-orbit deployment, power, communications, attitude control, propulsion and avionics. 

The avionics includes the command and data handling system (including the flight computer), the spacecraft clock that provides the precise timing we need for synchronization between the satellites, the gimbal drive electronics to steer the solar panels, and the control functions for all the deployment mechanisms to unfold the instrument once it reaches orbit. 

\subsubsection{Communications Links}

 The data volume collected by each POEMMA telescope is estimate to be 7.96 Gbits/day, where 7.22 Gbits/day is science data and the remainder is spacecraft, housekeeping and contingency. The latency of data reaching the ground can be up to 24 hours. The satellites have onboard storage for up to 7 days of data.

Each satellite carries three S-band omni-directional antennas. The S-band downlink requires 16 minutes of ground station contact per day while 125 minutes per day is available. Commands are normally uplinked during passes over ground stations, but the satellites also include the capability to receive ToO alerts and emergency commands linked via TDRSS through the S-band antennas.

\subsubsection{Avionics and Propulsion} 

The spacecraft avionics are designed to allow POEMMA to quickly slew its pointing by as much as $90^{\circ}$ in $500$~s. Additionally, the spacecraft propulsion systems will allow for adjusting the separation between the two satellites in the 525 km orbits. 

The avionics package on each spacecraft includes reaction wheels with the capability to achieve the $90^{\circ}$ in $500$~s slew rate. Magnetic torquers will be used to off-load the wheel angular momentum, and this can be augmented by a small amount of propulsion if needed. Thus, the re-pointing the satellites will have negligible impact on the fuel usage and the number of these maneuvers does not affect mission lifetime.  The propulsion system has capacity to perform both orbit maintenance to compensate for atmospheric drag for a five year mission and to adjust the satellite separation. The number of separation maneuvers that can occur during the mission depends on the distance and the timescale for the maneuvers and the available propulsion. The General Mission Analysis Tool (GMAT) was used to calculate the fuel usage for these separation changes, and the effects of atmospheric drag was taken into account. The results show that the POEMMA satellite orbits can be changed from a 300 km separation to a $25$~km separation $\sim 40$ times during the mission if the timescale of each maneuver is $\sim 1$~day to complete, assuming that each maneuver also increases the separation back to the original $300$~km after each ToO observation. If the duration for the initial satellite separation maneuver is reduced to $\sim 7$~hours and $1$~day to bring the satellites back to the $300$~km separation, then $\sim 12$ maneuvers can be performed over the mission lifetime. Note that the altitude variation for the spacecraft is $500$ -- $550$ km when performing these separation maneuvers and has minimal effects on EAS fluorescence measurement thresholds during the maneuvers.

The regulated monopropellant propulsion system for each POEMMA satellite is sized to correct the orbits for initial launch vehicle dispersion, maintain the orbits for each science formation, allow for the change of satellite separation to optimize ToO neutrino measurements, and to de-orbit the satellites at the end of the mission. Each satellite contains a propulsion tank with sufficient capacity for a 5-year mission with 10\% additional margin. A redundant suite of thrusters is included in each satellite. A Technical Readiness Level (TRL) of 9 was assessed for the POEMMA propulsion system based on the MDL design.

The avionics in each satellite consists of a prime and redundant command and data handling system based on a Rad 750 Processor with 25 GB of Memory Storage. A 100 Mhz Hz Oven-controlled crystal oscillators provide an accurate clock. It controls all the satellite functions, command and control of the science instrument,  storage and transmission of the data, and reception of commands.

\subsubsection{Power}

The 28 volt power system designed for the POEMMA observatories is designed for a 5-year mission (a 3-year primary mission and a 2-year extension). It uses 7.8 m$^2$ of triple-junction GaAs solar cells mounted on rigid honeycomb structure to produce 2,428 W at the start of life and 2,050 W at end-of-life. This satisfies the mission power budget of 2,030 W including a 30\% contingency. A 3-axis drive is used to maximize the solar array exposure without relying on the spacecraft attitude control system. 

The energy generated by the solar array is stored in a 145 AH lithium-ion battery pack that provides power when the spacecrafts are not in sunlight and for 60 minutes during the launch operations. This battery is designed for 27,639 charge-discharge cycles to about 20\% depth of discharge in each cycle. 

This power system design is based on high TRL ($\geq$ 7) components and devices with flight heritage. The design can comfortably generate enough power to meet the needs of the Instrument and Spacecraft (see Table 1 for power with contingencies for each unit). It
includes redundancy for critical elements to meet NASA Class B mission
requirements.

\subsubsection{Integration and Tests}

The POEMMA instrument integration will occur at a prime contractor to be selected, followed by the integration of the instrument with the spacecraft bus.  The testing of the payload can occur at NASA/GSFC. 

\begin{figure}[tbp]
    \postscript{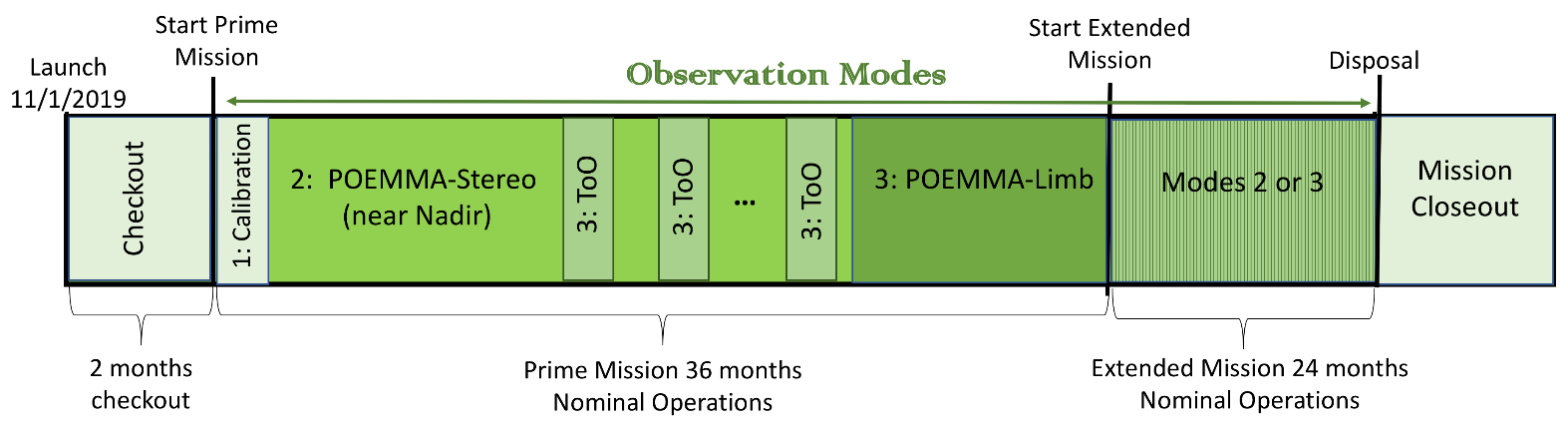}{0.99}
    \caption{Mission timeline (diagram not to scale).  \label{fig11}}
\end{figure}

\section{POEMMA Mission Concept}

POEMMA is a NASA Class B mission consisting of two identical telescopes deployed as two satellites by an Atlas V vehicle to LEO. The selected LEO has an altitude of 525~km and a  $28.5^\circ$ inclination angle with an orbital period of about 95 min. The POEMMA mission concept was developed at the MDL in late 2017. 
A timeline for the full project, beginning at the start of Phase A (Concept and Technology Development) to Phase E (Operations and Sustainment),  was developed at the MDL study \cite{POEMMAnasaReport}. From the start of Phase A to the launch (start of Phase D-3) of the POEMMA observatory the estimated time is about 5 years total. The on-orbit mission (Phase D-3, Phase E, and Phase F) corresponds to checkout, calibration, observations, and close out, over a prime mission of 3 years and an extended mission of 2 years to a total of 5 years, as shown in Figure \ref{fig11}. (Depending on the observatory performance, resilience, and science results, a longer mission could be achieved.) 

\begin{figure}[tbp]
    \postscript{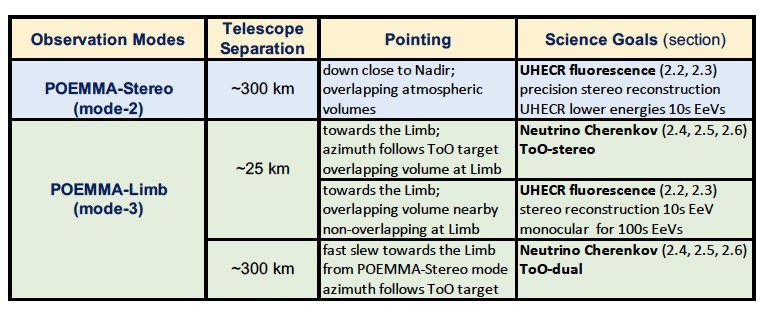}{1}
    \caption{POEMMA primary observation modes for UHECR science (\S2.2 and \S2.3) and cosmic neutrino science (\S2.4, \S2.5, and \S2.6). Atmospheric science (\S2.7) and observations of Meteors and nuclearites (\S2.8) can be done in all modes.}
    \label{fig13}
\end{figure}

An optimized strategy for time allocation between the different operating modes of POEMMA should be developed during Phase A informed by the accumulated knowledge from up-to-date ground observations of UHECRs and cosmic neutrinos. Further adjustments should be expected during the mission informed by POEMMA discoveries. Here we describe primary modes of operations that optimize the UHECR science and the cosmic neutrino ToO follow up (see Figure \ref{fig13}).

The prime 3-year mission has three basic phases: calibration (mode-1), POEMMA-Stereo (mode-2), and POEMMA-Limb (mode-3), (see Figure \ref{fig11}).  
During the extended 2-year mission, the satellites can operate in either mode-2 or -3 depending on the science achievements of the prime mission. 

Following on-orbit checkout after deployment, the two satellites will be oriented to point their instruments toward the zenith while separated by about 300 km. During this first (mode-1) observational phase, the instruments will be calibrated by making simultaneous observation of stars with well-known apparent luminosities (mode-1 in Figure \ref{fig11}). 

In the second observation phase (POEMMA-Stereo or mode-2), the two satellites will fly in a formation separated by $\sim$300 km pointing close to nadir to observe the same volume of the Earth's atmosphere, as depicted in the left panel of Figure \ref{fig2}. This is the primary satellite formation for the UHECR science shown in Figure \ref{fig11} and \ref{fig13}.  Once the UHECR science goals at the lower energy range of the mission are completed (i.e., sufficient statistics is accumulated around 10s of EeVs),  the satellites move to mode-3, also denoted POEMMA-Limb. In mode-3 the observatory is tilted away from nadir pointing at the limb of the Earth and the spacecraft move closer together towards a $\sim$ 25 km separation. Mode-3 (right panel of Figure \ref{fig2} and in the table of Figure \ref{fig13}) optimizes the search for cosmic neutrinos near the Earth's limb and increases the exposure to UHECRs at higher energies.  

The POEMMA-Limb pointing (mode-3) is optimal to observe Cherenkov from neutrinos from below the limb in ToO-stero mode.  Cherenkov signals observed by the PCC are reconstructed using spatial and temporal coincidence (or time-delay) between the satellites, which observe overlapping regions of the Earth's limb. Concurrently, UHECR observations are observed throughout the atmospheric volume by the PFC. The selection of the pointing azimuth for POEMMA-Limb observations is determined by the ToO strategy to follow-up recent transients as their positions rise or set below the Earth's limb. While in mode-3, during periods when no ToO is observable, the telescopes can be adjusted to point slightly crossed eye from each other, such that the UHECR stereo can be observed nearby and large statistics accumulated closer to the limb by observing non-overlapping volumes of the atmosphere. 

While flying in the mode-2 formation, the satellites will temporarily be re-oriented into mode-3 to point at ToOs following an alert of selected astrophysical transient events. 
For ToOs with predicted short neutrinos bursts coincidence, a fast follow up to mode-3 is achieved by swiftly pointing the telescopes toward the source position before it rises or after it sets below the Earth's limb. For this fast response (500s for 90$^\circ$ slew), the original satellite separation is maintained at $\sim$300 km corresponding to a POEMMA-Limb ToO-dual observation mode as in the table of Figure \ref{fig13}. In the ToO-dual mode the stereo vision of the Cherenkov signal is unlikely given the larger separation for most ToO source positions in the sky. 
For longer lasting ToOs, the observatory will continuously point in POEMMA-Limb mode following the source setting or rising position and the satellites can be brought closer together into the ToO-stereo configuration (to $\sim$25 km separation) as shown on the right of Figure \ref{fig2}. This brings both satellites within the Cherenkov light pool of a single upward EAS so that temporal coincidence reduces the night sky airglow background and lowers the energy threshold of the observed neutrinos.

\subsection{Launch Operations}

\begin{figure}[tbp]
    \postscript{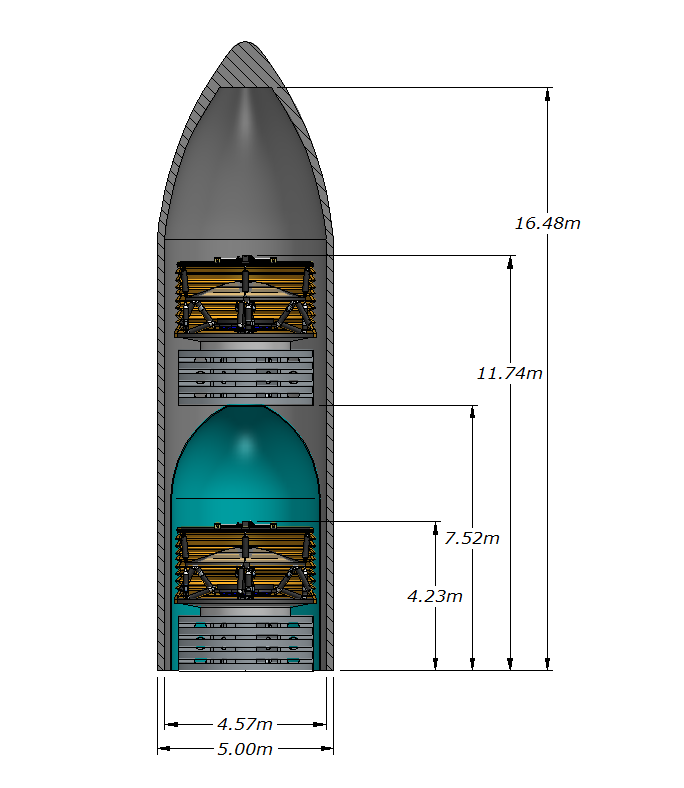}{0.6}
    \caption{Dual Launch Manifest in an Atlas V launch vehicle fairing. {Adapted from Ref. \cite{Olinto:2019mjh}}.
    \label{fig12}}
\end{figure}

The POEMMA satellites will be dual-manifested on an Atlas V using the long payload fairing as shown in Figure \ref{fig12}. The satellites will be launched into circular orbits at an inclination of 28.5 degrees and an altitude of 525 km, where they will remain until de-orbited for disposal in the southeastern Pacific at the end of the mission. 

The satellites are launched in a stowed configuration. Once on orbit, the corrector plate and focal surface will be deployed into their final position. As the instruments deploy, a telescoping shroud surrounding the optical bench will deploy. A solar array will be deployed from each spacecraft bus. These arrays are located behind the mirrors and are three-axis gimballed to track the sun.

\subsection{On-Orbit Operations}

POEMMA makes observations in umbra and in low moonlight conditions. The satellites will orbit the Earth with a period of 95 minutes, orbiting the Earth $\sim$15 times per day. As the satellites are leaving umbra the shutter doors are closed to protect the focal surface. They are opened again as the satellites approach umbra. There will be up to 30 door operations per day. When the shutter doors are open, the satellites are oriented so that the optical axes are over 90$^\circ$ away from the ram direction when possible to avoid the risk of atomic oxygen damage to the Schmidt correctors. During observations the attitude of the satellites must be maintained within 0.1$^\circ$ with knowledge of the attitude to within 0.01$^\circ$. Events will be time-tagged with a relative accuracy within 25 ns between the satellites as needed for ToO-stereo observations (in mode-3). 

In case the mission operators loose attitude control of one or both POEMMA satellites, the satellites will automatically go into safe mode and the shutter doors will be closed to prevent sunlight exposure and atomic oxygen damage to the exterior surfaces of the correctors.

The satellites are maneuvered on orbit by means of four reaction wheels (unloaded with torque rods) for attitude control and changes of attitude. The satellites also have eight hydrazine thrusters that are used for changing their orbital separation and for de-orbiting them at the end of the mission. In addition to making changes in separation, these thrusters are used to maintain accurate satellite separations during ToO-stereo observations  in order to tighten the temporal coincidence window between the satellites to 25 ns. The satellites have star trackers and sun sensors for accurate attitude knowledge.

\section{Technology Roadmap}

The POEMMA conceptual study relies primarily on simple and proven technology. No entirely new technologies need to be developed for the mission. However, POEMMA can take advantage of technologies as they mature. Those elements beneficial to POEMMA are listed here with a description of the desired maturation and rationale. POEMMA team members are actively working on the maturation of a number of these components through laboratory, sub-orbital (EUSO-SPB1 and EUSO-SPB2), and spaceflight (Mini-EUSO) testing. 

\subsection{Pathfinder Measurements}

The measurements by the previous, current, and upcoming pathfinder experiments provide crucial insight into the signatures of the various types of events that will occur inside the FoV of POEMMA. These measurements are important not only to estimate the optimal parameters for observation (e.g., trigger setting), but are also needed to improve the simulations that provide, for example, estimates of the energy threshold for different event classes with realistic night sky backgrounds included.

\textbf{EUSO-SPB1} measured the change of UV background level over an extended period of time over the ocean as well as for different cloud altitudes. It observed signatures of of direct cosmic ray hits in its Focal Surface and measured background triggers in searches for tracks from UHECR extensive air showers. Other measurements from EUSO-SPB1 demonstrated the effectiveness of the proposed on-board LED system (\ref{subsec:CalibrationSystems}) for monitoring the health and stability of the Focal Surface in the PFC. EUSO-SPB1 also provided preliminary data acquired with a SiPM from suborbital space \cite{Painter:2019a}.

\textbf{Mini-EUSO} vastly increased our understanding of the UV background from space over various ground and atmospheric conditions. Mini-EUSO also identified and quantified different artificial sources visible in our instrument such as city lights and fishing boats, further increasing our understanding of the anthropogenic background. The observation of atmospheric events (e.g. elves and meteors) by Mini-EUSO \cite{Bacholle:2020emk} confirms the high potential of POEMMA to make novel measurements of TLEs as discussed in \ref{subsec:AtmScience} with the proposed technology for the PFC.

\textbf{EUSO-SPB2} is the pathfinder instrument that is the closest to the POEMMA design. It employs the same optical design, utilizing not only the PFC but also the PCC.  For development simplification, the two are split into two independent Schmidt telescopes rather than combined in one focal surface. 

The EUSO-SPB2 fluorescence telescope will, for the first time, measure the signal of an EAS from a UHECR via fluorescence detection from suborbital space and therefore qualify the proposed detector design in a near space environment. It will also use multiple (three) PDMs for the first time. Their data will also further increase the understanding of the UV background emission of the Earth's atmosphere for the UHECR observation. This telescope will point in nadir and will have sensitivity to bright upward-going EASs for an exploratory search over the ocean.  It will be accompanied by an IR camera to characterize clouds and cloud top heights. 

EUSO-SPB2 will also, for the first time, use the PCC technology, implemented in the Cherenkov Telescope, to search for nanosecond scale optical signals by observing $\pm10^\circ$ above and below the Earth's limb. The Cherenkov light from UHECR passing above the limb will produce signatures with the same properties as those from Earth-skimming tau neutrinos, and by observing them, EUSO-SPB2 will validate the technology for this type of observation with current optics and electronics. Currently, there are no background measurements available on the timescale (tens of ns) or in the wavelength range (200-1000nm) pertinent to detection of the Cherenkov signal. EUSO-SPB2 will provide these measurements. Specifically, in addition to observing steady backgrounds and potential artificial sources, EUSO-SPB2 will begin to identify possible background events, quantifying for example: 
\begin{enumerate}
    \item the reflection from downward going UHECR,
    \item the atmospherically refracted Cherenkov signal from above the limb UHECR which appears to originate from below the limb.
\end{enumerate}
These measurements will help to further define the optimal experimental parameters, sensitivity, and energy threshold of space based detectors such as POEMMA. EUSO-SPB2 will also investigate whether the PCC design can be used to study atmospheric electricity without compromising the neutrino search.  In addition, a Target-of-Opportunity operation mode is planned.  If a strong candidate for transient neutrino emission is slightly below the Earth's limb during an operation period, this telescope could be pointed at this object by rotating the entire gondola in azimuth and the telescope in elevation.

\subsection{Mechanical}

The POEMMA shutter doors must operate during the entire mission and are the highest mission risk. This risk can be reduced during early phases of the project through iterative design and testing cycles. Fortunately, this sub-system is well suited for ground testing by off-loading its mass in the 1g ground environment and cycling through the operation cycle in a large thermal-vacuum (T/V) chamber. The reliability of the drive train can be tested by itself over an accelerated period in a small T/V chamber for the full number of cycles anticipated for the mission. Testing will verify the performance of the design, workmanship and motor operations reducing the likelihood of malfunctions of the shutter during the mission. Final verification can be a full up cycle test in a T/V chamber together with deployment of the focal surface, corrector plate and light shield to verify the operation of the full mechanical system. 

There are engineering solutions to ensure the structural integrity of the instrument in the stowed and deployed positions will survive launch and meet observing requirements. Additional design effort is required to identify the best approach and can be addressed early in the preliminary design phase. 

\subsection{Optics}

The baseline POEMMA optics meets science requirements with no issue for manufacturing. There are a number of technology advances that can increase the performance or reduce risk that can further improve the design. 

Space qualified UV coatings resistant to atomic oxygen are under development for a number of projects. An effective coating on the POEMMA corrector plate would mitigate AO damage, enabling freer selection of the ToO observations. 

Of the coatings we have investigated, we concluded that Al2O3 is the best choice. An Al2O3 coating has been investigated by \cite{Cooper2008,Minton2010} These authors also describe coating PMMA using an approach based on atomic layer deposition (ALD). They report test results from atomic oxygen exposures showing that Al2O3 coatings provide excellent protection from atomic oxygen in LEO. Based on the testing that was done on the ground, it is our opinion that this coating is at or near TRL level 4. What remains is to investigate the surface roughness, transmission, and reflection losses of a PMMA sample coated with a sufficiently thick Al2O3 film to provide protection from AO during the POEMMA mission.  An SiO2 ALD coating is also a candidate for an AO protective coating.

An advanced split mirror design (bi-focal) for the  observations near the Earth's limb can enhance the ability to identify Cherenkov events by splitting the image between two pixels for an internal coincidence test to tag these events. This will further improve immunity to non-correlated spurious signals that may be generated in the focal surface SiPM sensors and front-end electronics. Discrimination of the Cherenkov events from the vast majority of recorded background luminous events is performed by a detailed time-amplitude analysis of two replicas of the event recorded by a unique combination of the bi-focal pair of pixels.
Performance of the bi-focal optics design will be tested on the EUSO-SPB2 balloon-born telescope \cite{Scotti:2020vpf}, where a Cherenkov camera will be operating at the fairly low threshold allowing the instrument to record a large amount of various events associated with SiPM noise, night sky air glow, and other stray light events such as lightning, airplanes, ground flashers, and other background sources.

A new optics design, POEMMA360, with a 360$^\circ$ azimuth FoV for limb observations is currently under study.  The goal is to significantly improve the sensitivity of the detection of the diffuse neutrino emission while also having the inherent ability to detect transient neutrino sources in a ToO mode with high sensitivity. 

In order to increase the light collection ability of the optics, a larger primary mirror would be needed. The desire to reduce the throughput falloff with increasing field of view, Figure \ref{fig9} (right) also pushes the mirror diameter to be larger.  The initial POEMMA optics design included a 6.7 m diameter mirror with twelve, rigid deployable petals, which are needed to accommodate the structure within the launch vehicle fairing. However, the mechanical requirements for the deployment of the mirror proved to be overly complex and led to a significant increase in mass and cost of the instrument. A new POEMMA optical design was developed that  assumed a monolithic fixed mirror, not one that deploys additional segments to increase the area.  This reduces instrument mass and the complexity of the instrument and fits inside the anticipated launch fairing but at the cost of vignetting.  One technology that could be employed to provide a larger lighter mirror that can fit in the launch vehicle and easily deploy is an inflatable membrane mirror 
\cite{Banik,MacEwen2013,Membrane,Freeland1998, Freeland1993, Freeland1996, Cortes-Medellin2016, Walker2014}.  

Fortunately, the tolerances on the mirror are relatively loose, which makes membrane mirror technology feasible for POEMMA.  The POEMMA instruments do not require diffraction limited mirrors. One can quickly estimate the surface accuracy necessary by setting the maximum allowed displacement of a ray at the focal surface, then determining the associated slope error at the mirror.   For example, for ray errors at the focal surface under 0.5 mm, slope errors at the mirror would be on the order of 1 arcminute.  This is not an unrealistic tolerance for membrane mirrors.  Additionally, coatings of $>$90\% average reflectivity are also practical.

Though additional engineering is required, there is a practical conceptual deployment concept for a membrane mirror based instrument.  The circular aperture mirror surface would be sealed to a semi-rigid circular ring, which would be sealed to another membrane, perhaps conical in shape, that seals to the edge of the corrector plate.  The mirror/ conical support would be folded for launch and deployment would be achieved through low pressure inflation of the mirror and structure. Additional research and engineering are needed to verify performance modeling and manufacturing limits.

\subsection{Focal surface: SiPMs}

The two POEMMA focal surface sections, the PFC and the PCC,  were studied during the IDL study. The IDL study  concluded that the MAPMT-based PFC section could be built with existing technology. The PCC SiPM-based section would benefit greatly from technology development in the following areas:
\begin{enumerate}
\item	Space qualification of appropriate SiPM arrays.

\item Packaging schemes for those arrays that allow the FSU to conform to the focal surface.

\item Further development and space flight qualification of the front-end ASICs that interface  with the SiPMs.

\item  Investigation of the diffusion of charge from the substrate to the surface in SiPMs. An energetic charged particle passing through the SiPM will create an ionized track extending into the substrate. Charge from this track can diffuse back to the surface causing multiple avalanche photodiodes within the struck pixel to fire, thus magnifying the background count rate from ionizing particles.

\item Investigation the use of SiPMs for UHECR fluorescence detection. The PDE of SiPMs now rival that of hi-QE PMTs in the $300\text{--}500$~nm range. However, the wavelength response of SiPM extends to $\sim$1000 nm, leading to huge increase in background rates due to the nature of dark-sky background \cite{Krizmanic:2020shl}. UV filters that only have bandpass in the $300\text{--}500$~nm nm range would eliminate the increase due the significant dark-sky airglow background above 500 nm.
\end{enumerate}

Investment in the development of large-scale SiPM arrays for space-based applications will benefit other future missions that depend on small, single photon sensitive, and fast photon detectors, e.g., gamma-ray instruments and instruments that study transient luminous events (TLE) in the upper atmosphere. All three of the development areas are therefore actively being studied in the laboratory and with sub-orbital testing. 

Members of POEMMA are currently participating in a NASA Super-Pressure Balloon program (SPB) with the EUSO collaboration. The first SPB flight (EUSO-SPB1) included an early prototype of a SiPM-based system. It provided extensive information about the operation of SiPM at the low temperatures and pressures of the upper atmosphere and also the background UV environment. Funding for a second EUSO-SPB2 flight has now been approved as part of the APRA program. EUSO-SPB2 is specifically planned to serve as a pathfinder for POEMMA. The EUSO-SPB2 payload, currently in preparation, will include two Schmidt telescopes. One will feature a focal surface instrumented with MAPMTs. The other will include a focal surface section that is tiled with SiPM arrays, which are read-out by full-functionality prototype FSUs. A separate 3-year APRA proposal was funded by NASA in 2019 to develop a Cherenkov Camera concept and prototype for EUSO-SPB2 and POEMMA. During those three years, the interface between SiPM arrays and ASICS will be developed and optimized for operation in the upper atmosphere and LEO. The prototype camera will undergo extensive laboratory testing and field tests with a pulsed laser~\cite{Hunt:2015a} and other light sources before integration into the EUSO-SPB2 gondola.

\subsection{Numerical Simulations}

We have simulated many aspects of the POEMMA science, but much still can be learned from end-to-end simulations of high-altitude and upward-moving EASs and further refining the associated POEMMA sensitivity. 
For the UHECR simulations, further development of EAS reconstruction in both the stereo and limb-viewing mode will improve the fidelity of quantifying POEMMA's performance, leading to potential improvements in UHECR EAS reconstruction and measurements. For example, including the EAS time development in the stereo reconstruction should improve stereo reconstruction efficiency, while also improving the angular and $X_{max}$ resolution as compared to that for the pure geometrical reconstruction we have initially used and present in this paper.  Further improvements in the modeling of the monocular performance are underway, including the improvements in the reconstruction of observed EASs using reduced time sampling. Our current modeling of the \tauon induced EAS Cherenkov signal uses a baseline end-to-end simulation. This is currently being improved under the $\nu$SpaceSim effort, a neutrino modeling APRA-funded program \cite{nuSpaceSim} that was initiated by this POEMMA work and has increased the fidelity of the modeling of the \tauon induced EASs.  These improvements in the fluorescence and Cherenkov signal modeling will help optimize the observing strategy including best Earth-observing angles, satellite separations, and ToO strategies.

Theoretical models of extremely energetic transient events are currently being developed by many authors given the recent capabilities in multi-messenger observations. A careful survey of the most likely neutrino emitters together with the brightest neutrino time-scales will help prioritize a sequence of ToO targets to maximize the discovery potential of POEMMA.

\section{Summary}

POEMMA is a unique NASA Probe-class mission designed to answer fundamental open questions in the multi-messenger domain over the next decade.  POEMMA can  enable charged-particle astronomy over the full sky by observing a large sample of UHECRs with POEMMA's ability to measure the UHECR composition, thus providing a critical understanding on the effects of magnetic deflection from the source(s). POEMMA will also potentially open the time-domain neutrino astronomy above 20 PeV, a current inaccessible energy range that will be probed with POEMMA's full-sky sensitivity.

POEMMA will:
\begin{enumerate}
    \item{Elucidate the origin of ultra-high-energy cosmic rays by measuring the composition, sky distribution, and energy spectrum of UHECRs at the highest energies using precision stereo fluorescence EAS measurements,} 
    \item{Have unprecedented all-flavor UHE neutrino sensitivity via stereo fluorescence measurements of the EASs sourced from charge-current and neutral current atmospheric neutrino interactions,}
    \item{Have the ability to increase by the exposure to UHECRs above 100 EeV via tilted-mode UHECR fluorescence observations,}
    \item{Have the ability to observe cosmic VHE cosmic neutrinos from transient astrophysical sources be measuring the beamed Cherenkov radiation from Earth-interacting  neutrinos via the induced upward EASs from \tauon decay,}
    \item{Study fundamental physics with the most energetic cosmic particles including the measurement of the proton-proton cross section at a center-of-mass energy of $\sim$320 TeV for $E_{\rm CR} = 50$ EeV,}
    \item{Search for signatures from Super-Heavy Dark Matter from VHE and UHE neutrinos and UHE photons,}
    \item{Have unparalleled exposure to atmospheric transient events, including TLEs, TEBS, etc, while also surveying meteor populations and having unparalleled sensitivity for the detection of nuclearites.}
    
\end{enumerate}

Community driven white-papers on the UHECR science \cite{Sarazin:2019fjz} and astrophysical neutrino science \cite{Ackermann:2019ows, Ackermann:2019cxh} describe the scientific challenges that POEMMA will address.
The multi-messenger domain began over the last decade and it will flourish during the 2020s with the capability provided by POEMMA. 

\acknowledgments

The conceptual design of POEMMA was supported by NASA Probe Mission Concept Study grant NNX17AJ82G for the 2020 Decadal Survey Planning. Contributors to this work were supported in part by NASA awards 16-APROBES16-0023, 17-APRA17-0066, NNX17AJ82G, NNX13AH54G, 80NSSC18K0246, 80NSSC18K0473, 80NSSC19K0626, and 80NSSC18K0464, US Department of Energy grant DE-SC0010113, and the U.S. National Science Foundation (NSF Grant PHY-1620661).
MB is supported by the {\sc Villum Fonden} under project no.~29388.
The authors from University of Torino acknowledge support from Compagnia di San Paolo within the project 'ex-post-2018.
The Czech authors are supported by the Ministry of Education, Youth and Sports of the Czech Republic.
GMT is supported by PAPIIT IN111621/DGAPA/UNAM.

\section{APPENDIX I: Acronym List}

\noindent AMS: Atmospheric Monitoring System 



\noindent ASIC: Application-Specific Integrated Circuit

\noindent DP: Data Processor 

\noindent EAS: Extensive Air Shower

\noindent EC: Elementary Cells 

\noindent EFL: Effective Focal Length

\noindent EPD: Effective Pupil Diameter 

\noindent FPGA: Field Programmable Gate Array 

\noindent FSU: Focal Surface Units 


\noindent GTU: Gate Time Units 



\noindent JEM-EUSO: Joint Experiment Missions for Extreme Universe Space Observatory 


\noindent LEO: Low Earth Orbit

\noindent MAPMT: Multi-Anode Photo-Multiplier Tubes

\noindent Mini-EUSO: Multi-wavelength Imaging New Instrument for Extreme Universe Space Observatory 


\noindent PCC: POEMMA Cherenkov Camera

\noindent PDE: Photon Detection Efficiency 

\noindent PFC: POEMMA Fluorescence Camera


\noindent PSF: Point Spread Function


\noindent SiPM: Silicon Photomultipliers 

\noindent ToO: Target of Opportunity

\noindent TRL: Technical Readiness Level

\section{APPENDIX II: Optics Specification}

Table~\ref{OpticsTable} details the parameters of the optical design of the POEMMA Schmidt telescopes.

\begin{table}
\centering
\caption{{\bf Optics Specifications:}}
\label{tab-2}  \
\begin{tabular}{lll}
\hline
\hline
R & = spherical base radius of curvature \\
k & = conic constant \\
A to F & =  4th to 14th rotation symmetric polynom. coeff. \\
\hline
{\bf Surface diameters} & and relative locations (vertex distances)  \\
\hline
Stop Diameter & = 3,300 mm\\
Stop to Corrector & =  889.6 mm\\
Corrector center thickness &  = 12.25 mm\\
Corrector diameter & = 3,300 mm\\
Corrector to Mirror & =  4,032.8 mm\\
Mirror diameter & = 4,000 mm\\
Corrector to Focal Surface & =  2,105.9 mm\\
Focal Surface to Mirror & = 1,926.9 mm\\
Focal Surface diameter & = 1,610 mm \\
\hline
{\bf Surface Form} & \\
\hline
External Corrector Surface: & \\
\hline
R & =1,825.7 mm \\
A & = -5.97872 $10^{-10}$/mm$^3$ \\
B & = 1.06661 $10^{-15}$/mm$^5$ \\
C & = -9.89055 $10^{-22}$/mm$^7$ \\
D & = 4.62547 $10^{-28}$/mm$^9$ \\
E & = -1.07407 $10^{-34}$/mm$^{11}$ \\
F & =  9.81265 $10^{-42}$/mm$^{13}$ \\
\hline
Internal Corrector Surface: & \\
\hline
R & = 1,897.1 mm \\
A & = -6.01297 $10^{-10}$/mm$^3$  \\
B & =  1.08541 $10^{-15}$/mm$^5$  \\
C & = -1.00484 $10^{-21}$/mm$^7$ \\
D & =  4.71169 $10^{-28}$/mm$^9$ \\
E & = -1.09889 $10^{-34}$/mm$^{11}$ \\
F & =  1.01227 $10^{-41}$/mm$^{13}$ \\
Edge thickness of corrector &   $\sim$ 34 mm      \\
\hline
Mirror &\\
\hline   
R = & 3,991.7 mm  \\
k = & 1.77565 $10^{-2}$ \\
\hline
Focal Surface  & \\
\hline 
R = &2,074.4 mm \\
\hline
\hline
\label{OpticsTable}
\end{tabular}
 \end{table}

\vfill
\eject

\end{document}